\documentclass[preprintnumbers, floatfix, letterpaper, twocolumn,aps,prd,epsfig,nofootinbib,natbib,longbibliography]{revtex4-1}

\usepackage{graphicx}
\usepackage{epstopdf}
\usepackage{latexsym}
\usepackage{amssymb}
\usepackage{amsmath}
\usepackage{color}
\usepackage{mathrsfs}
\usepackage{xparse}
\usepackage{bbding}
\usepackage{enumitem}
\usepackage{pifont}
\usepackage[center]{subfigure}
\usepackage[
            pdfstartview=FitH,
            bookmarksnumbered=true,
            bookmarksopen=true,
            colorlinks,
            linkcolor=blue,
            anchorcolor=green,
            citecolor=blue
            ]{hyperref}
\begin{document}

  \renewcommand\arraystretch{2}
 \newcommand{\bq}{\begin{equation}}
 \newcommand{\eq}{\end{equation}}
 \newcommand{\bqn}{\begin{eqnarray}}
 \newcommand{\eqn}{\end{eqnarray}}
 \newcommand{\nb}{\nonumber}
 \newcommand{\lb}{\label}
 \newcommand{\cb}{\color{blue}}
    \newcommand{\cc}{\color{cyan}}
        \newcommand{\cm}{\color{magenta}}
\newcommand{\rc}{\rho^{\scriptscriptstyle{\mathrm{I}}}_c}
\newcommand{\rd}{\rho^{\scriptscriptstyle{\mathrm{II}}}_c} 
\NewDocumentCommand{\evalat}{sO{\big}mm}{%
  \IfBooleanTF{#1}
   {\mleft. #3 \mright|_{#4}}
   {#3#2|_{#4}}%
}
\newcommand{\PRL}{Phys. Rev. Lett.}
\newcommand{\PL}{Phys. Lett.}
\newcommand{\PR}{Phys. Rev.}
\newcommand{\CQG}{Class. Quantum Grav.}


\title{Properties of the spherically symmetric  polymer black holes}

\author{Wen-Cong Gan${}^{a}$}
\email{Wen-cong$\_$Gan1@baylor.edu}

\author{Nilton O. Santos${}^{b}$}
\email{Nilton.Santos@obspm.fr}

\author{Fu-Wen Shu${}^{a, c, d}$}
\email{shufuwen@ncu.edu.cn}

\author{Anzhong Wang${}^{a}$ \footnote{Corresponding author}}
\email{Anzhong$\_$Wang@baylor.edu; Corresponding author}

\affiliation{${}^{a}$ GCAP-CASPER, Physics Department, Baylor University, Waco, Texas 76798-7316, USA\\
${}^{b}$ Sorbonne Universit\'e, UPMC Universit\'e Paris 06, LERMA, UMRS8112 CNRS, Observatoire de Paris-Meudon, 5, Place Jules Janssen, F-92195 Meudon Cedex, France\\
 ${}^{c}$ Department of Physics, Nanchang University,
No. 999 Xue Fu Avenue, Nanchang, 330031, China\\
${}^{d}$ Center for Relativistic Astrophysics and High Energy Physics, Nanchang University, No. 999 Xue Fu Avenue, Nanchang 330031, China}

\date{\today}

\begin{abstract}

 In this paper we systematically study a recently proposed  model of  spherically symmetric polymer  black/white holes by Bodendorfer, Mele, and M\"unch (BMM), which generically possesses  five free parameters. However, we find that, out of these five parameters, only three independent combinations of them are physical and uniquely determine the local and global properties of the spacetimes. After exploring the whole 3-dimensional (3D) parameter space, we show that the model has very rich physics, and depending on the choice of these parameters, various possibilities exist, including: (i) spacetimes that have the standard black/white hole structures, that is, spacetimes that are free of spacetime curvature singularities and possess  two asymptotically flat regions, which are connected by a transition surface (throat) with a finite and nonzero geometric radius. The black/white hole masses measured by observers in the two asymptotically flat regions are all positive, and the surface gravity of the black (white) hole is positive (negative). In this case, there also exist possibilities in which the two horizons coincide, and the corresponding surface gravity vanishes identically. (ii) Spacetimes that have wormholelike structures, in which the two masses measured in the two asymptotically flat regions are all positive, but no horizons exist, neither a trapped (black hole) horizon nor an anti-trapped (white hole) horizon. (iii) Spacetimes that still possess curvature singularities, which
can be either hidden inside  trapped regions or naked. However, such spacetimes correspond to only some limit cases. In particular,  the necessary (but not sufficient) condition is that at least one of the two ``polymerization" parameters vanishes.  These results are not in conflict to the Hawking-Penrose singularity theorems, as the effective energy-momentum tensor, purely geometric and resulted from the ``polymerization'' quantization, satisfies none of the three (weak, strong or dominant) energy conditions in any of the two asymptotically flat regions for any choice of the three independent free parameters, although they can hold at the throat and/or at the two horizons for some particular choices of them. In addition, it is true that quantum gravitational effects are mainly concentrated in the region near the throat, however, in this model even for solar mass black/white holes, such effects can be still very large at the black/white hole horizons, again depending on the choice of the parameters. Moreover, in principle the ratio of the two masses (for both of the black/white hole and wormhole spacetimes) can be arbitrarily large.

\end{abstract}

\maketitle

\section{Introduction}
 \renewcommand{\theequation}{1.\arabic{equation}}\setcounter{equation}{0}
 
There are few beacons on the road to the quantum theory of gravity. Among them singularities in classical general relativity (GR) are always the key one that any quantum theory of gravity needs to address  properly. It is generally believed that spacetime singularities can be resolved once quantum gravity effects are taken into considerations.  One of most successful and heuristic examples is the resolution of the big bang singularity in cosmology with the use of an effective tool developed by loop quantum cosmology (LQC) in  {the} past few years (see e.g. \cite{Singh:2009mz,Ashtekar:2011ni}).

Inspired by the remarkable achievements made in LQC, attempts to extend the approaches developed in LQC to black hole singularities, the ones inside black hole interiors, have recently attracted considerable attention in the loop quantum gravity (LQG) community, see, for example, \cite{Ashtekar:2005qt,Modesto:2005zm,Bohmer:2007wi,Campiglia:2007pb,Brannlund:2008iw,Modesto:2008im,Chiou:2008nm,Chiou:2008eg,GP08,GP13,Joe:2014tca,Corichi:2015xia,Dadhich:2015ora,Olmedo:2017lvt,Cortez:2017alh,AP17,BMM18,BCDHR18,AOS18a,AOS18b,CR18,MBM19,Alesci:2019pbs,Assanioussi:2019twp,Bodendorfer:2019nvy,CDLV19,AAN20,AO20,ZMSZ20,GOP20,KSWe20,Liu:2020ola,Agullo20,Bojowald:2018xxu}  and references therein  {(See also \cite{CR17,RMD18,MDR19} for a somehow different approach).} Among these studies, most attentions was paid to the Schwarzschild black hole. This is on one hand because it is the simplest black hole in GR, and on the other hand it is because the interior of the Schwarzschild black hole is isometric to the Kantowski-Sachs cosmological model.  Actually, it is this similarity that stimulates ones to borrow similar techniques from LQC to deal with singularities in the Schwarzschild black hole.
In fact, the Kantowski-Sachs spacetime can be written in the form, 
\bq
\lb{eq1.1}
ds^2 = - N_{\tau}^2d\tau^2 + \frac{p_b^2}{\left|p_c\right|L_0^2} dx^2 + \left|p_c\right|d^2\Omega,
\eq
where $d^2\Omega \equiv d^2\theta + \sin^2\theta d^2\phi$, and  $L_0$ is the length of the fiducial cell with $x \in(0, L_0)$. The quantities $b, \; c, \; p_b$ and $p_c$ represent the dynamical variables with the commutation relations,
\bq
\lb{eq1.2}
\left\{c, p_c\right\} = 2 G \gamma, \quad \left\{b, p_b\right\} = G \gamma,
\eq
where $b$ and $c$ are the conjugate momenta of the canonical variables $p_b$ and $p_c$, 
$G$ denotes the Newtonian constant, and $\gamma$ the Barbero-Immirzi parameter, arising in the passage from classical to quantum theory. Its value is generally fixed to be $\gamma \simeq 
0.2375$ using black hole entropy considerations \cite{KM04}. Choosing (classically) the lapse function $N_{\tau}$ as
\bq
\lb{eq1.3}
N^{\text{cl}}_{\tau} = \frac{\gamma \; \text{sgn}\left(p_c\right)\left|p_c\right|^{1/2}}{b},
\eq
the corresponding Hamiltonian is given by
\bqn
\lb{eq1.4}
H_{\text{cl}} = - \frac{1}{2G\gamma}\left[2cp_c + \left(b + \frac{\gamma^2}{b}\right)p_b\right].
\eqn

A key procedure in constructing effective quantum geometry which solves the classical singularity is the so-called ``polymerization'' \cite{Thiemann08} in the LQG literature, which is characterized by two quantum parameters $\delta_b$ and $\delta_c$ for spherical spacetimes \cite{Ashtekar20}. It is related to the fact that in LQG, there is minimal area gap  {$\Delta_{pl}$, which is nonzero and given by $\Delta_{pl} \equiv 4\sqrt{3} \pi \gamma \ell^2_{Pl}$, } where $ \ell_{Pl}$ denotes the Planck length. The basic idea is that the effective quantum theory can be achieved by replacing the canonical variables ($b, c$) in the phase space with their regularized ones, 
\bqn
\lb{eq1.5}
b &\rightarrow&\frac{\sin(\delta_b b)}{\delta_b},\quad c \rightarrow\frac{\sin(\delta_c c)}{\delta_c},
\eqn
where $\delta_b$ and $\delta_c$ are the so-called  ``polymerization scales,'' which control the onset of quantum effects. With the above replacement, the effective  Hamiltonian is given by   \cite{AOS18b},
\bqn
\lb{eq1.6}
H_{\text{eff}} &=& - \frac{1}{2G\gamma}\Bigg[2 \frac{\sin(\delta_c c)}{\delta_c}\left|p_c\right| \nb\\
&& + \left(\frac{\sin(\delta_b b)}{\delta_b} + \frac{\gamma^2 \delta_b}{\sin(\delta_b b)}\right)p_b\Bigg].
\eqn
Clearly,  as $\delta_b$ and $\delta_c$ approach $0$, the classical limit is recovered. When quantum effects are supposed to become relevant, the above replacement effectively cures the classical divergence, suggesting that the polymerization scales are at the Planck one.

However, due to the lack of the full theory of quantum gravity, a complete  route map on the choice of $\delta_b$ and $ \delta_c$ is still absent. In the literatures there are many different choices. Generally speaking they can be divided into the following three broad classes:

\begin{itemize}
 
\item $\mu_0$-scheme: In this approach, the two quantum parameters $\delta_b$ and $ \delta_c$ are simply taken as constants. This is the case studied, for example, in  \cite{Ashtekar:2005qt,Modesto:2005zm,Modesto:2008im,Assanioussi:2019twp,Bodendorfer:2019nvy,Campiglia:2007pb}.

\item \textit{Generalized} $\mu_0$-scheme: In this approach, the quantum parameters $\delta_b$ and $\delta_c$ are considered as the Dirac observables, i.e., they are phase space variables, but are constants along the effective trajectories
of the system  \cite{AOS18a,AOS18b,AO20,Corichi:2015xia,Olmedo:2017lvt}.

\item $\bar \mu$-scheme: In this approach, the two quantum parameters $\delta_b$ and $\delta_c$ are the phase space functions, and  their functional dependence on the canonical variables depends on the specific ways to carry out the quantization, which have been explored  in detail in \cite{Chiou:2008nm,Bohmer:2007wi,Brannlund:2008iw,Dadhich:2015ora,Cortez:2017alh,Chiou:2008eg,Joe:2014tca,Alesci:2019pbs}.

\end{itemize}

  \begin{widetext}
 
  \begin{table}
\caption{\label{tab:table-mu}%
Three broad classes of the choices of the quantum parameters $\delta_b, \delta_c$. The parameters $\alpha$ and $\beta$ are constants with the values $\alpha=1$ or $\beta=1$ or $\alpha \beta=1$. $r_0$ is the Schwarzschild radius, $L_0$ is the fiducial length to be specified. $\gamma$ is the Barbero-Immirzi parameter, $\Delta_{pl} =4\sqrt{3}\pi \gamma l_{pl}^{2}$ is LQG area gap. $m=GM$, where $M$ is the (classical) Schwarzschild black hole mass.  }
\begin{tabular}{|l|c|c|c|c|c|}  \hline
\multicolumn{2}{|c|}{\bf  $\mu_0$ scheme}
&\multicolumn{2}{|c|}{\bf \textit{generalized} $\mu_0$ scheme} &\multicolumn{2}{|c|}{\bf  $\bar \mu$ scheme}\\  \hline
 \multicolumn{2}{|c|}{$\delta_b,\; \delta_c =$ Const.} &$\delta_b=\alpha \frac{\sqrt{\Delta_{pl}}}{r_0}$, $\delta_c=\beta \frac{\sqrt{\Delta_{pl}}}{L_0}$& $\delta_b=( \frac{\sqrt{\Delta_{pl}}}{\sqrt{2\pi}\gamma^2 m})^{1/3}$, $\delta_c=\frac{1}{2}( \frac{\gamma\Delta_{pl}^2}{4\pi^2 m})^{1/3} $ &   $\delta_b= \frac{\sqrt{\Delta_{pl}}}{p_b}$, $\delta_c= \frac{\sqrt{\Delta_{pl}}}{p_c}$     &    $\delta_b= \frac{\sqrt{\Delta_{pl}}}{p_c}$, $\delta_c= \frac{\sqrt{p_c \Delta_{pl}}}{p_b}$ \\ \hline
 \multicolumn{2}{|c|}{\cite{Ashtekar:2005qt,Modesto:2005zm,Modesto:2008im,Assanioussi:2019twp,Bodendorfer:2019nvy,Campiglia:2007pb} }  & \cite{Corichi:2015xia,Olmedo:2017lvt}  & \cite{AOS18a,AOS18b,AO20}  & \cite{Chiou:2008nm,Chiou:2008eg}  & \cite{Chiou:2008nm,Bohmer:2007wi,Brannlund:2008iw,Dadhich:2015ora,Cortez:2017alh,Chiou:2008eg,Joe:2014tca}     \\ \hline
\end{tabular}
\lb{Table3}
\end{table}
 
  \end{widetext}

 Table \ref{tab:table-mu} summarizes these studies.
 In some of these schemes, the fiducial structure may appear in the final results \cite{Ashtekar:2005qt,Modesto:2005zm,Campiglia:2007pb}, while in other approaches, the quantum effects could be large even in the semiclassical region \cite{Corichi:2015xia,Olmedo:2017lvt,Chiou:2008nm,Bohmer:2007wi,Cortez:2017alh}.   See \cite{Ashtekar20,AO20,Bouhmadi-Lopez:2019hpp,Bojowald:2019dry,Bojowald20} for the debates over these  issues.

In addition, 
in classical Hamiltonian mechanics, a canonical transformation 
\bq
\lb{eq1.7a}
(q_i,\; p_i) \rightarrow (Q_i,\; P_i),
\eq
 is always allowed, and does not change the physics of the system,
where $Q_i = Q_i(q_k, p_k; t)$, $P_i = P_i(q_k, p_k; t)$,  $q_i = (b, c)$, and $p_i = (p_b, p_c)$ \cite{GPS02}.  However,   the 
polymerization (\ref{eq1.5}) depends on the choice of the canonical variables, and different canonical variables in general lead to different effective theories.
It was exactly along this vein, Bodendorfer, Mele, and M\"unch (BMM)  considered the following transformation \cite{BMM19,BMM20}, 
\bq
\lb{eq1.7}
v_1 \equiv \frac{1}{24}\left|p_c\right|^{3/2}, \quad v_2 \equiv -\frac{1}{8} p_b^2,
\eq
for which the corresponding conjugate momenta are denoted by $P_1$ and $P_2$, respectively.   Then, instead of Eq.(\ref{eq1.5}), now the 
polymerizations are  carried out via the replacements \cite{BMM19},
\bqn
\lb{eq2.12}
P_1 &\rightarrow&\frac{\sin(\lambda_1 P_1)}{\lambda_1},\quad P_2 \rightarrow\frac{\sin(\lambda_2 P_2)}{\lambda_2},
\eqn
where $\lambda_1$ and $\lambda_2$ play the same role as   $\delta_b$ and $\delta_c$.    
 In this approach, the polymerization scales ($\lambda_1, \lambda_2$) are taken as constants, but as pointed out in \cite{BMM20}, 
 this choice of polymerization scales does not correspond to $\mu_0$-scheme in terms of the variables ($b, \; c$), instead, when translated back to ($b, \; c$), they correspond to a specific $\bar \mu$-scheme. 

 It must be noted that the BMM model is based on a set of new canonical variables ($v_i, P_i$). Although the canonical transformation (\ref{eq1.7a}) is always allowed classically,   the corresponding loop quantization has not been carried out yet in terms of these new   variables.  As a result,  it is not clear what are relations of such effective theory  [obtained  by simply the replacement of  Eq.(\ref{eq2.12})] to LQG. Therefore, to be distinguished with the effective theory obtained from LQG by taking only  the  leading order of quantum corrections into account, we refer such black holes as  polymer black holes. Additional  questions related  to  this issue can be found  from \cite{Ashtekar:2005qt,APSV07}.

 {With the above caveat in mind, in this paper, we shall systematically study the local and global properties of the model proposed in \cite{BMM19}. In  particular, we find that, out of the five parameters appearing in the model, 
only three independent combinations  of them are physically relevant, and uniquely determine the properties of the spacetimes. 
In this 3D phase space, there exist regions, in which the solutions can represent two asymptotically flat regions connected 
by a throat with a finite and nonzero geometric radius, and the masses read off in these two asymptotically flat regions are all positive. In such case, a black/white horizon  exists or not also depending on the choice of the
three free parameters. When they do exist, the surface gravity at the black (white) hole horizon can be  positive (negative). When they do not exist,   the
spacetimes have wormhole structures. In all these solutions, spacetime curvature singularities are absent, which does not contradict to the Hawking-Penrose singularity theorems \cite{HE73}, as now
the effective energy-momentum tensor does not satisfy any   of the three  energy conditions in the two asymptotically flat regions, despite the fact that the masses measured by observers in these two asymptotical  regions are all positive. 
This is mainly due to the fact that the relativistic Komar energy density \cite{Komar59} is still positive in a large region of the spacetime. 
The violation of the three energy conditions in the  asymptotically flat regions is a generic feature of the model, independent of the choice of the parameters of the solutions. 
Spacetime curvature singularities can occur, but  the necessary (not sufficient) condition is  at least    one of the two ``polymerization" parameters vanishes. 
 In addition, although  it is true that quantum gravitational effects are mainly concentrated in the region near the throat,  in this model such effects still can be very large at the black/white 
hole horizons even for solar mass black/white holes, again depending on the choice of the free parameters. Moreover,  in principle the ratio of the two masses (for both of the black/white 
hole and wormhole spacetimes) can be arbitrarily large. }

It should be noted that,  despite the fact that in this paper we consider only a particular model, we believe the main conclusions should hold for more general cases.
 In particular, the Schwarzschild solution is the unique vacuum solution
of GR with a single parameter---the black hole mass, according to the Birkhoff theorem \cite{Birkhoff23}. However, due to the polymerization process, two more free parameters, $\delta_b$ and $\delta_c$ (or in the present case, $\lambda_1$ and  $\lambda_2$), are introduced. So, 
the resulted spacetimes should be characterized physically by only  three  free parameters,  {although the two polymerization parameters   may be completely fixed, when the quantization is
carried out explicitly, such as in the case considered in \cite{AOS18a,AOS18b}. } Clearly, in order for this to be  consistent with the Birkhoff theorem, effective matter must be present, purely due to the quantum geometric 
effects. In addition, to be in harmony with the Hawking-Penrose singularity theorems \cite{HE73}, the effective energy-momentum tensor necessarily violates the weak/strong energy conditions.

The rest of the paper is organized as follows: In Sec. \ref{Section.II}, we first review the model built in  \cite{BMM19} and then write the corresponding solutions in terms of only three  independent combinations of the original five parameters, which are denoted by ${\cal{D}},\; {\cal{C}}, \; x_0$, defined explicitly in Eq.(\ref{eq2.3}). Then, we study the model in detail over the 
whole parameter space in Secs. \ref{Section.III} - \ref{Section.V}, respectively, for $\Delta > 0$, $\Delta = 0$, and $\Delta < 0$, as in each case the spacetimes have quite different properties, where $\Delta$ is defined by Eq.(\ref{eq2.3b}). 
The main results in each of these sections are summarized, respectively,  in  Tables \ref{Table1} - \ref{Table3}.  {The paper is ended} up in Sec. \ref{Section.VI}, in which we summarize our main conclusions. 
An appendix is also included, in which the general expressions of the energy density and pressures of the effective energy-momentum tensor are given explicitly.

 Before proceeding further, we would like to point out that some  solutions of the current model were lately studied in \cite{BL20}, including their perturbations and the associated quasinormal modes of 
 massless scalar field perturbations, electromagnetic perturbations, and axial gravitational perturbations.  {In particular, the authors }  found that    the corresponding 
 quasinormal frequencies of perturbations with different spins share the same qualitative
tendency with respect to the change of the quantum parameters involved  in this model. For more details, we refer readers to \cite{BL20}.

\section{Spherically symmetric polymer  black holes}
 \label{Section.II}
 \renewcommand{\theequation}{2.\arabic{equation}}\setcounter{equation}{0}

Studying spherically symmetric spacetimes inside black holes, Bodendorfer, Mele, and M\"unch recently obtained the following spherically 
symmetric black hole solutions \cite{BMM19},
\begin{equation}
\label{eq2.1}
d\bar s^2= - \frac{\bar a(x)}{L_0^2}d\bar t^2+ \frac{{\cal{L}}_0^2}{\bar a(x)} dx^2 + \bar b^2(x)d\Omega^2,
\end{equation}
where   {${\cal{L}}_0 = \sqrt{n}$, } $x \in (-\infty, \infty)$, and
\begin{eqnarray}
\label{eq2.2}
\bar a(x)&=&n\left(\frac{\lambda_2}{\sqrt{n}}\right)^4 \left(1 + \frac{n x^2}{\lambda_2^2}\right)\left(1 - \frac{3CD}{2\lambda_2\sqrt{1 + \frac{n x^2}{\lambda_2^2}}}\right)\nonumber\\
&& \times \left[\frac{\lambda_2^6}{16C^2\lambda_1^2n^3}\left(\frac{\sqrt{n} x}{\lambda_2} + \sqrt{1 + \frac{n x^2}{\lambda_2^2}}\right)^6 + 1\right]^{-2/3} \nb\\
&& \times \left(\frac{1}{3C^2D\lambda_1^2}\right)^{2/3}\left(\frac{\sqrt{n} x}{\lambda_2} + \sqrt{1 + \frac{n x^2}{\lambda_2^2}}\right)^2,\nb\\
\bar b(x)&=&\frac{\sqrt{n}\left(3C^2D\lambda_1^2\right)^{1/3}}{\lambda_2}\nb\\
&& \times \frac{ \left[\frac{\lambda_2^6}{16C^2\lambda_1^2n^3}\left(\frac{\sqrt{n} x}{\lambda_2} + \sqrt{1 + \frac{n x^2}{\lambda_2^2}}\right)^6 + 1\right]^{1/3}}
      {\frac{\sqrt{n} x}{\lambda_2} + \sqrt{1+ \frac{n x^2}{\lambda_2^2}}}, 
\end{eqnarray}
where $\lambda_1, \; \lambda_2, \; n, \; C$ and $D$ are real constants with $n > 0$.  

As shown in \cite{BMM19}, there are two  independent Dirac observables, $F_Q$ and $\bar{F}_Q$, which are constants
along the trajectories of the effective dynamics, and their on-shell values are given by,
\bqn
\lb{eq2.9}
F_Q &=& \left(\frac{3D}{2}\right)^{4/3} \left(\frac{C}{\sqrt{n}}\right), \nb\\
 \bar{F}_Q &=& \frac{3CD\sqrt{n}}{\lambda_2^2}\left(3DC^2\lambda_1^2\right)^{1/3}.
\eqn
It can be shown that both of them are invariant under a fiducial cell rescaling. As a result, the integration constants $C$ and $D$ are independent. In fact, at the limits, $ x \rightarrow \pm \infty$, we have
\bq
\lb{eq2.10}
\bar a(x) \propto \begin{cases}
1 - \frac{F_Q}{\bar b}, & x  \rightarrow  \infty,\cr
1 - \frac{\bar{F}_Q}{\bar b}, & x  \rightarrow  -\infty.\cr
\end{cases}
\eq
Thus, they are essentially related to the black and white hole masses via the relations,
\bqn
\lb{eq2.11}
\bar M_{BH} &=& \frac{1}{2}F_Q = \left(\frac{3D}{2}\right)^{4/3} \left(\frac{C}{2\sqrt{n}}\right),\nb\\
 \bar M_{WH} &=&  \frac{1}{2} \bar{F}_Q = \frac{3CD\sqrt{n}}{2\lambda_2^2}\left(3DC^2\lambda_1^2\right)^{1/3}.
\eqn

 Introducing the quantities,
\bqn
\lb{eq2.3}
 {\cal{D}} &\equiv& \frac{3CD}{2\sqrt{n}}, \quad  {\cal{C}} \equiv \left(16C^2\lambda_1^2\right)^{1/6}, \quad x_0 \equiv \frac{\lambda_2}{\sqrt{n}},
\eqn
we find that the metric (\ref{eq2.1}) takes the form,  
\bqn
\lb{eq2.4}
d\bar s^2&=& \left(\frac{3D}{16}\right)^{2/3}d{s}^2   \nb\\
&\equiv&  \left(\frac{3D}{16}\right)^{2/3}  \left(- {a}(x) d{t}^2+ \frac{dx^2}{{a}(x)} + {b}^2(x)d\Omega^2\right), ~~~~
\eqn
with $t \equiv (\sqrt{n}/{L_0})\left(16/3D\right)^{2/3} \bar t$, and 
\bqn
\lb{eq2.5}
{a}(x) = \frac{\left(x^2 - \Delta\right)XY^2}{\left(X + {\cal{D}}\right)Z^2},\quad {b}(x) =  \frac{Z}{Y},
\eqn
where 
\bqn
\lb{eq2.6}
X &\equiv& \sqrt{x^2 + x_0^2}, \quad Y \equiv x + X, \nb\\
Z &\equiv& \left(Y^6 + {\cal{C}}^6\right)^{1/3}, 
\eqn
and 
\bqn
\lb{eq2.3b}
 \Delta &\equiv& {\cal{D}}^2 - x_0^2 = \frac{9C^2D^2 - 4\lambda_2^2}{4n}.
\eqn

 \begin{figure}[h!]
\includegraphics[height=4.8cm]{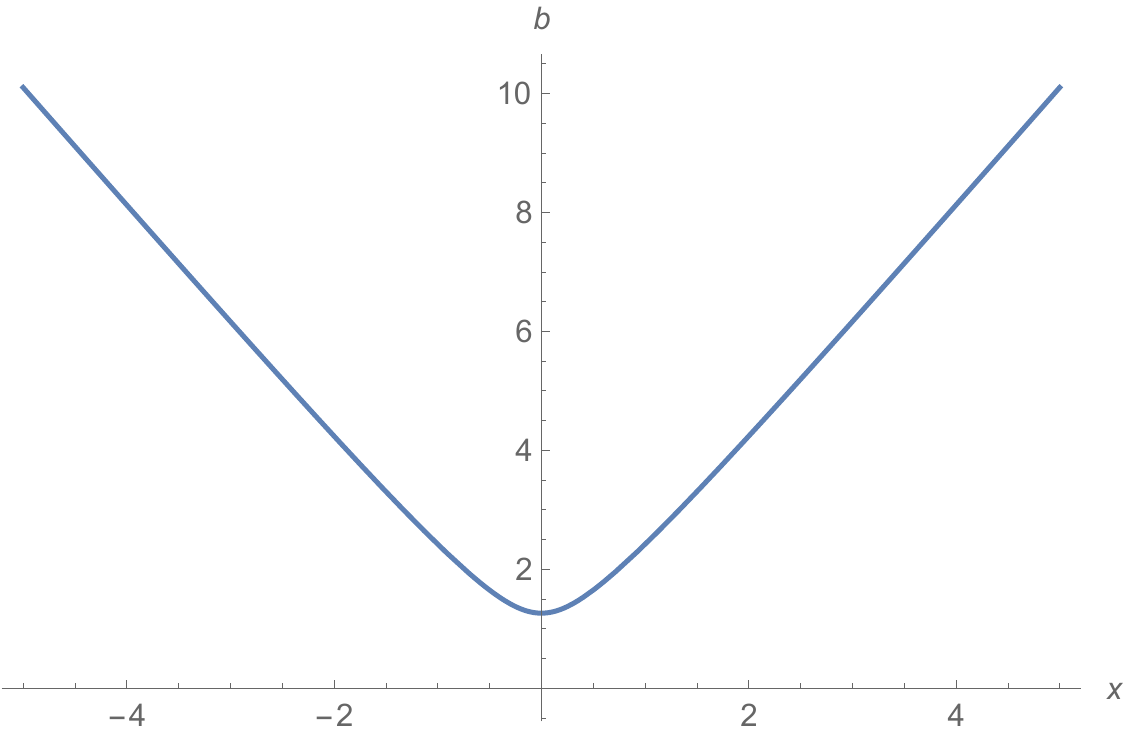}\\
{(a)}\\
\vspace{.5cm}
\includegraphics[height=4.8cm]{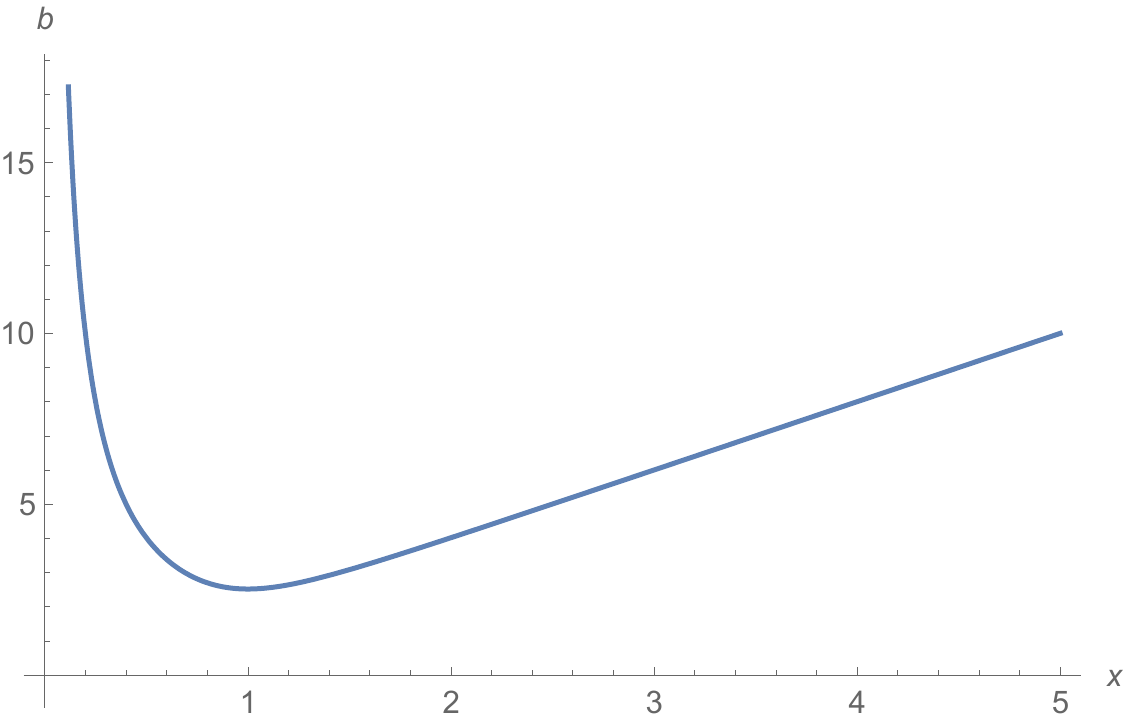}\\
{(b)}\\
\vspace{.5cm}
\includegraphics[height=4.8cm]{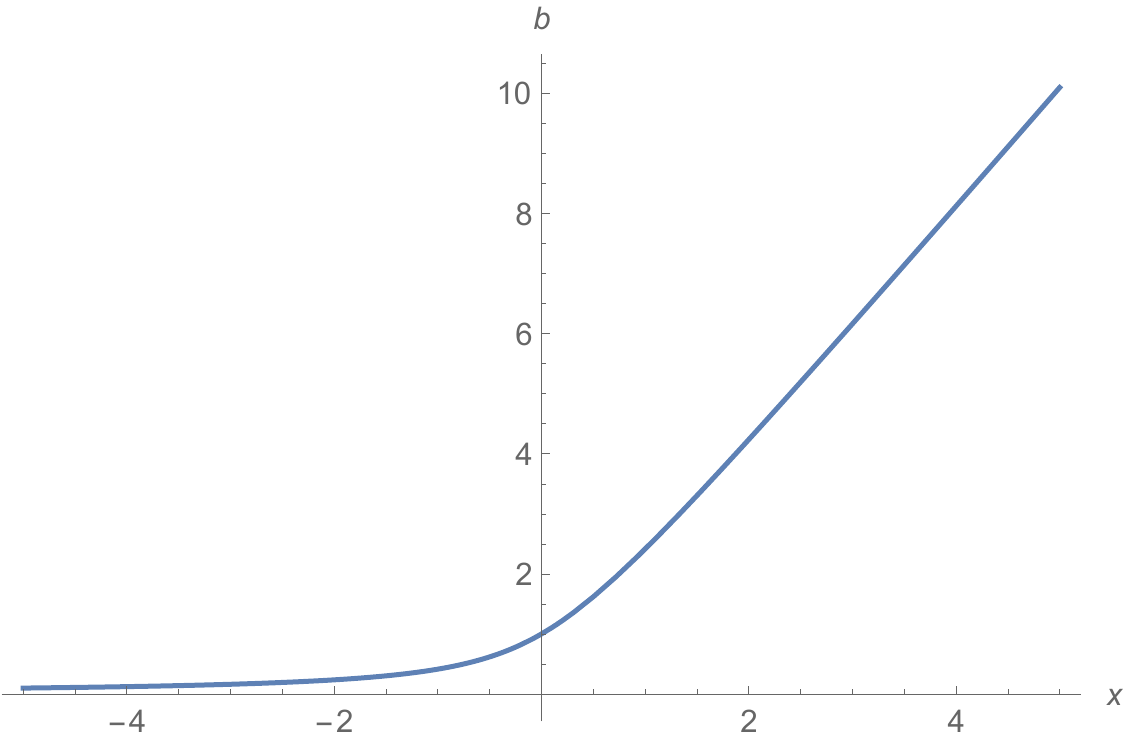}\\
{(c)}\\
\vspace{.5cm}
\caption{The  geometric radius $b(x)$ vs $x$. (a) Upper panel: ${\cal{C}} x_0 \not = 0$. When plotting this curve, we had set  $x_0=1, \; \mathcal{C}=1$.
 (b) Middle panel:   ${\cal{C}} \not=0,\;  x_0 = 0$. When plotting this curve, we had set 
 ${{\mathcal{C}}} =2$. (c) Bottom panel:  ${\cal{C}} =0,\;  x_0 \not = 0$. When plotting this curve, we had set $x_0=1$. } 
\label{fig1a-c}
\end{figure}

Since $ds^2$ is related to $d\bar s^2$ only by a conformal constant factor $\left({3D}/{16}\right)^{2/3}$ \footnote{Under this rescaling, 
the Ricci and Kretschmann scalars are scaling, respectively,  as $R = \left({3D}/{16}\right)^{2/3}\bar{R}$ and $K = \left({3D}/{16}\right)^{4/3}\bar{K}$.}, 
without loss of generality, we shall consider only the spacetimes described by $ds^2$.
Then, we can see that only three independent combinations of the five parameters $\lambda_1, \; \lambda_2, \; n, \; C$, and $D$ 
appear in the metric coefficients, as defined  by Eq.(\ref{eq2.3}). 

 It is remarkable to note that in GR, due to the Birkhoff theorem \cite{Birkhoff23}, the black hole mass is the only free parameter. However, in LQG,
 due to the  polymerizations (\ref{eq2.12}),  two new parameters $\lambda_1$ and $\lambda_2$ are introduced, so now the solutions generically depend on three free parameters.
 When setting $\lambda_1 = \lambda_2 =0$ (or ${\cal{C}} = x_0 = 0$), the above solutions reduce precisely to the Schwarzschild solution with ${\cal{D}}$ as the black hole mass. 

One of our goals in this paper is to understand their physical and geometrical meanings. 
To this goal, let us first note the following: 

\begin{itemize}

\item  Since $x \in (-\infty, \infty)$, from Eq.(\ref{eq2.6}) we find that 
\bq
\lb{eq2.7}
X \ge x_0, \quad Y > 0, \quad Z > {\cal{C}}^2.
\eq

\item In \cite{BMM19,BMM20} it was assumed that 
\bq
\lb{eq2.7a}
{\cal{D}} > 0, \quad \Delta > 0, 
\eq
so that two metric horizons always exist at $x^{\pm}_H \equiv \pm \sqrt{\Delta}$, and the asymptotic limits of Eq.(\ref{eq2.10}) are always true (See also \cite{BL20}). 

\item The solutions were  initially  derived only in the region
$ - x^{-}_H < x < x^{+}_H$, in which the spacetime is homogeneous, and the Killing vector $\xi \equiv \partial_t$ is spacelike. The horizon at $x = x^{+}_H$ is referred to as the black hole horizon, while the one at 
$x = x^{-}_H$ is referred to as the white hole horizon, although in between them, no spacetime singularities exist  {at all} \cite{AOS18a, AOS18b}. However, following the standard process of extensions, one can easily extend the solutions beyond 
these horizons to the regions $|x| > \sqrt{\Delta}$. In the extended regions $ x < x^{-}_H$ and $x > x^+_H$, the metrics will take the same form as that given by Eqs.(\ref{eq2.4})-(\ref{eq2.6}), but now 
the   Killing vector   {$\partial_t$} becomes timelike.

\end{itemize}

In this paper, we shall not impose the conditions (\ref{eq2.7a}), except that we still  assume that $C$ and $D$ are real. In particular, since $C, \; D, \; n, \; \lambda_1$, and $\lambda_2$ are arbitrary constants, in principle, they 
can take any real values. However, since $ds^2= \left({3D}/{16}\right)^{2/3}d\bar{s}^2$, we shall assume that $D = 0$ holds only in the limiting sense. In addition, the two constants $\lambda_1$ and  $\lambda_2$ originate from the 
polymerization (\ref{eq2.12}),
so we also assume that $\lambda_1\lambda_2 \not= 0$,   {and consider the case $\lambda_1\lambda_2 = 0$ only as some limit cases, as to be explained explicitly below.
 Recall that we also assumed $n >0$ in order to have the metric be real. }

Then,   the geometric radius $b(x)$  and the ranges of $x$ all depend on the choices of the two parameters $x_0$ and ${\cal{C}}$, which are shown explicitly in Fig. \ref{fig1a-c}.
In particular, when ${\cal{C}} x_0 \not = 0$, we find that $x \in(-\infty, \infty)$, and a minimal point (the throat) of $b(x)$ always exists, with $b(\pm\infty) = \infty$,
as shown by the upper panel of Fig. \ref{fig1a-c}. 
When ${\cal{C}} \not=0,\;  x_0 = 0$, the range of $x$ is
restricted to  $x \in(0, \infty)$ with  $b(0) = \infty$ and $b(\infty) = \infty$. In this case, a minimum (throat) of $b(x)$ also exists, as shown explicitly in the middle  panel of Fig. \ref{fig1a-c}]. 
When ${\cal{C}} =0,\;  x_0 \not = 0$,
the range of $x$ is $x \in(-\infty, \infty)$, but now $b(x)$ is a monotonically increasing function of $x$ with $b(-\infty) = 0$ and $b(\infty) = \infty$, and a throat does not exists [cf. the bottom panel of Fig. \ref{fig1a-c}].

 In this paper, we shall study the main properties of  these spherical  polymer  black hole solutions. In particular, 
 we shall pay particular attention to  the locations of the throat and horizons, and the asymptotic behaviors of the spacetimes.   
 
 To these purposes, let us first notice that
the effective energy-momentum tensor $T_{\mu\nu}$, defined as $T_{\mu\nu} \equiv \kappa^{-1} G_{\mu\nu}$, can be cast in the form,
\bq
\lb{eq3.3}
T_{\mu\nu} = \rho u_{\mu}u_{\nu} + p_r  v_{\mu}v_{\nu} + p_{\theta}\left(\theta_{\mu}\theta_{\nu} + \phi_{\mu}\phi_{\nu}\right),
\eq
where 
\bqn
\lb{eq3.4}
 u_{\mu}^{+}  &=& -a^{1/2}(x)\delta_{\mu}^{t}, \quad  v_{\mu}^{+}  =  a^{-1/2}(x)\delta_{\mu}^{x},\nb\\
 \theta_{\mu}  &=&  b^{1/2}(x)\delta_{\mu}^{\theta}, \quad  \phi_{\mu}  =  b^{1/2}(x)\sin\theta \delta_{\mu}^{\phi},\;   ( a > 0),~~~~
\eqn
and
\bqn
\lb{eq3.5}
 \kappa\rho^{+}  &=& -\frac{1}{b^2}\Big[b(x) \left(2 a b '' + a 'b'\right)+a {b '}^2-1\Big], \nb\\
 \kappa p_r^{+}  &=& \frac{1}{b^2}\Big[ba 'b' +a {b'} ^2-1\Big], \nb\\
 \kappa p_{\theta}  &=& \frac{1}{2b}\Big[ba '' +2 ab '' +2 a 'b '\Big],\;  ( a > 0),
\eqn
with $\kappa \equiv 8\pi G/c^4$,  $a' \equiv da(x)/dx$, and so on. 

 It should be noted that in writing down Eqs.(\ref{eq3.4}) and (\ref{eq3.5}) we had assumed that $a(x) > 0$, as already indicated in these equations, so the coordinate $t$ is timelike. 
However, if a (black/white) horizon exists,  across this horizon $a(x)$ becomes negative,
and  the two coordinates $t$ and $x$ exchange  their roles. Then,  in the region $a(x) < 0$, 
the effective energy-momentum tensor can be still cast in  the form (\ref{eq3.3}), but now with
\bqn
\lb{eq3.4b}
 u_{\mu}^{-}  &=& |a|^{-1/2}\delta_{\mu}^{x}, \quad  v_{\mu}^{-}  =  -  |a|^{1/2}\delta_{\mu}^{t},\nb\\
  \kappa\rho^{-}  &=& -\frac{1}{b^2}\Big[ba 'b' +a {b'} ^2-1\Big],  \quad\quad ( a < 0), \nb\\
 \kappa p_r^{-}  &=& \frac{1}{b^2}\Big[b(x) \left(2 a b '' + a 'b'\right)+a {b '}^2-1\Big],
\eqn
while $\theta_{\mu},\; \phi_{\mu}$, and $p_{\theta}$ are still given by Eqs.(\ref{eq3.4}) and (\ref{eq3.5}).

 It should be also noted that, although the effective energy-momentum tensor in both of the regions $a > 0$ and 
$a < 0$ is written in the same form given by Eq.(\ref{eq3.3}), the physical interpretations  of the quantities $\rho^{\pm}$ and $p_r^{\pm}$ are different. In particular, the energy density $\rho^{+}$ in the region 
$a >0$ is measured by the observers who are moving along $dt$-direction, while their $x, \; \theta$, and $\phi$ coordinates  are fixed. The quantity $p_r^{+}$ is the principal pressure
 along the $dx$-direction measured by these observers. On the other hand, the energy density $\rho^{-}$ in the region 
$a < 0$ is measured by the observers who are moving along $dx$-direction, while their $t, \; \theta$, and $\phi$ coordinates  are fixed. In addition, the quantity $p_r^{-}$  now is
 the principal pressure along the $dt$-direction. Thus, in general such defined $\rho^{\pm}$ and $p_r^{\pm}$ are not continuous across the horizons. One way to avoid such discontinuities is
 to adopt the Eddington-Finkelstein coordinates, and then define a new set of observers, with respect to whom the energy density and principal pressure along the radial direction 
 are continuous across these horizons.  However, since in this paper we are mainly concerned with the energy conditions of ``the effective
 (quantum) matter," \footnote{ As mentioned above, the BMM model has not been obtained from quantizations of gravity yet, but rather
obtained by  simply applying the ``polymerization" (\ref{eq2.12}) to the corresponding classical Hamiltonian. So, it is not clear whether  these effects are indeed due to quantizations of gravity or not. In the 
rest of this paper, whenever we mention ``quantum gravitational effects" or ``quantum geometric effects" of this model, we always understand them as ``polymerization effects"  without any further explanations.
In the same sense, the expression  ``quantum black holes" of this model really means  polymer  black holes.} the current considerations are sufficient.

 In addition, although this effective energy-momentum tensor is purely due to the polymerization (\ref{eq2.12}),  and is not related to any real matter fields,  it does provide important information on
how the spacetime singularity is avoided, and the deviation of the spacetimes from the classical one, as in GR the geometry is uniquely determined by the Schwarzschild spacetime, in
which the spacetime is vacuum, and a spacetime curvature singularity is always present at the center of the black hole. In fact, this kind of singularities inevitably occurs in GR, as longer 
as the corresponding matter fields satisfy some  energy conditions, as follows directly from the Hawking-Penrose singularity theorems \cite{HE73}.

 The commonly used three energy conditions are {\it the weak, dominant, and strong energy conditions} \cite{HE73}. For   $T_{\mu\nu}$ given by Eq.(\ref{eq3.3}), they can be expressed as follows:
 The weak energy condition (WEC) is satisfied, when
\bq
\lb{eq3.13}
(i)\; \rho \geq 0, \quad (ii) \;  \rho+p_r \geq 0, \quad (iii) \; \rho+p_{\theta} \geq 0,
\eq
while the dominant energy condition (DEC) is satisfied, provided that
\bq
\lb{eq3.16}
(i)\; \rho \geq 0, \quad (ii) \;  - \rho \le  p_r \le \rho,  \quad (iii) \; -\rho \le p_{\theta} \le \rho.
\eq
The  strong energy condition (SEC) requires,
\bq
\lb{eq3.15}
 (i) \;  \rho+p_r \geq 0, \; (ii) \; \rho+p_{\theta} \geq 0,  \; (iii) \; \rho+p_r + 2p_{\theta}\geq 0.
\eq

Moreover, without causing any confusions, in the rest of this paper we shall absorb  $\kappa$ into $\rho, \; p_r$ and $ p_{\theta}$,
i.e., 
\bq
\lb{eq3.5a}
 \kappa\left(\rho,  \; p_r,\; p_{\theta}\right) \rightarrow \left(\rho,  \; p_r,\; p_{\theta}\right).
 \eq
 To study these solutions  in more details, let us consider the cases $\Delta > 0$, $\Delta = 0$ and $\Delta < 0$, separately,  { in the following three sections.}

\section {Spacetimes with $\Delta > 0$}
\label{Section.III}
 \renewcommand{\theequation}{3.\arabic{equation}}\setcounter{equation}{0}
 
From Eq.(\ref{eq2.3b}) we find that this case corresponds to 
\bq
\lb{eq3.6}
  \left|\lambda_2\right| < \frac{3}{2} \left|CD\right|.
\eq
 However, depending on the choice of the integration constants $C$ and $D$,   {there are still the possibilities, } ${\cal{D}} > 0$,  
 and ${\cal{D}} < 0$, provided that  $\Delta =  {\cal{D}}^2 - x_0^2 > 0$. In each of these cases, the physics of the corresponding solutions is quite different, so in the following
  let us consider them case by case.
 
 \subsection{${\cal{D}} > 0$}\lb{sec3-1}

 In this case, we have $CD > 0$, and $\Delta =  {\cal{D}}^2 - x_0^2 > 0$ implies,
 \bq
 \lb{eq3.6b}
 \beta \equiv \frac{\cal{D}}{\left|x_0\right|} > 1.
 \eq
Then, we find that there are two 
 asymptotically flat regions, corresponding to $x \rightarrow \pm \infty$,
 respectively. They are connected by a throat located at
 \bqn
 \lb{eq3.7}
 b_{m} &\equiv&   2^{1/3} {\cal{C}},\quad
x_{m} = \frac{1}{2{\cal{C}}}\left({\cal{C}}^2 -x_0^2\right),
 \eqn
 where $ b_{m} \equiv b(x = x_{m})$ and $b'(x = x_{m}) = 0$ [cf. Fig. \ref{fig1a-c}(a)].
 It is interesting to note that $x_{m}$ can be positive, zero or negative, depending on the choice of the two parameters ${\cal{C}}$ and 
 $x_0$ (or $\lambda_1, \; \lambda_2,\; n$ and $C$).  
 
  On the other hand, in the current case the white and black hole horizons always exist, and are located, respectively,  at 
\bq
\lb{eq3.7a}
x^{\pm}_H = \pm \sqrt{{\cal{D}}^2 - x_0^2}.
\eq
Clearly, there exist the possibilities in which  $|x_m| \le x_H^{+}$,  or $|x_m| > x_H^{+}$. When $|x_m| \le x_H^{+}$, 
 the throat  is located in the region between  the black and white hole horizons, in which we have 
$a(x) \le 0$, so the corresponding energy density and radial principal pressure  {in the region containing the throat } are given by $\rho^{-}$ and $p_r^{-}$. When $|x_m| > x_H^{+}$,
 the throat  is located in the region where $a(x) > 0$, so  the corresponding energy density and radial principal pressure
 {at the throat } are given by $\rho^{+}$ and $p_r^{+}$, respectively.


 \begin{figure}[h!]
\includegraphics[height=4.8cm]{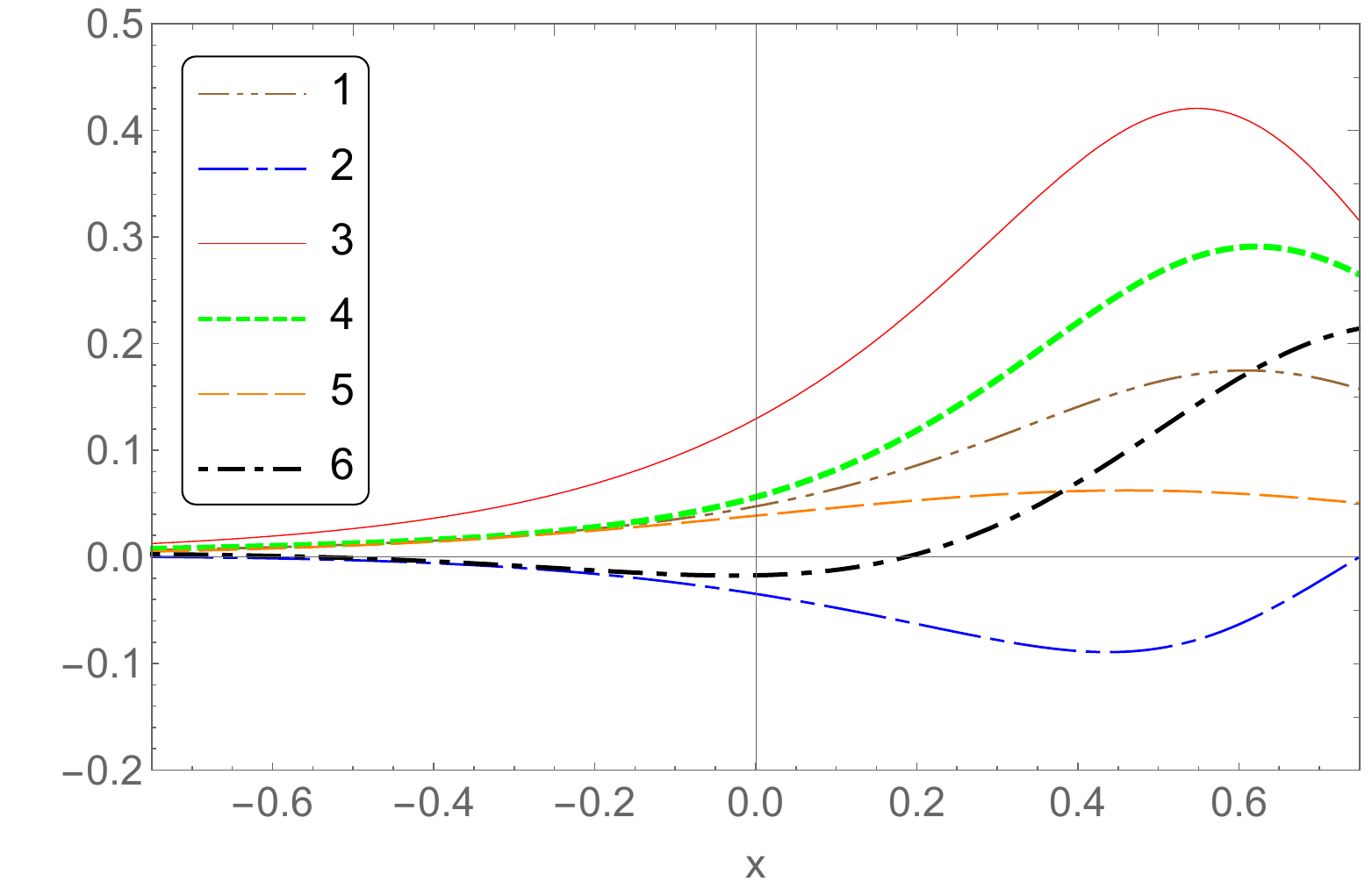}\\
(a)\\
\includegraphics[height=4.8cm]{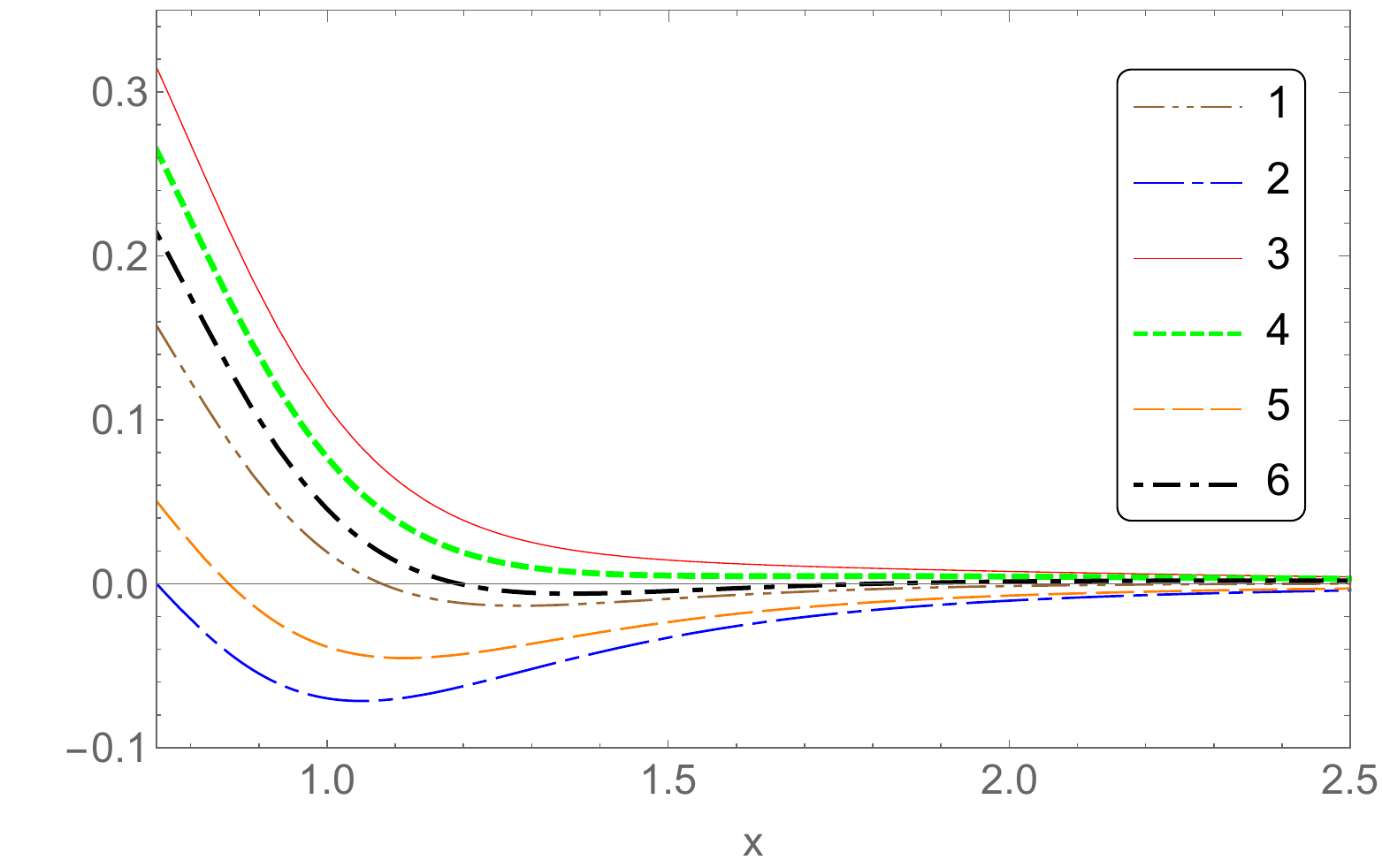}\\
 (b)  \\
\includegraphics[height=4.8cm]{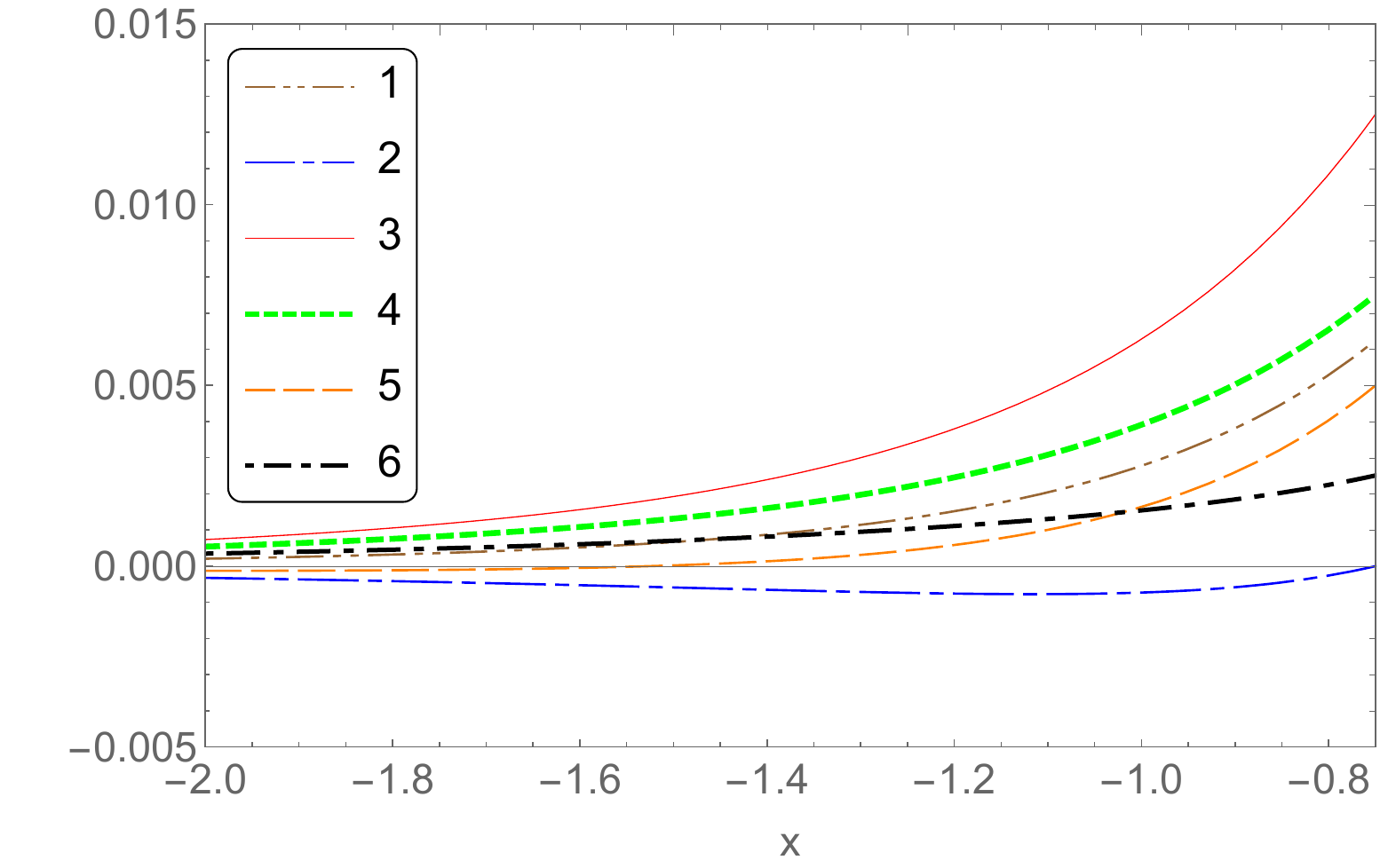}\\
(c)  \\
\caption{Case $\Delta > 0,\; {\cal{D}} > 0, \; \left|x_m\right| < x_H^+,\;  \beta =  1+ \frac{\left(\alpha -1\right)^2}{2\alpha},  \alpha \not= 1$: The  physical quantities, $\rho$, $(\rho + p_r)$,  $(\rho - p_r)$, $(\rho + p_{\theta})$, $(\rho - p_{\theta})$, and  $(\rho + p_r + 2p_{\theta})$, represented, respectively, by Curves 1 - 6,  vs $x$: When plotting these curves, we had set  $\alpha=2$, $\beta=5/4$, $x_0=1$, so that  the condition (\ref{eq3.14-3a}) is satisfied, for which we have  $x_m=x_{H}^{+}= - x_{H}^{-} = 0.75$.  Panel  (a): the physical quantities in the region  between the white and black horizons, $ x_H^{-} \le x \le x_{H}^{+}$. Panel (b):  the physical quantities in the region outside  the black  horizon, $ x \ge x_H^{+} = 0.75$.  Panel (c): the physical quantities in the region outside  the white  horizon, $ x \le x_H^{-} = -0.75$.
} 
\label{fig2rho}
\end{figure} 


 \begin{figure}[h!]
\includegraphics[height=4.8cm]{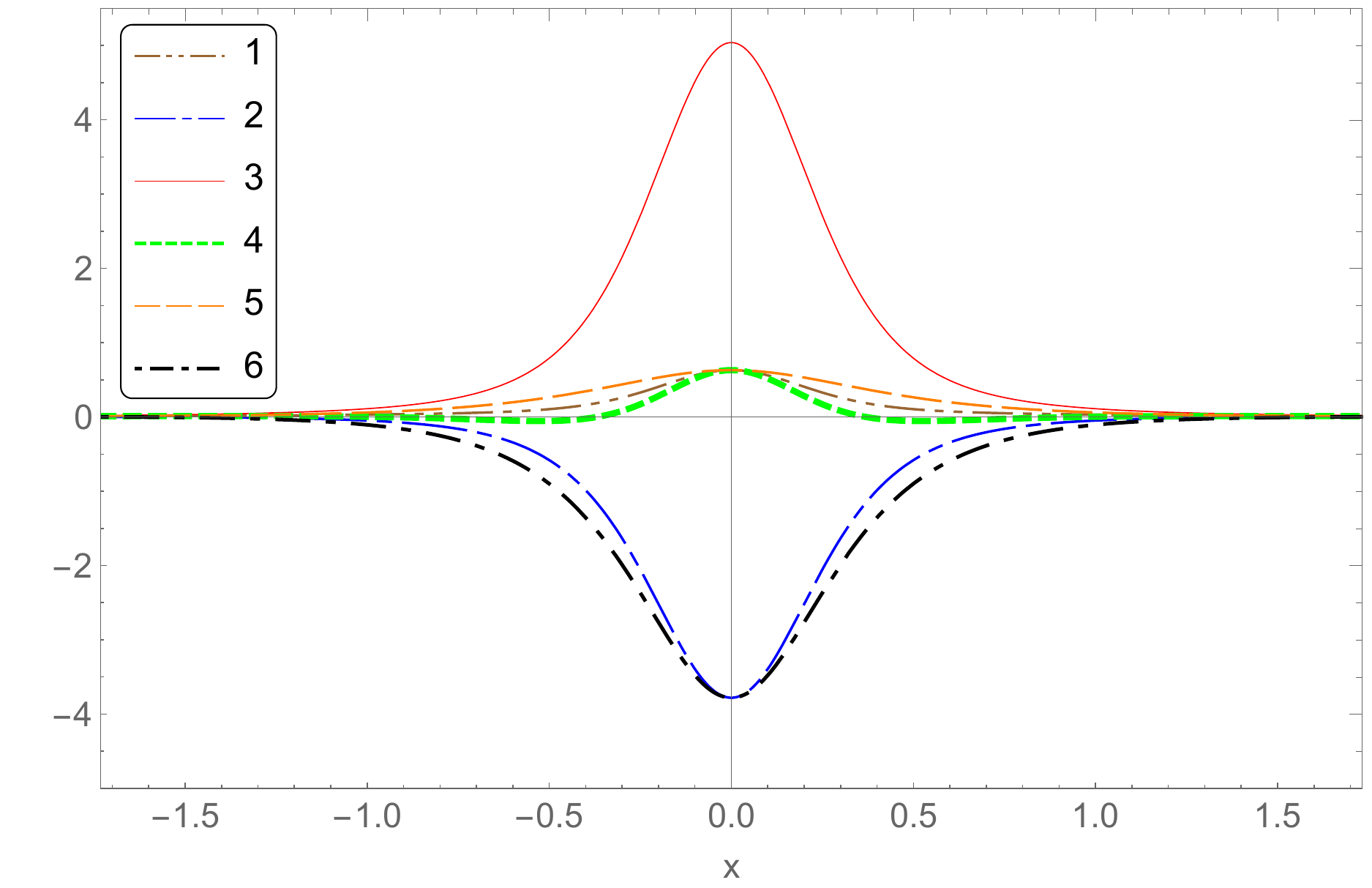}\\
(a)  \\
\includegraphics[height=4.8cm]{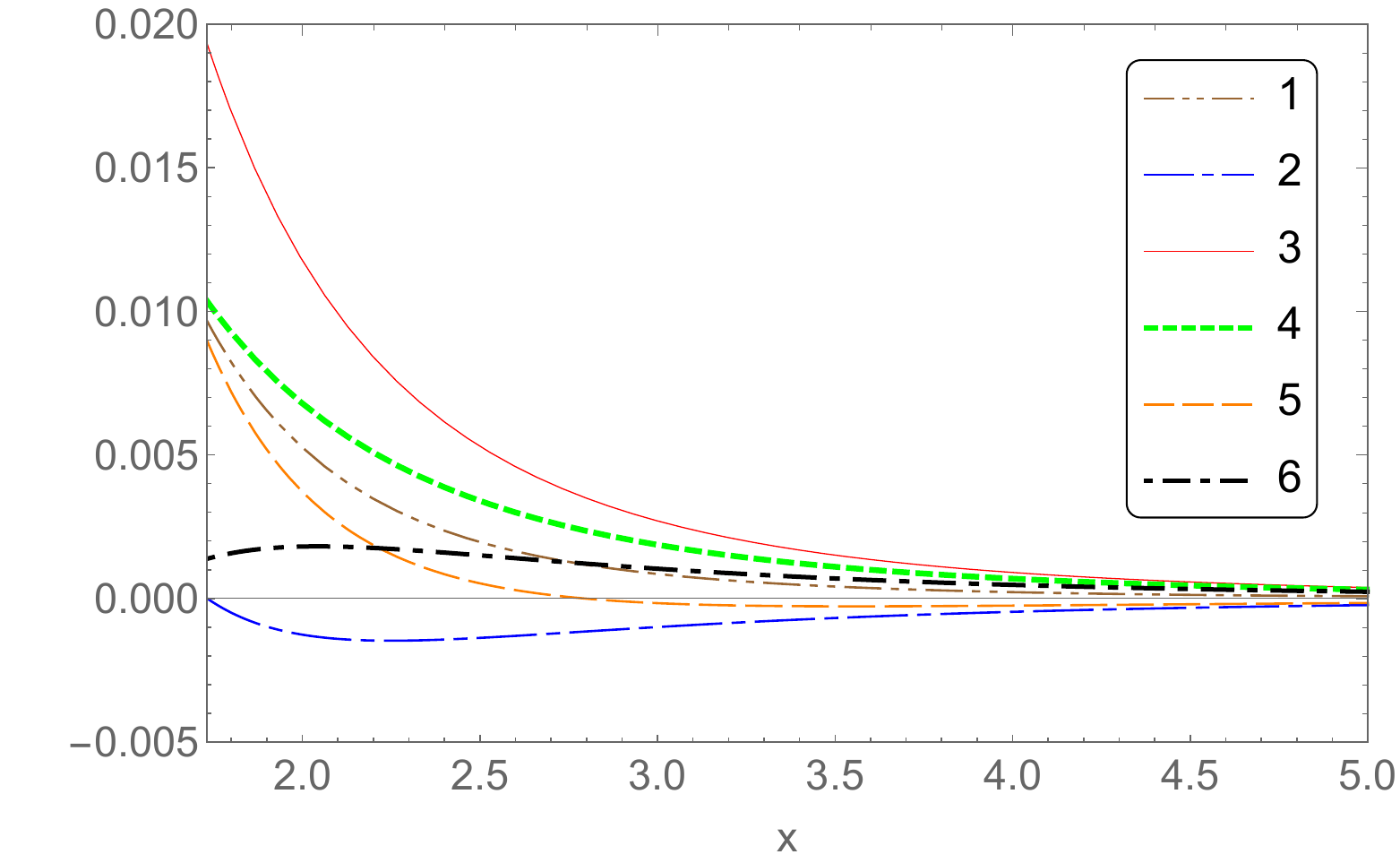}\\
(b)\\
\includegraphics[height=4.8cm]{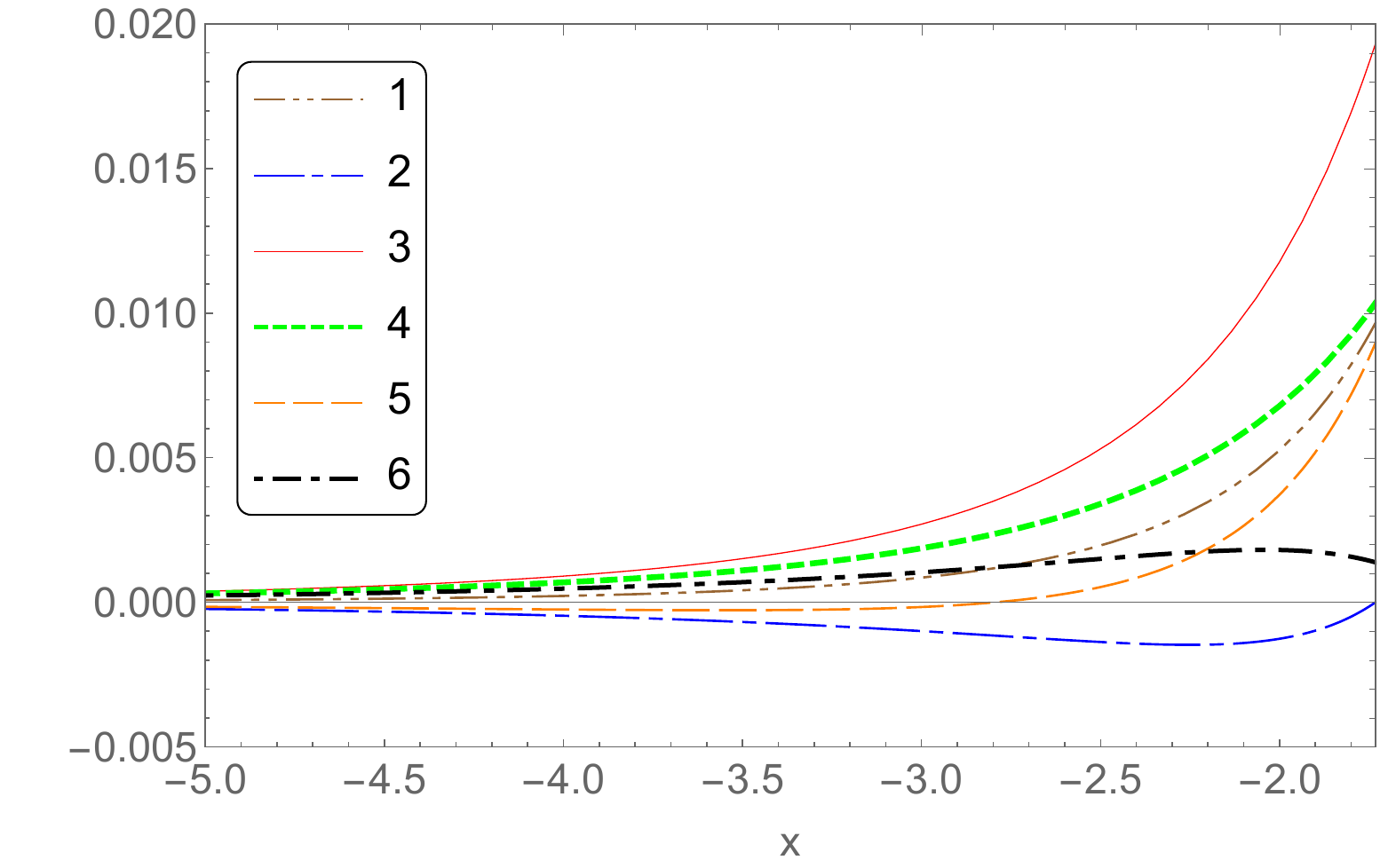}\\
(c)   \\
\caption{Case $\Delta > 0,\; {\cal{D}} > 0, \; \left|x_m\right| < x_H^+,\;  \beta \not=  1+ \frac{\left(\alpha -1\right)^2}{2\alpha}$: The  physical quantities, $\rho$, $(\rho + p_r)$,  $(\rho - p_r)$, $(\rho + p_{\theta})$, $(\rho - p_{\theta})$, and  $(\rho + p_r + 2p_{\theta})$, represented, respectively, by Curves 1 - 6,  vs $x$: When plotting these curves, we had set  $\alpha=1$, $\beta=2$, $x_0=1$, $x_{H}^{\pm}=\pm \sqrt{3}$, $x_m=0$. None of the three energy conditions is satisfied
at the throat, although all of them are satisfied at the two horizons $ x = x_{H}^{\pm}$.  Panel  (a): the physical quantities in the region  between the white and black horizons, $ x_H^{-} \le x \le x_{H}^{+}$. Panel (b):  the physical quantities in the region outside  the black  horizon, $ x \ge x_H^{+} = \sqrt{3}$.  Panel (c): the physical quantities in the region outside  the white  horizon, $ x \le x_H^{-} = -\sqrt{3}$.} 
\label{fig2rhoB}
\end{figure}

 \subsubsection{$x_{H}^{-} \le x_m \le x_{H}^{+}$}
   
  In this case, we find that  $|x_m| \le x_{H}^{+}$ implies
 \bqn
 \lb{eq3.8b-1}
 && (i) \;\;  \alpha = 1,  \quad \quad {\text{or}}\\
  \lb{eq3.8b-2}
&&  (ii) \;\; \beta \geq  1+ \frac{\left(\alpha -1\right)^2}{2\alpha},
 \eqn 
 where $ \alpha  \equiv {{\cal{C}}}/{\left|x_0\right|} > 0$. Since now the throat is located inside the black hole horizon, in which we have $a(x) < 0$, 
 we need to use Eq.(\ref{eq3.4b}) to calculate the effective energy density $\rho$ and pressure $p_r$ at the throat,
 and find that  
  \bqn
 \lb{eq3.8-2}
  \rho  &= &\frac{1}{2^{2/3} \mathcal{C} ^2} ,\nb\\
  p_r  &= & -\frac{\mathcal{C}  (12 \mathcal{D} -5 \mathcal{C} )-5 x_{0}^2}{2^{2/3} \mathcal{C} ^2 \left(x_{0}^2+\mathcal{C} ^2\right)},\nb\\
  p_{\theta}  &= & \frac{{\left(x_{0}^2+\mathcal{C} ^2\right)^3} -{4 \mathcal{D}  x_{0}^2 \mathcal{C} ^3}}{2^{2/3} \mathcal{C} ^2{\left(x_{0}^2+\mathcal{C} ^2\right)^3}}.
 \eqn
 Then,   we find that at the throat WEC is satisfied for
\bqn
\lb{eq3.14-1}
 && (a)\;\;  \beta \le  1+ \frac{\left(\alpha -1\right)^2}{2\alpha},  \quad\quad \text{or}\\
\lb{eq3.14-2}
&&  (b) \;\;  \beta \le \frac{1}{2}\alpha.
 \eqn
Combining Eqs.(\ref{eq3.8b-1})-(\ref{eq3.8b-2}) with Eqs.(\ref{eq3.14-1})-(\ref{eq3.14-2}) and considering Eq.(\ref{eq3.6b}),  we find that  their common solutions are
\bqn
\lb{eq3.14-3a}
   \beta =  1+ \frac{\left(\alpha -1\right)^2}{2\alpha}, \quad \alpha \not= 1,
\eqn
 which leads to $x_m=x_{H}^{+}$.

On the other hand,  SEC is also satisfied in the domain given by Eq.(\ref{eq3.14-3a}),
while   DEC requires
\bqn\lb{eq3.15-2}
&& (a)\;\; 0<\alpha<2 \beta   , \quad \beta \le \frac{\alpha^2 +1}{2\alpha}\le \frac{3}{2}\beta,   \;\;\;  \text{or} ~~~~~~
\\
\lb{eq3.15-3}
&& (b) \;\; \; 2 \beta \le \alpha<3 \beta, \quad \beta \ge \frac{1+\alpha^2}{3\alpha}.
 \eqn
 Combining Eqs.(\ref{eq3.8b-1})-(\ref{eq3.8b-2}) with Eqs.(\ref{eq3.15-2})-(\ref{eq3.15-3}),  we find that  their common solution is also
 given by Eq.(\ref{eq3.14-3a}).

Therefore, at the throat none of the  three energy conditions is satisfied, except for the case in which the throat coincides with the black hole horizon, $x_m = x_H^+$, which is a direct result of the condition  Eq.(\ref{eq3.14-3a}).
In Fig. \ref{fig2rho}, we show this case, from which one can see that the three energy conditions are satisfied indeed only at the throat.  
In Fig. \ref{fig2rhoB}, we show the case that does not satisfy the condition  Eq.(\ref{eq3.14-3a}), from which one can see that none of the three energy conditions is satisfied  at the throat ($x_m = 0$).

 In addition, if we consider the limit to the black hole horizon from outside of it, then
 we have $\rho = \rho^{+}$ and $p_r = p_r^+$, and  the energy density and pressures are  given, respectively, by
  \bqn
\lb{eq3.17a}
\rho &=& - p_r=-\frac{Y^3}{XZ^8}\Bigg[\Big( 32 \mathcal{D} ^5 \mathcal{C} ^6+10 \mathcal{D}  x_{0}^{10}-160 \mathcal{D} ^3 x_{0}^8\nb\\
&& +672 \mathcal{D} ^5 x_{0}^6-1024 \mathcal{D} ^7 x_{0}^4+2 \mathcal{D}  x_{0}^4 \mathcal{C} ^6+512 \mathcal{D} ^9 x_{0}^2\nb\\
&& -24 \mathcal{D} ^3 x_{0}^2 \mathcal{C} ^6   \Big) \sqrt{\Delta}  + 32 \mathcal{D} ^6 \mathcal{C} ^6-x_{0}^{12}+50 \mathcal{D} ^2 x_{0}^{10}\nb\\
&& -400 \mathcal{D} ^4 x_{0}^8+1120 \mathcal{D} ^6 x_{0}^6-1280 \mathcal{D} ^8 x_{0}^4+10 \mathcal{D} ^2 x_{0}^4 \mathcal{C} ^6\nb\\
&& +512 \mathcal{D} ^{10} x_{0}^2-40 \mathcal{D} ^4 x_{0}^2 \mathcal{C} ^6+\mathcal{C} ^{12} \Bigg],
\eqn
\bqn
\lb{eq3.17b}
p_{\theta}&=&\frac{Y^2}{2X^2 Z^8}\Bigg[\Big( 128 \mathcal{D} ^7 \mathcal{C} ^6+2 \mathcal{D}  \mathcal{C} ^{12}+10 \mathcal{D}  x_{0}^{12}\nb\\
&& -160 \mathcal{D} ^3 x_{0}^{10}+672 \mathcal{D} ^5 x_{0}^8-1024 \mathcal{D} ^7 x_{0}^6-12 \mathcal{D}  x_{0}^6 \mathcal{C} ^6\nb\\
&& +512 \mathcal{D} ^9 x_{0}^4+88 \mathcal{D} ^3 x_{0}^4 \mathcal{C} ^6-192 \mathcal{D} ^5 x_{0}^2 \mathcal{C} ^6 \Big)\sqrt{\Delta} \nb\\
&& + 128 \mathcal{D} ^8 \mathcal{C} ^6+2 \mathcal{D} ^2 \mathcal{C} ^{12}-x_{0}^{14}+50 \mathcal{D} ^2 x_{0}^{12}\nb\\
&& -400 \mathcal{D} ^4 x_{0}^{10}+1120 \mathcal{D} ^6 x_{0}^8+2 x_{0}^8 \mathcal{C} ^6-1280 \mathcal{D} ^8 x_{0}^6\nb\\
&& -40 \mathcal{D} ^2 x_{0}^6 \mathcal{C} ^6+512 \mathcal{D} ^{10} x_{0}^4+168 \mathcal{D} ^4 x_{0}^4 \mathcal{C} ^6\nb\\
&& -256 \mathcal{D} ^6 x_{0}^2 \mathcal{C} ^6-x_{0}^2 \mathcal{C} ^{12}  \Bigg].
\eqn
It can be shown that each of the three energy conditions is satisfied provided that $\beta > 1$, which is precisely the condition $\Delta > 0$, as shown in Eq.(\ref{eq3.6b}). 
In addition,  the surface gravity of the black hole is given by,
 \bqn 
\lb{eq3.18a}
\kappa_{BH} &\equiv& \frac{1}{2}a'(x =\sqrt{\Delta}) =  \frac{Y^2\left|x_0\right|^7}{2Z^5}\nb\\
&& \times \Bigg[\sqrt{\beta^2 - 1}\; \Big(32\beta^6- 48 \beta^4 + 18\beta^2 - 1 + \alpha^6\Big) \nb\\
&& +2\beta\left(16\beta^6 -32 \beta^4 + 19\beta^2 - 3\right)\Bigg],
\eqn
which is also always positive for  $\beta > 1$.  

 \begin{widetext}
 
     \begin{figure}[h!]
 \begin{tabular}{cc}
\includegraphics[height=4.8cm]{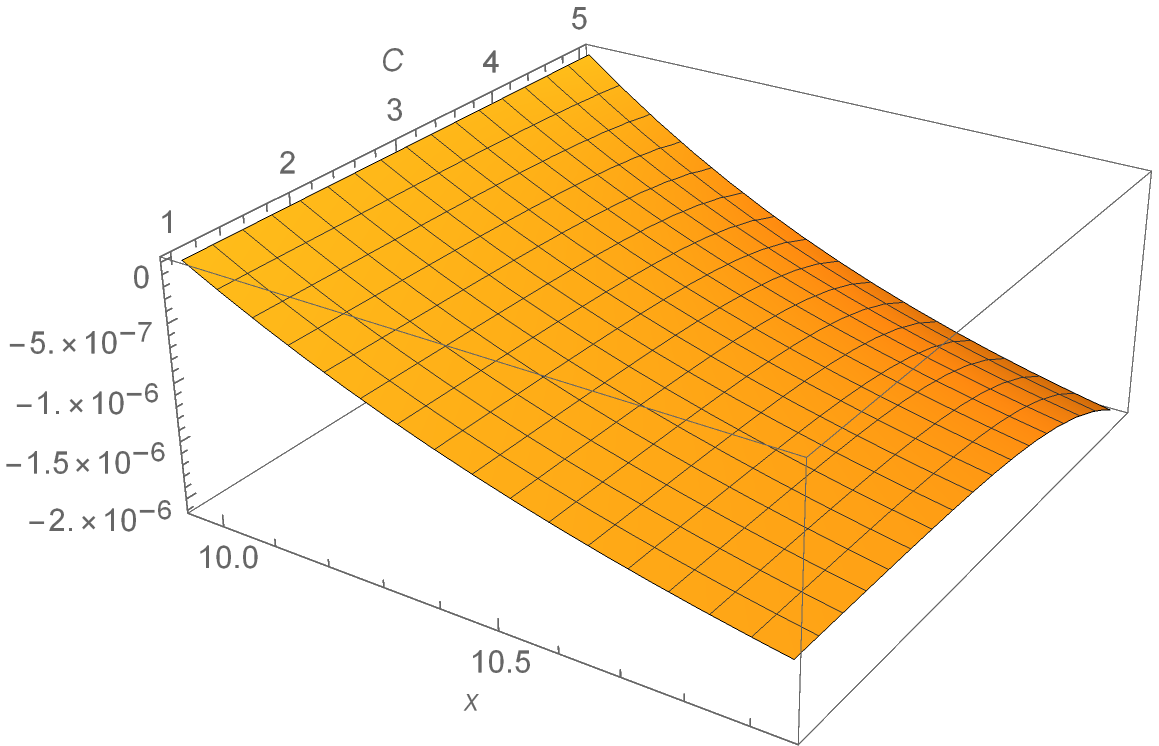}&
\includegraphics[height=4.8cm]{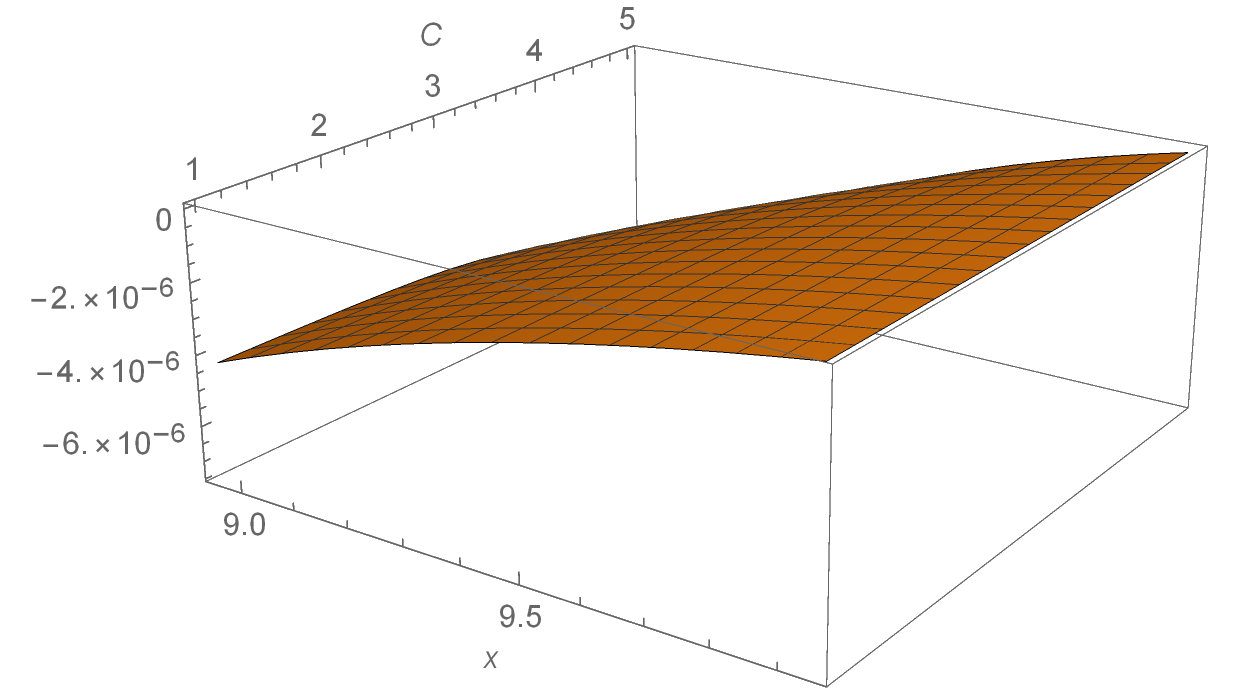}\\
	(a) & (b)  \\[6pt]
\includegraphics[height=4.8cm]{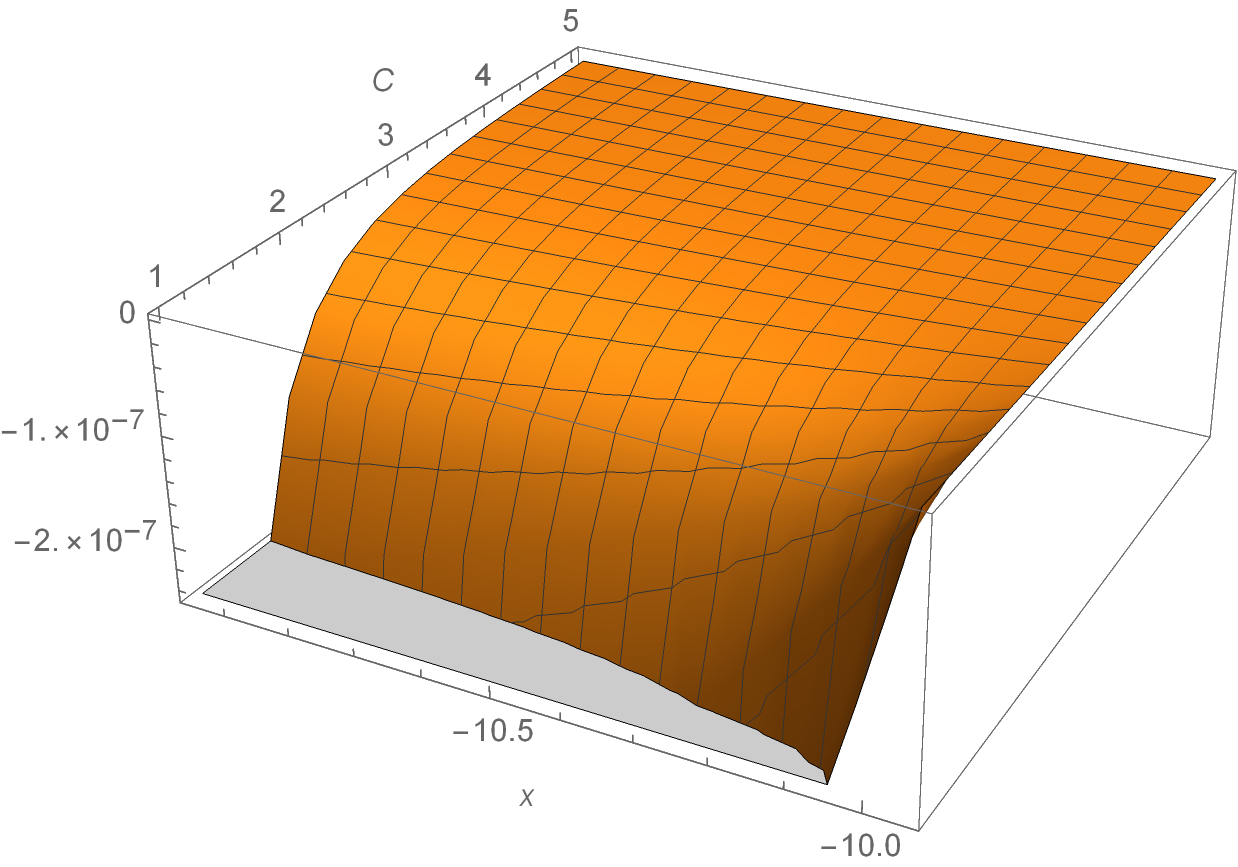}&
\includegraphics[height=4.8cm]{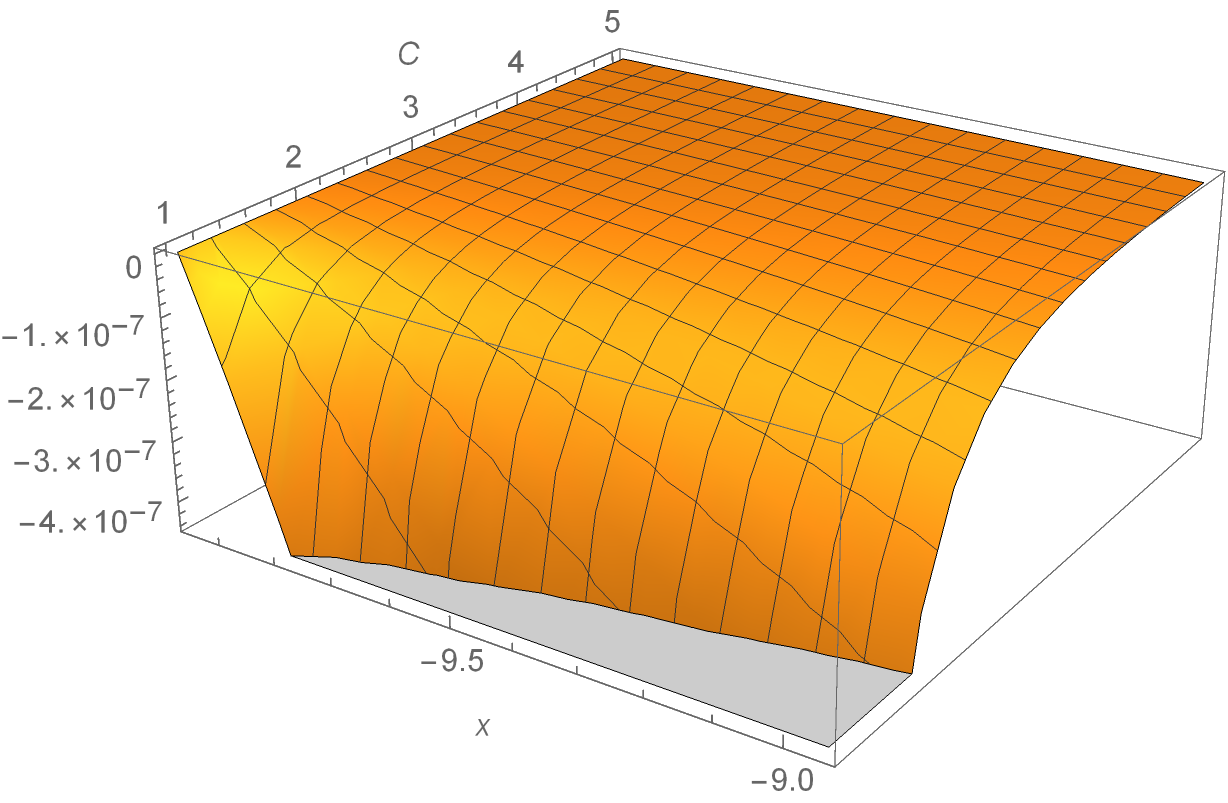}\\
	(c) & (d)  \\[6pt]
	\end{tabular}
\caption{The  physical quantity $(\rho + p_r)$ vs the radial coordinate $x$ and the parameter $\mathcal{C}$: (a) outside the black hole horizon; (b) inside the black hole horizon;
 (c) outside the white hole horizon; and (d) inside the white hole horizon.
Graphs are plotted with $x_{0}=1,\; \mathcal{D}=10$, for which the horizons are at $x^{\pm}_H \approx \pm 10$.} 
\lb{fig-horizon}
\end{figure} 

\end{widetext}
 
At the white hole horizon $(x =-\sqrt{\Delta})$, taking  the  limit from the outside of it, so that $\rho = \rho^{+}$ and $p_r = p_r^+$, we find that 
\begin{widetext}
\bqn
\lb{eq3.19}
\rho &=& -p_r=-\frac{Y}{\mathcal{D}  Z^{8} }
\Bigg(\Big[128 \mathcal{D} ^7 \mathcal{C} ^6+2 \mathcal{D}  \mathcal{C} ^{12}-12 \mathcal{D}  x_{0}^{12}+280 \mathcal{D} ^3 x_{0}^{10}-1792 \mathcal{D} ^5 x_{0}^8\nb\\
&&+4608 \mathcal{D} ^7 x_{0}^6-2 \mathcal{D}  x_{0}^6 \mathcal{C} ^6-5120 \mathcal{D} ^9 x_{0}^4+48 \mathcal{D} ^3 x_{0}^4 \mathcal{C} ^6+2048 \mathcal{D} ^{11} x_{0}^2-160 \mathcal{D} ^5 x_{0}^2 \mathcal{C} ^6 \Big]\sqrt{\Delta}-128 \mathcal{D} ^8 \mathcal{C} ^6\nb\\
&&-2 \mathcal{D} ^2 \mathcal{C} ^{12}-x_{0}^{14}+72 \mathcal{D} ^2 x_{0}^{12}-840 \mathcal{D} ^4 x_{0}^{10}+3584 \mathcal{D} ^6 x_{0}^8-6912 \mathcal{D} ^8 x_{0}^6+14 \mathcal{D} ^2 x_{0}^6 \mathcal{C} ^6+6144 \mathcal{D} ^{10} x_{0}^4\nb\\
&&-112 \mathcal{D} ^4 x_{0}^4 \mathcal{C} ^6-2048 \mathcal{D} ^{12} x_{0}^2+224 \mathcal{D} ^6 x_{0}^2 \mathcal{C} ^6+x_{0}^2 \mathcal{C} ^{12}
\Bigg), \nb\\
p_{\theta} &=&\frac{Y^2}{2 \mathcal{D} ^2  Z^{8}} \Bigg(\Big[128 \mathcal{D} ^7 \mathcal{C} ^6+2 \mathcal{D}  \mathcal{C} ^{12}+10 \mathcal{D}  x_{0}^{12}-160 \mathcal{D} ^3 x_{0}^{10}+672 \mathcal{D} ^5 x_{0}^8-1024 \mathcal{D} ^7 x_{0}^6-12 \mathcal{D}  x_{0}^6 \mathcal{C} ^6\nb\\
&&+512 \mathcal{D} ^9 x_{0}^4+88 \mathcal{D} ^3 x_{0}^4 \mathcal{C} ^6-192 \mathcal{D} ^5 x_{0}^2 \mathcal{C} ^6  
\Big]\sqrt{\Delta}   -128 \mathcal{D} ^8 \mathcal{C} ^6-2 \mathcal{D} ^2 \mathcal{C} ^{12}+x_{0}^{14}-50 \mathcal{D} ^2 x_{0}^{12}+400 \mathcal{D} ^4 x_{0}^{10}\nb\\
&&-1120 \mathcal{D} ^6 x_{0}^8-2 x_{0}^8 \mathcal{C} ^6+1280 \mathcal{D} ^8 x_{0}^6+40 \mathcal{D} ^2 x_{0}^6 \mathcal{C} ^6-512 \mathcal{D} ^{10} x_{0}^4-168 \mathcal{D} ^4 x_{0}^4 \mathcal{C} ^6+256 \mathcal{D} ^6 x_{0}^2 \mathcal{C} ^6+x_{0}^2 \mathcal{C} ^{12} \Bigg).
\eqn

It can be shown that  for $\beta >1$, all  the three energy conditions are satisfied at the white hole horizon. Moreover, at this white hole horizon, the surface gravity  is given by,
\bqn
\lb{eq3.19a}
\kappa_{WH} &\equiv& \frac{1}{2}a'(x = - \sqrt{\Delta})\nb\\
&=& -\frac{Y^2}{2Z^5}
 \times \Bigg[\Big(32 \mathcal{D} ^6-x_{0}^6+18 \mathcal{D} ^2 x_{0}^4-48 \mathcal{D} ^4 x_{0}^2+\mathcal{C} ^6
\Big)\sqrt{\Delta}
-32 \mathcal{D} ^7+6 \mathcal{D}  x_{0}^6-38 \mathcal{D} ^3 x_{0}^4+64 \mathcal{D} ^5 x_{0}^2
\Bigg],
\eqn
which is always negative   when the condition (\ref{eq3.6b}) holds. 

 \end{widetext}

In Figs. \ref{fig2rho} and  \ref{fig2rhoB}, we  {also show the physical quantities near the two horizons, and find} that all
  the three energy conditions are indeed satisfied  at these horizons, no matter whether   Eq.(\ref{eq3.14-3a}) is satisfied or not. 
  From these figures we can see that $\rho+p_{r}$ is the key quantity to determine the energy conditions. In particular,  it is zero only at the two horizons and negative at other locations.
   Thus, the energy conditions are normally satisfied only at horizons. To show this more clearly, we plot $\rho+p_{r}$ vs $x$ and the parameter $\mathcal{C}$ in Fig. \ref{fig-horizon}, from which we can 
   see that even with different choices of the free parameter,  $\rho+p_{r}$ is non-negative only on the two horizons.

 In addition,  as $x \rightarrow \pm \infty$, we find that 
\bqn
\lb{eq3.20}
\rho(x) &=& 
\begin{cases}
\frac{\mathcal{D}  x_{0}^2}{8 x^5}+ {\cal{O}}\left(\epsilon^6\right), & x \rightarrow \infty,\cr
-\frac{\mathcal{D}  x_{0}^6}{8 x^5 \mathcal{C} ^4}+ {\cal{O}}\left(\epsilon^6\right), &  x \rightarrow - \infty,\cr
\end{cases}\nb\\
p_r(x) &=&
\begin{cases}
-\frac{x_{0}^2}{4 x^4}+\frac{\mathcal{D}  x_{0}^2}{8 x^5}+ {\cal{O}}\left(\epsilon^6\right), &  x \rightarrow \infty,\cr
-\frac{x_{0}^6}{4 x^4 \mathcal{C} ^4}-\frac{\mathcal{D}  x_{0}^6}{8 x^5 \mathcal{C} ^4}+ {\cal{O}}\left(\epsilon^6\right), &  x \rightarrow - \infty,\cr
\end{cases}\nb\\
p_{\theta}(x) &=&
\begin{cases}
\frac{x_{0}^2}{4 x^4}-\frac{\mathcal{D}  x_{0}^2}{4 x^5}, &  x \rightarrow \infty,\cr
\frac{x_{0}^6}{4 x^4 \mathcal{C} ^4}+\frac{\mathcal{D}  x_{0}^6}{4 x^5 \mathcal{C} ^4}+ {\cal{O}}\left(\epsilon^6\right), &  x \rightarrow - \infty,\cr
\end{cases}\nb\\
\eqn
where $\epsilon \equiv 1/|x|$.  Thus, in these two asymptotically flat regions, none of these three energy conditions  holds.
On the other hand,  at these limits, we also have,
\bqn
\lb{eq3.21}
a(x) &=& \begin{cases}
 \frac{1}{4}\left(1-\frac{2\mathcal{D} }{b}\right) +  {\cal{O}}\left(\epsilon^2\right), & x  \rightarrow  \infty,\cr
 \frac{x_{0}^4}{4 \mathcal{C} ^4}\left(1-\frac{\left(2\mathcal{D} \mathcal{C} ^2/x_{0}^2\right)}{ b}\right) +  {\cal{O}}\left(\epsilon^2\right), & x  \rightarrow  -\infty,\cr
\end{cases}\nb\\
b(x) &\simeq&  \begin{cases}
2x, & x  \rightarrow  \infty,\cr
-2 \left(\mathcal{C} ^2 /x_{0}^{2}\right) x, & x  \rightarrow  -\infty,\cr
\end{cases}
\eqn
from which we find that   the masses of the black and white holes are given, respectively, by 
\bqn
\lb{eq3.22}
M_{BH} &=& \mathcal{D},\quad
 M_{WH} =   \frac{\mathcal{D} \mathcal{C} ^2}{x_{0}^2}.
\eqn

 \begin{figure}[h!]
\includegraphics[height=4.8cm]{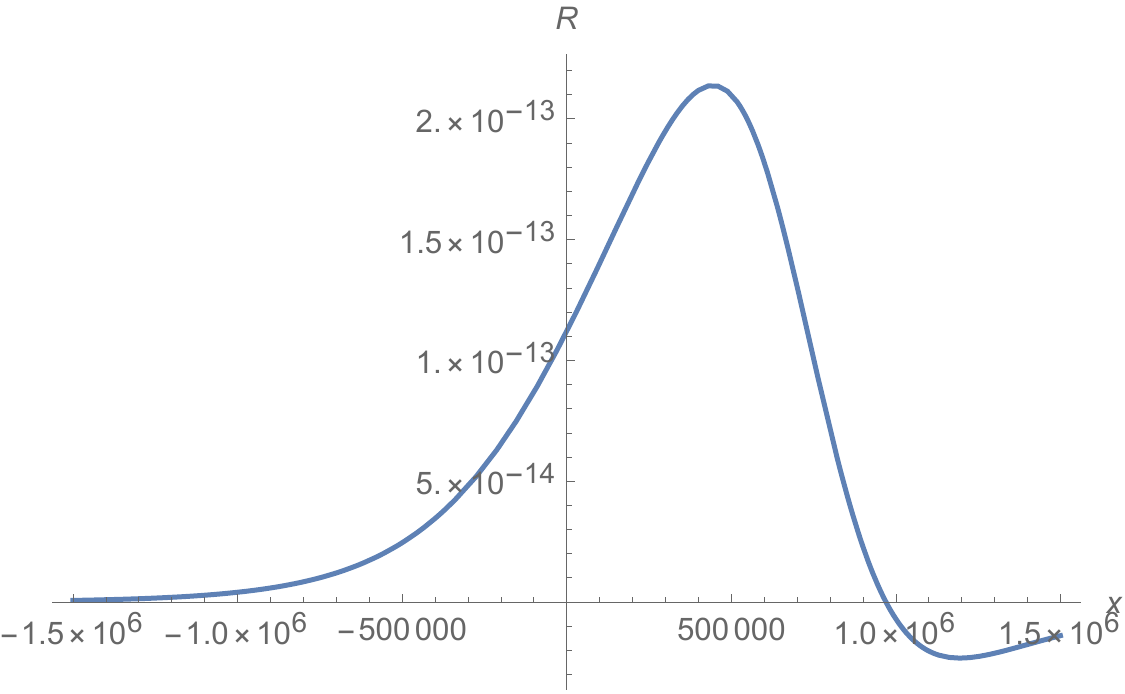}\\
(a)\\
\includegraphics[height=4.8cm]{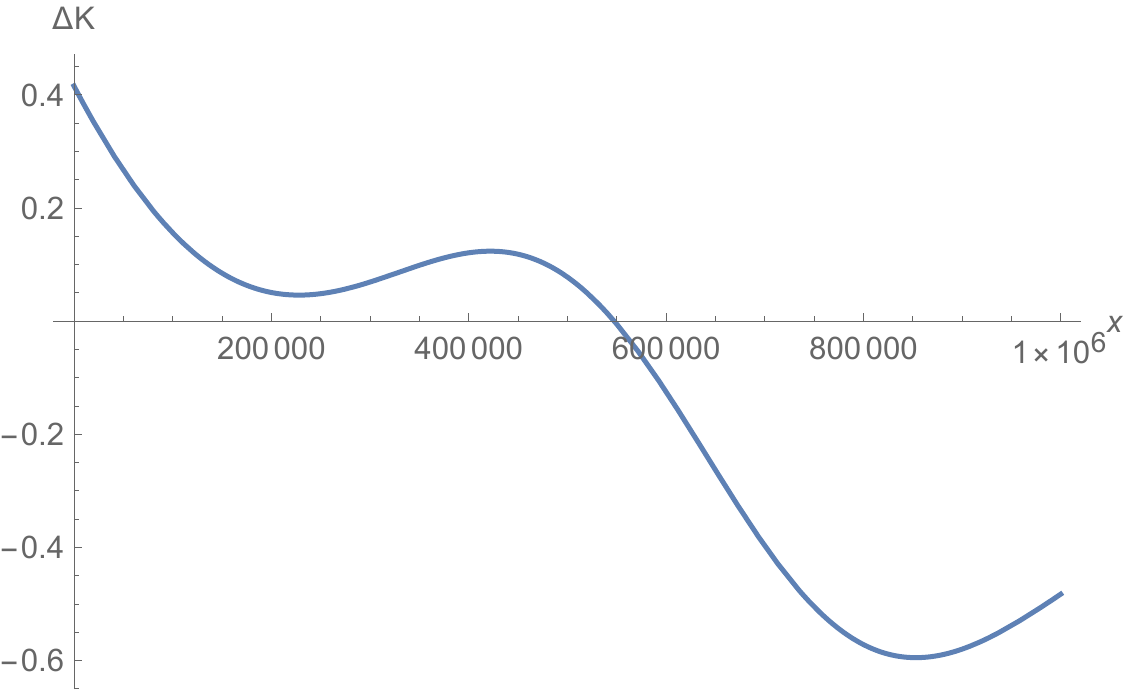}\\
 (b)  \\
\includegraphics[height=4.8cm]{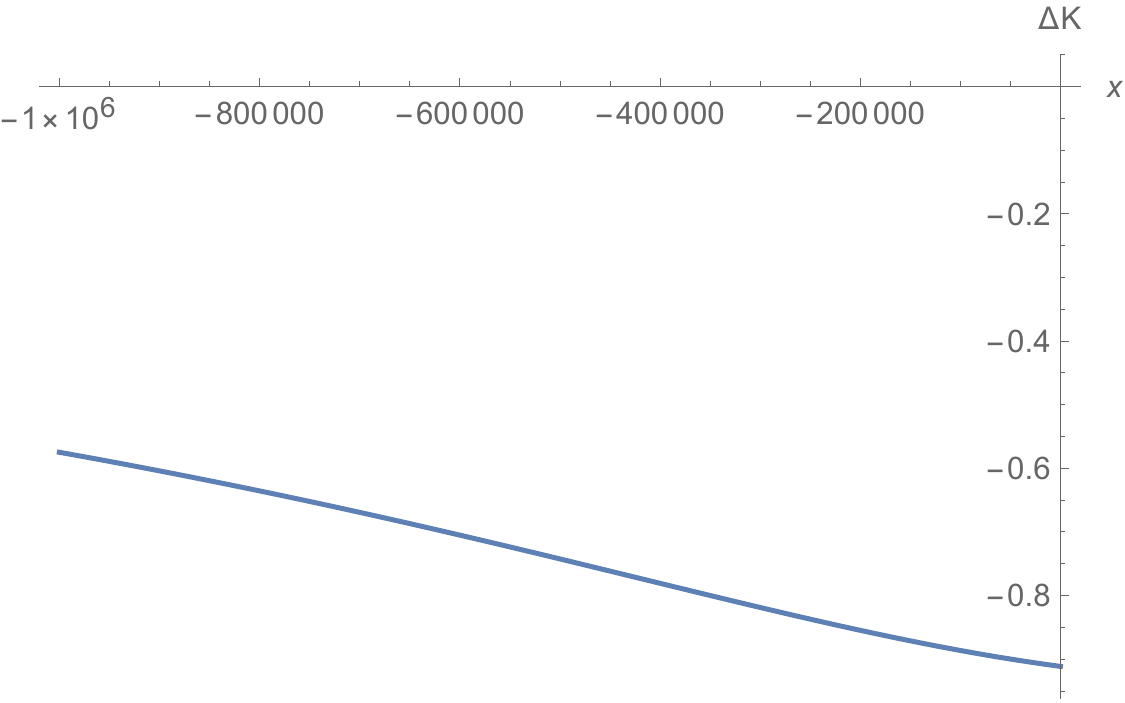}
\\
(c)   \\
\caption{Case $\Delta > 0,\; {\cal{D}} > 0, \; \left|x_m\right| \le x_H^+,\;  \beta =  1+ \frac{\left(\alpha -1\right)^2}{2\alpha},  \alpha \not= 1$:
The quantities $R$ and $\Delta  {\cal{K}}$ vs $x$. 
 Here we choose $\mathcal{C}=2\times 10^{6},\; x_0 =10^{6},\; \mathcal{D}=\frac{5}{4}\times 10^{6}$,  for which the horizons are located at 
$x_H^{\pm} = \pm 0.75\times 10^{6}$, and the throat  is at $x_{m}=  x_H^+$, while the black and white hole masses are
$M_{BH}=\frac{5}{4}\times 10^6\; M_{Pl}$ and $M_{WH} = 5 \times 10^{6}\; M_{Pl}$, respectively. } 
\label{fig3-1}
\end{figure}

 \begin{figure}[h!]
\includegraphics[height=4.8cm]{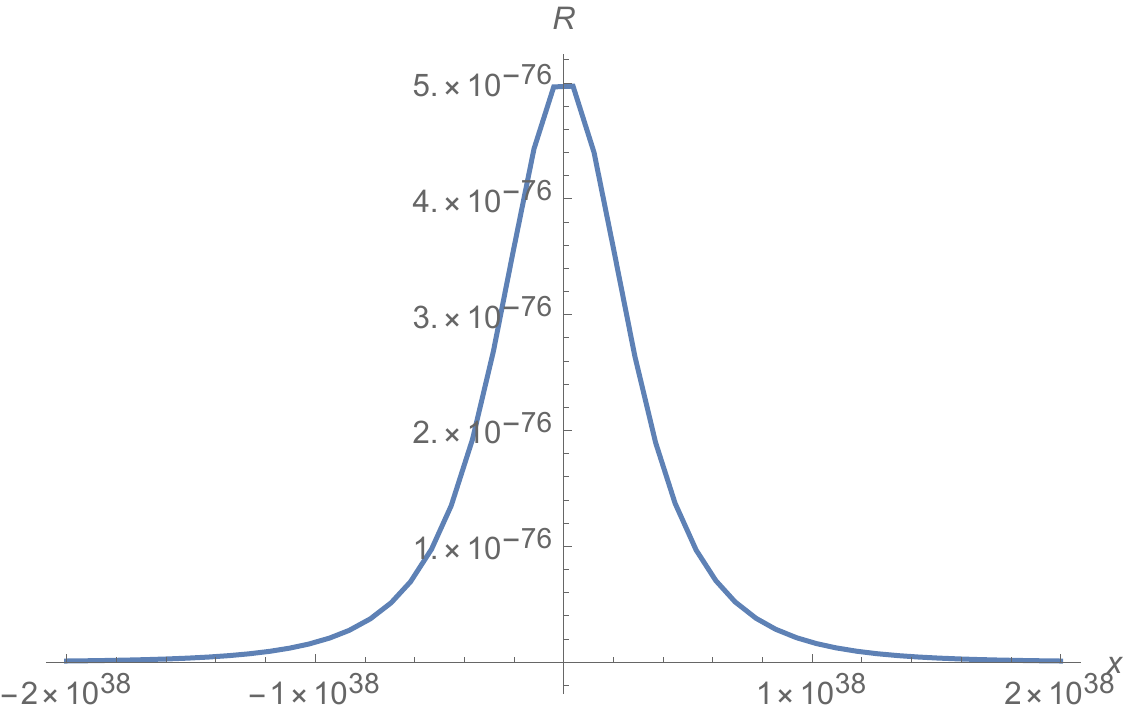}\\
(a)\\
\includegraphics[height=4.8cm]{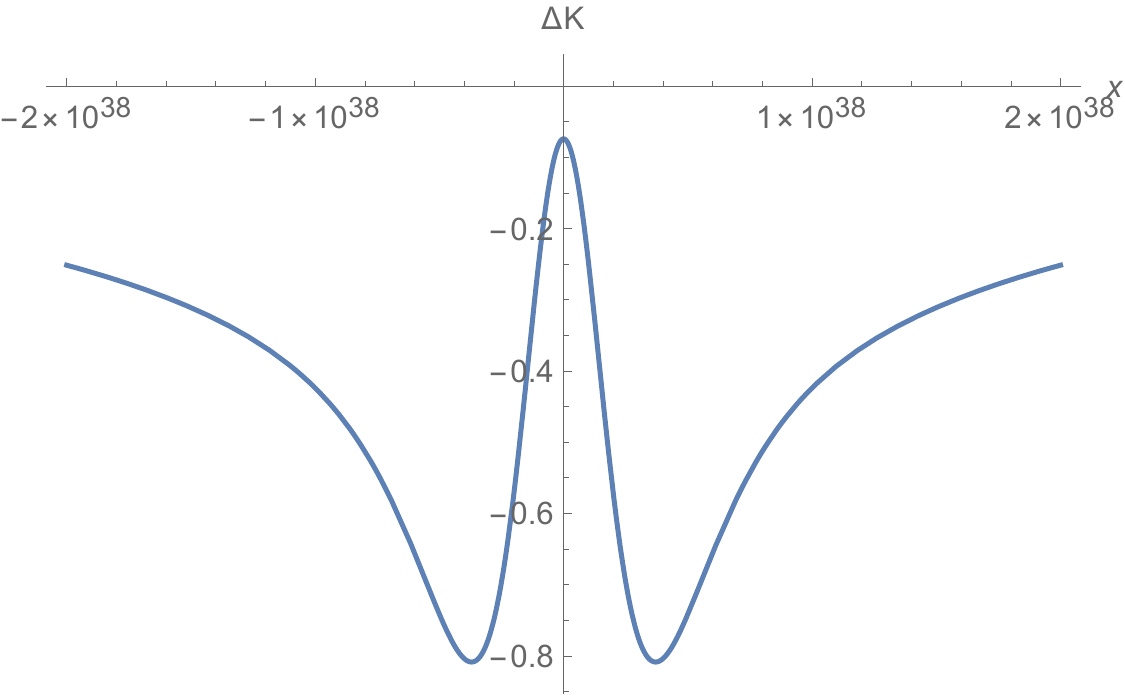}\\
 (b)  \\
\caption{Case $\Delta > 0,\; {\cal{D}} > 0, \; \left|x_m\right| < x_H^+,\;  \beta \not=  1+ \frac{\left(\alpha -1\right)^2}{2\alpha}$:
The quantities $R$ and $\Delta  {\cal{K}}$ vs $x$. 
 Here we choose $\mathcal{C}=10^{38},\; x_0 =10^{38},\; \mathcal{D}=2\times 10^{38}$,  for which the horizons are located at 
$x_H^{\pm} = \pm \sqrt{3}\times 10^{38}$, and the throat  is at $x_{m}=0$, while the black and white hole masses are
$M_{BH}=2\times 10^{38}\; M_{Pl}$ and $M_{WH} = 2 \times 10^{38}\; M_{Pl}$, respectively. } 
\label{fig3-1b}
\end{figure}

 To study the quantum gravitational effects further,
 let us turn to consider the Ricci scalar $R$ and the relative difference $\Delta  {\cal{K}}$ of the Kretschmann scalar,
 defined by
 \bq
 \lb{eq3.25b}
\Delta {\cal{K}} \equiv \frac{{  {\cal{K}}-\cal{K}}^{GR} }{{\cal{K}}^{GR}},
\eq
where ${\cal{K}}^{GR}$ denotes the Kretschmann scalar of the Schwarzschild solution, given by,
\bqn
\lb{3.25ba}
{\cal{K}}^{GR} &\equiv& R_{\alpha\beta\mu\nu}R^{\alpha\beta\mu\nu} =
 \begin{cases}
\frac{48 M_{BH}^2}{b^6(x)}, & x > x_{m},\cr
\frac{48 M_{WH}^2}{b^6(x)}, & x < x_{m}.\cr
\end{cases} ~~~~~
\eqn
In GR, we have $R^{GR} = 0$, But due to the quantum geometric effects, clearly now we have $R \not = 0$. Therefore, both quantities, $R$ and $\Delta  {\cal{K}}$, will describe the 
deviations of the quantum black holes from the  classical one.  Before proceeding further, we would like to point out that Eqs.(\ref{eq3.25b}) and (\ref{3.25ba}) are applicable 
 when the two horizons and asymptotic regions  exist. In some particular cases, this is not true, and a proper modification for $\Delta {\cal{K}}$ is needed, as to be shown below.

In addition, another important quantity is the  scalar 
\bq
\lb{eq3.25ba}
C_{\mu\nu\alpha\beta} C^{\mu\nu\alpha\beta} =  {\cal{K}}^2 + \frac{1}{3}R^2 - 2R_{\mu\nu} R^{\mu\nu},
\eq
where $C_{\mu\nu\alpha\beta}$ denotes the Weyl tensor.  
However, for the sake of simplicity, in the following we shall consider only the quantities $\Delta {\cal{K}}$ and $R$, which are
sufficient for our current purpose.


   In Fig. \ref{fig3-1}, the quantities $R$ and $\Delta  {\cal{K}}$ are plotted  in the  { region  between} the two horizons ($x_H^{\pm} =\pm 0.75\times 10^{6}$),
   from which it can be seen that the deviation from GR are still large near these two horizons, although the curvature decays rapidly  when away from them in both directions. 
 In particular,  for $M_{BH}=2\times 10^6\; M_{Pl}$ and $M_{WH} = 32 \times 10^{6}\; M_{Pl}$, near the horizons we find that $R(x_H^{+}) \lesssim  10^{-13},\;  R(x_H^{-}) \lesssim  10^{-14}$,  
 and $\left|\Delta {\cal{K}}(x_H^{+})\right| \lesssim 0.50$,  $ \left|\Delta {\cal{K}}(x_H^{-})\right| \lesssim  0.65$, respectively.  This is because now the throat coincides with the black hole horizon
 ($x_m =  x_H^+ = 0.75\times 10^{6}$), and to keep the throat open,  the quantum effects at this point must be strong enough.

  {In Fig. \ref{fig3-1b}, we plot $R$ and $\Delta  {\cal{K}}$ in the region that covers the throat ($x_{m}=0$)  as well as the two horizons ($x_H^{\pm} = \pm \sqrt{3}\times 10^{38}$). 
Thus, in the current case the throat is located far away from both of the two horizons. But, the deviations of the curvature near the two horizons  are still large. 
 In particular,   we find that $R(x_H^{+}) \lesssim  10^{-76},\; R(x_H^{-}) \lesssim   10^{-76}$, and $\left|\Delta {\cal{K}}(x_H^{+})\right|\lesssim 0.2$  and
 $\left|\Delta {\cal{K}}(x_H^{-})\right|\lesssim  0.2$ for solar mass $M_{BH}=2\times 10^{38}\; M_{Pl}$ and $M_{WH} = 2 \times 10^{38}\; M_{Pl}$. } 
   {Therefore,  in the current model the quantum gravitational effects can be  still large near the horizons even for astrophysical black holes.}
  More detailed analyses show that this is due 
 to the fact that in the current case both  $x_0$ and ${\cal{C}}$ are large
  { ($x_0 = \mathcal{C}=10^{38}$)}. Since large $x_0$ and  $\mathcal{C}$ implies large $\lambda_1$ and  $\lambda_2$, as one can see from the relations  ${\cal{C}} \equiv \left(16C^2\lambda_1^2\right)^{1/6}$  and $x_0 \equiv \frac{\lambda_2}{\sqrt{n}}$. As mentioned above, the two parameters $\lambda_1$,  $\lambda_2$ control quantum gravitational corrections. In particular, large $\lambda_1$ and $\lambda_2$ will lead to large quantum effects.


 \begin{widetext}

 \begin{figure}[h!]
 \begin{tabular}{cc}
\includegraphics[height=4.8cm]{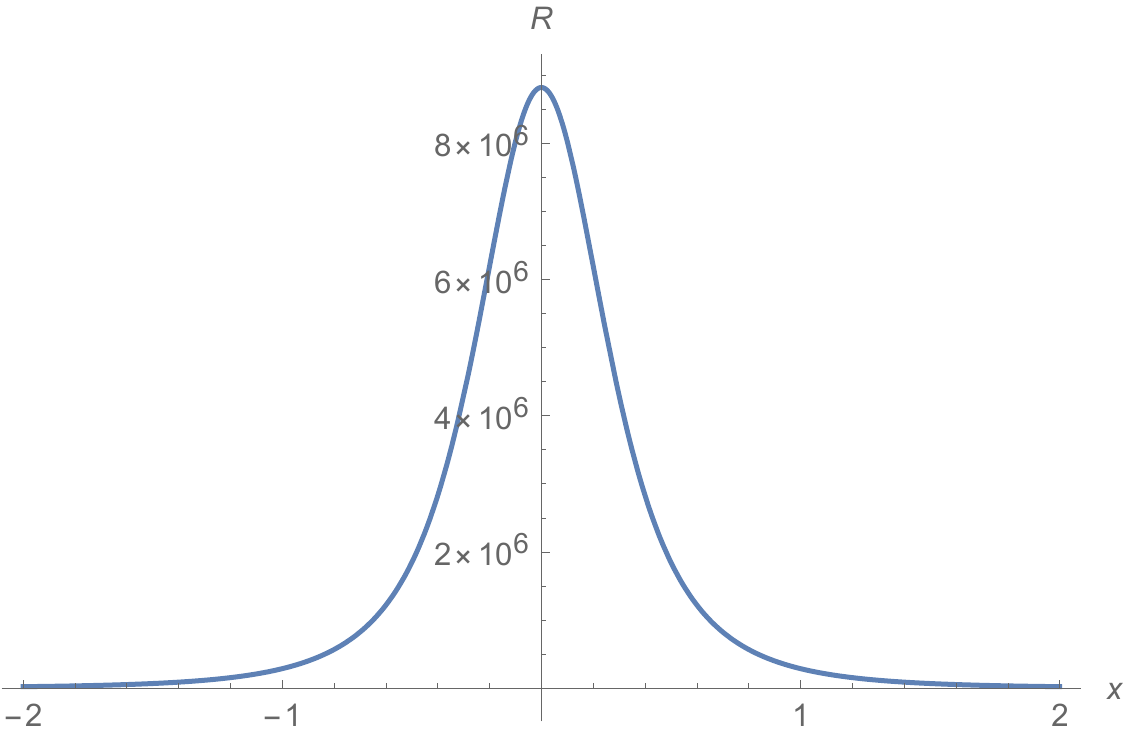}&
\includegraphics[height=4.8cm]{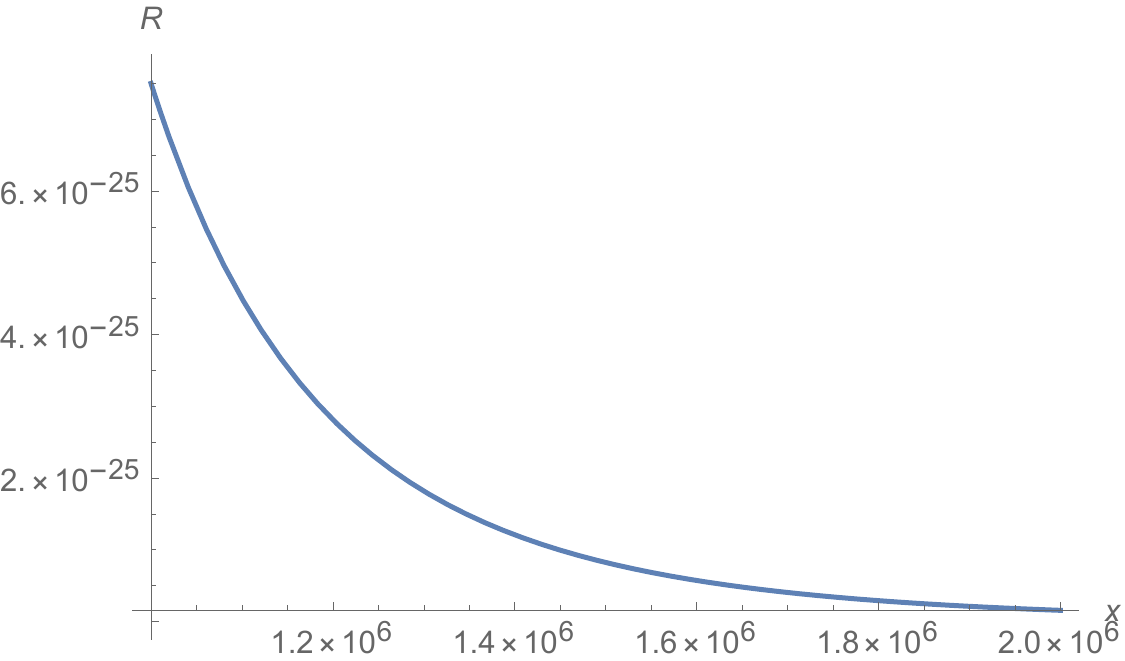}\\
(a) & (b)  \\[6pt]
\includegraphics[height=4.8cm]{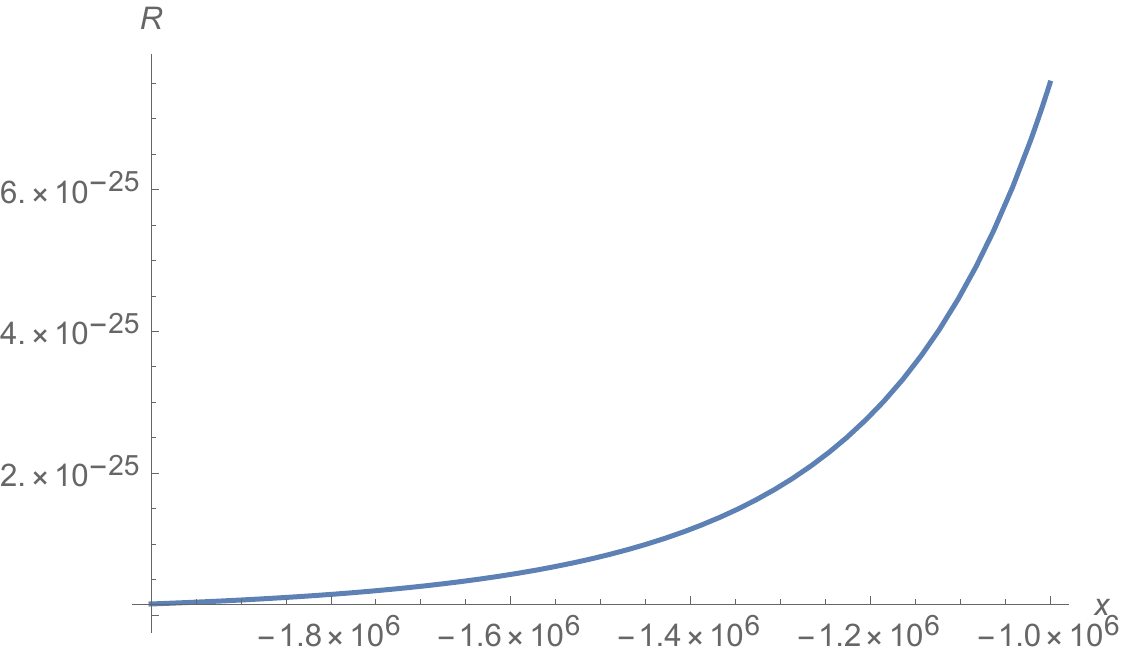}&
\includegraphics[height=4.8cm]{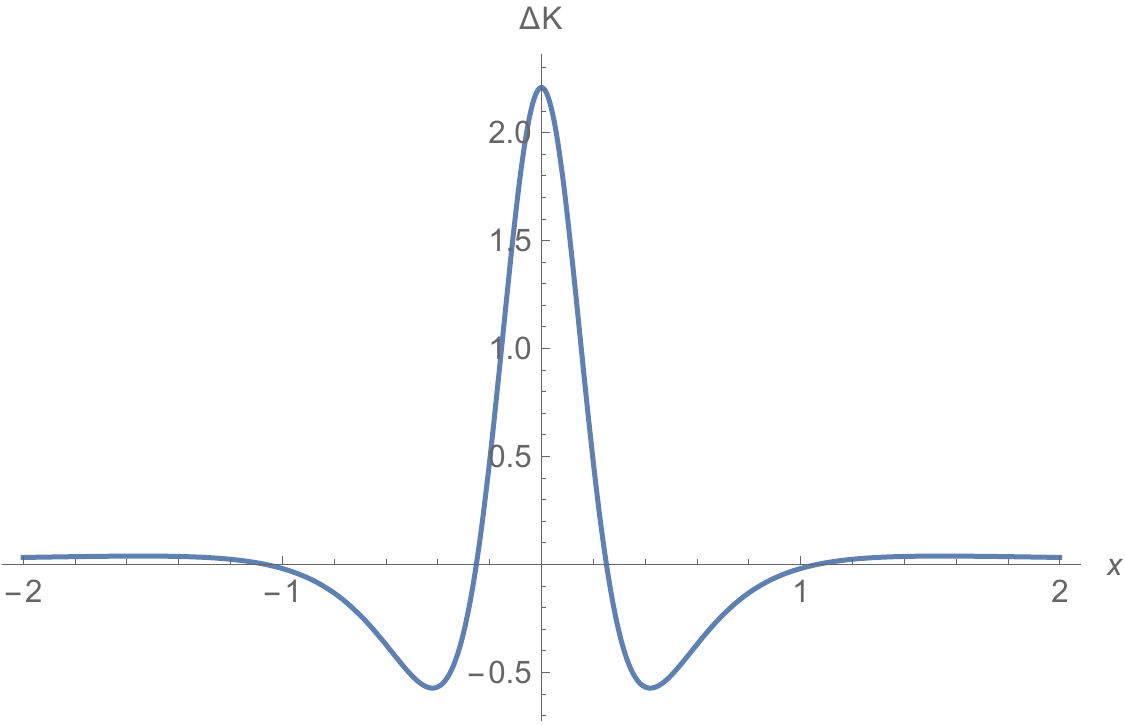}\\
(c) & (d)  \\[6pt]
\includegraphics[height=4.8cm]{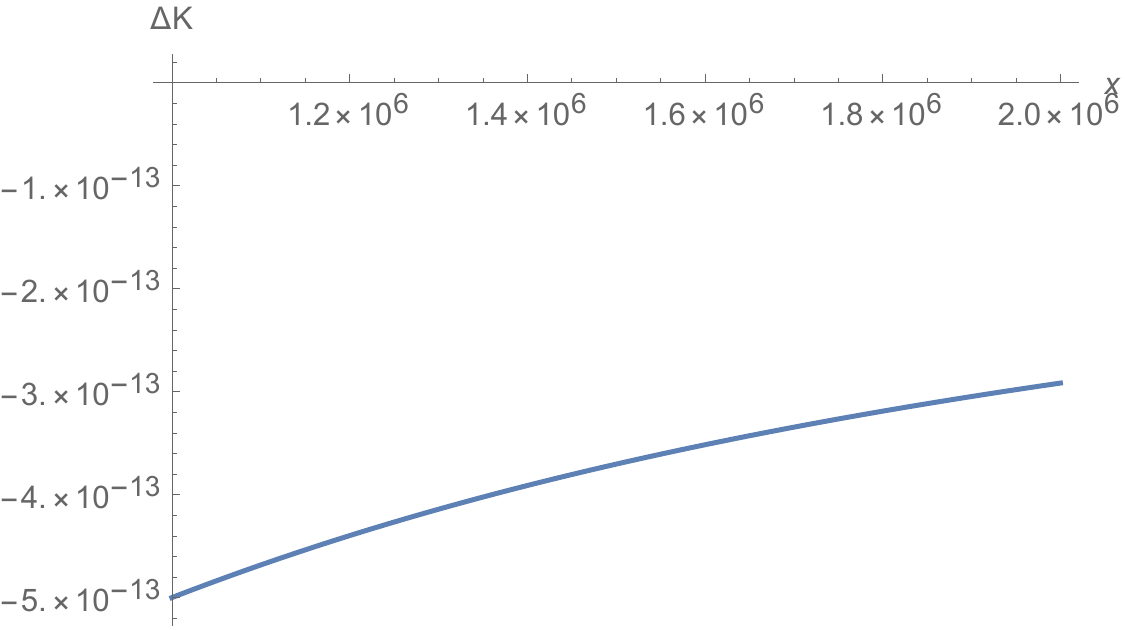}&
\includegraphics[height=4.8cm]{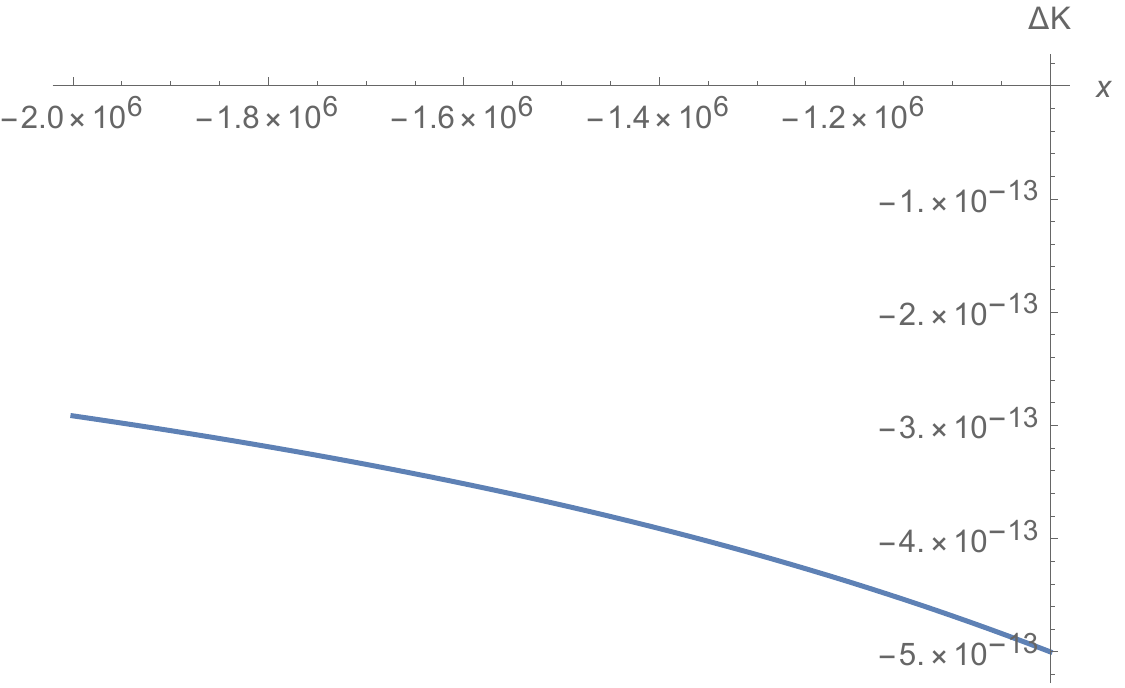}\\
(e) & (f)  \\[6pt]
\end{tabular}
\caption{Case $\Delta > 0, \; {\cal{D}} > 0,\;  |x_m| \le x_H^{+}, \beta \not=   1+ \frac{\left(\alpha -1\right)^2}{2\alpha}$: The quantities $R$ and $\Delta  {\cal{K}}$ vs $x$. 
 Here we choose $\mathcal{C}=1,\; x_0 =1,\; \mathcal{D}=2\times 10^{6}$,  for which the throat  is at $x_{m} =0$ and  the black/white hole horizons are located at 
$x_H^{\pm} \approx \pm 2\times 10^{6}$, respectively. The black and white hole masses are
$M_{BH}= M_{WH} = 2 \times 10^{6}\; M_{Pl}$. } 
\label{fig3-1c}
\end{figure}

\end{widetext}

Therefore, to have negligible  quantum gravitational effects, we must consider the cases where $\lambda_1$ and $\lambda_2$  are effectively small.
 In Fig. \ref{fig3-1c}, we plot $R$ and $\Delta  {\cal{K}}$ in the region between the two horizons for $\mathcal{C}=1,\; x_0 =1,\; \mathcal{D}=2\times 10^{6}$,  for which the horizons are located at 
$x_H^{\pm} \approx \pm 2\times 10^{6}$, and the throat  is at $x_{m} =0$, while the black and white hole masses are $M_{BH}= M_{WH}  = 2\times 10^6\; M_{Pl}$. 
 From this figure we can see that now   the deviations from GR decays rapidly when away from the throat in both directions, and near the two horizons the quantum effects already become extremely small. 
 In fact, near the two horizons now we find that $R(x_H^{+}) \lesssim 10^{-25},\; R(x_H^{-}) \lesssim  10^{-25}$, and $\left|\Delta {\cal{K}}(x_H^{+})\right| \lesssim 10^{-13}$ and $\left|\Delta {\cal{K}}(x_H^{-})\right|\lesssim  10^{-13}$.  
 Therefore, in the current case,   the quantum gravitational effects  are mainly concentrated  in the neighborhood of the throat. 


 \begin{widetext}

 \begin{figure}[h!]
 \begin{tabular}{cc}
\includegraphics[height=4.8cm]{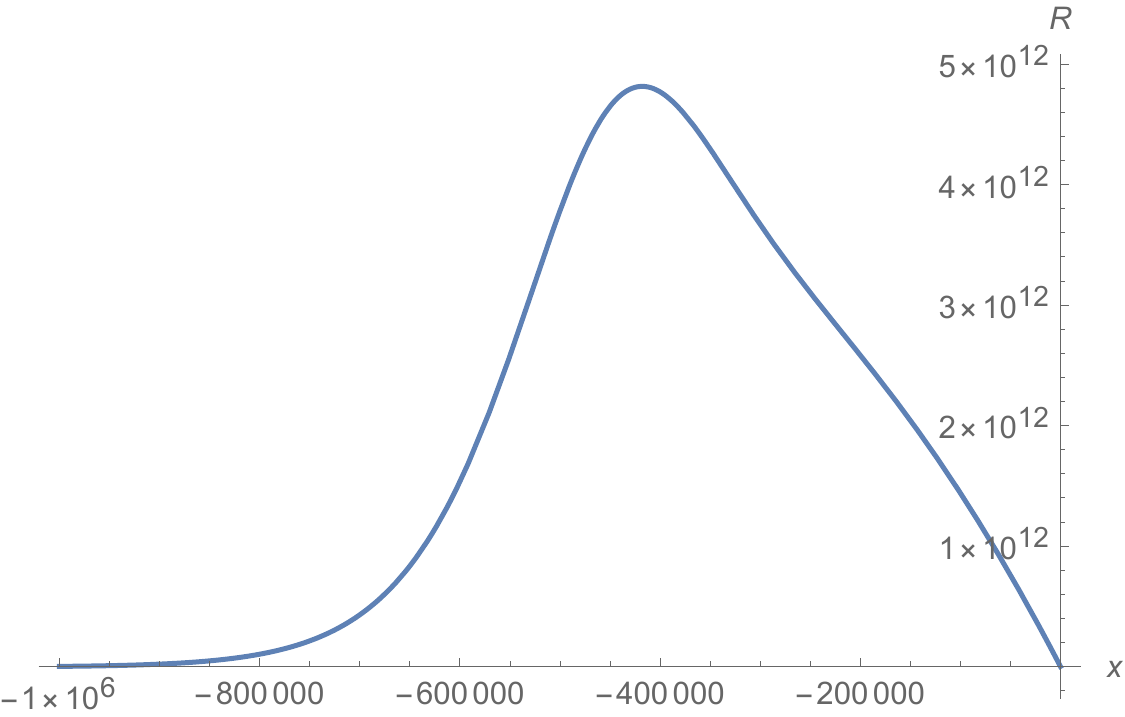}&
\includegraphics[height=4.8cm]{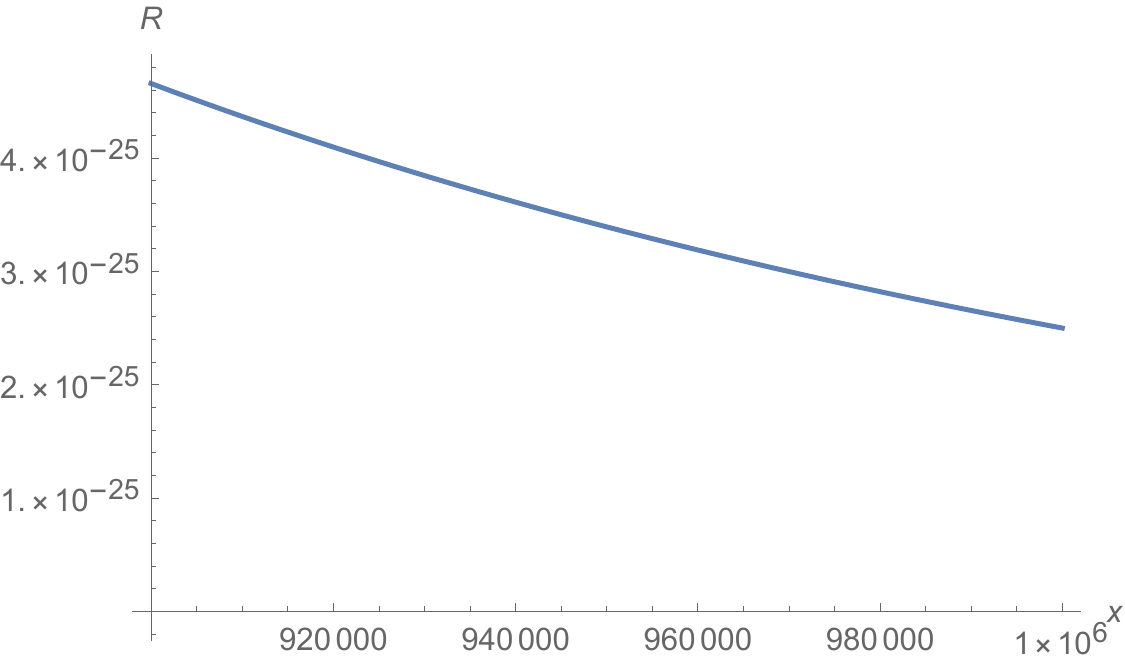}\\
(a) & (b)  \\[6pt]
\includegraphics[height=4.8cm]{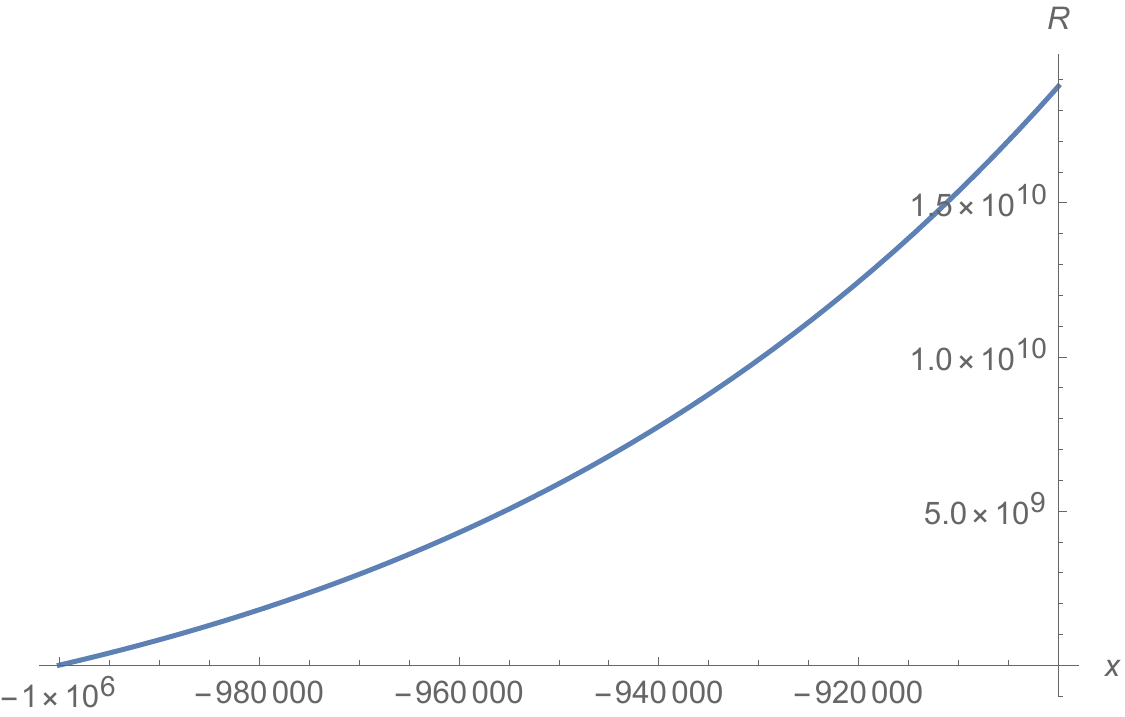}&
\includegraphics[height=4.8cm]{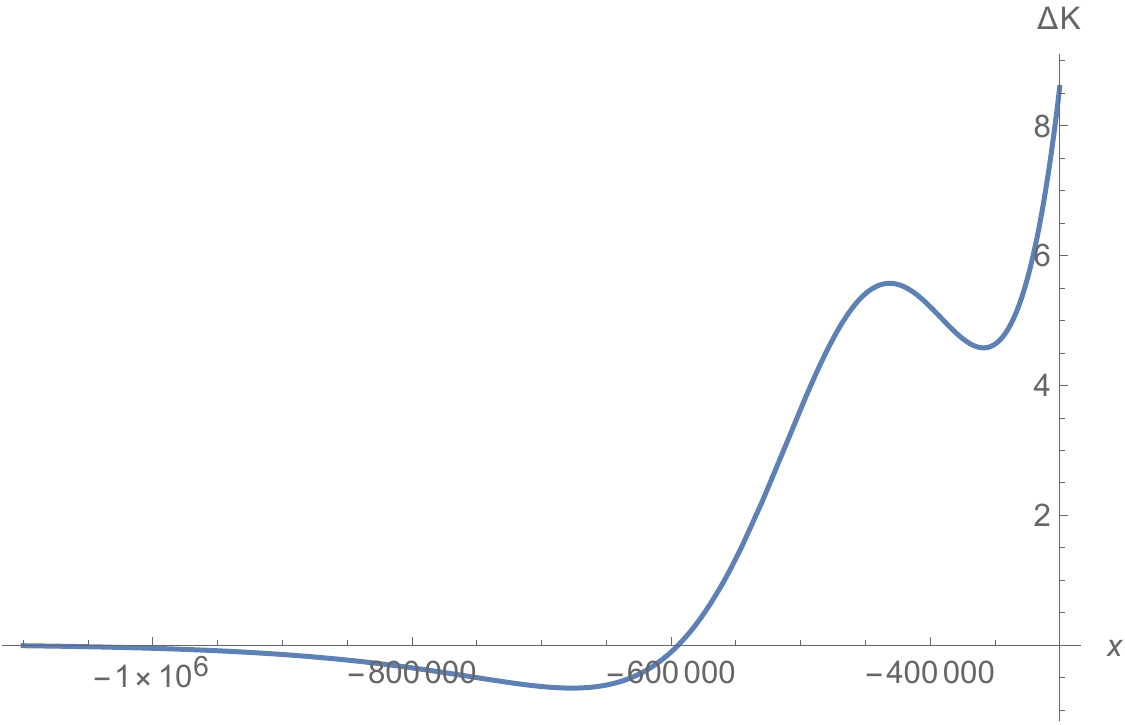}\\
(c) & (d)  \\[6pt]
\includegraphics[height=4.8cm]{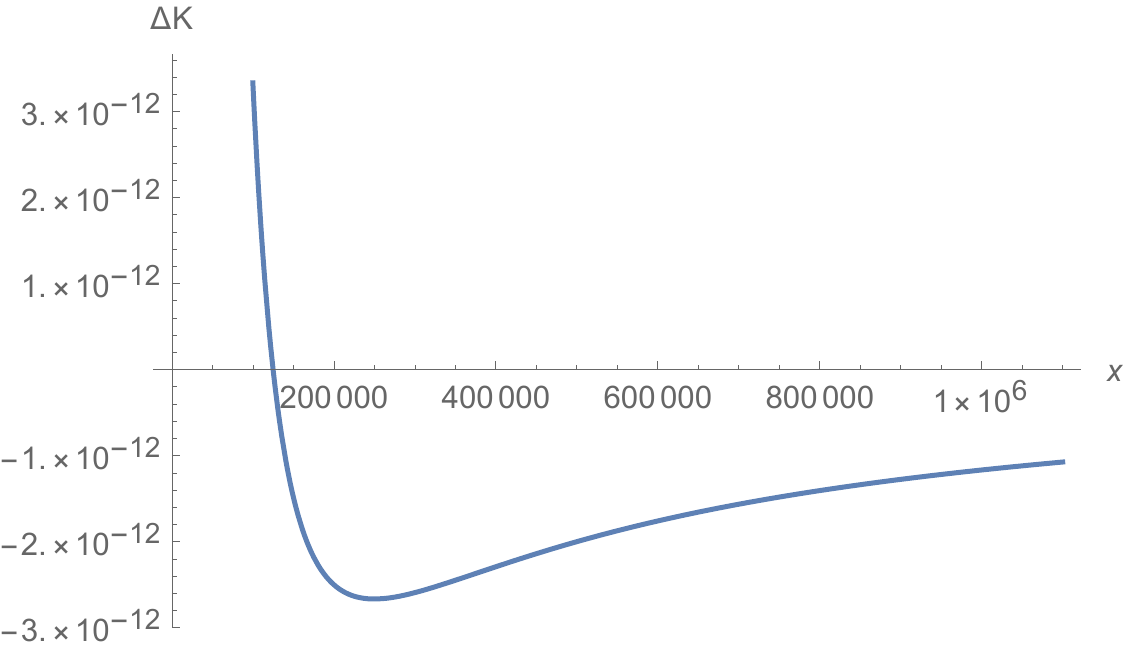}&
\includegraphics[height=4.8cm]{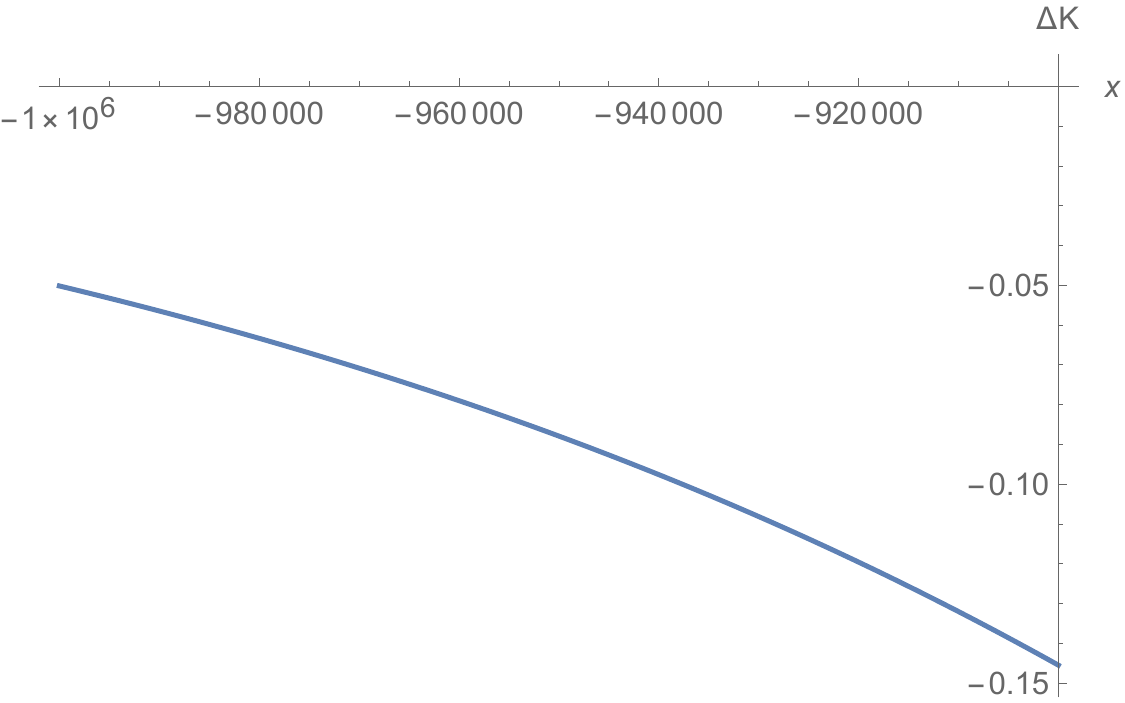}\\
(e) & (f)  \\[6pt]
\end{tabular}
\caption{Case $\Delta > 0, \; {\cal{D}} > 0,\;  |x_m| \le x_H^{+}, \beta \not=   1+ \frac{\left(\alpha -1\right)^2}{2\alpha}$: The quantities $R$ and $\Delta  {\cal{K}}$ vs $x$. 
 Here we choose $\mathcal{C}=10^{-6},\; x_0 =1,\; \mathcal{D}=10^{6}$,  for which the throat  is at $x_{m} \approx -\frac{1}{2}\times 10^{6}$ and  the black/white hole horizons are located at 
$x_H^{\pm} \approx \pm  10^{6}$, respectively. The black and white hole masses are
$M_{BH} = 10^{6}\; M_{Pl}$, $M_{WH} =  10^{-6}\; M_{Pl}$. } 
\label{fig7q}
\end{figure}

\end{widetext}


On the other hand, in Fig. \ref{fig7q} we plot $R$ and $\Delta  {\cal{K}}$ in the region between the two horizons for $\mathcal{C}=10^{-6},\; x_0 =1,\; \mathcal{D}= 10^{6}$,  for which the horizons are located at 
$x_H^{\pm} \approx \pm  10^{6}$, and the throat  is at $x_{m} \approx -\frac{1}{2}\times 10^{6}$, while the black and white hole masses are $M_{BH} = 10^{6}\; M_{Pl}$, $M_{WH} =  10^{-6}\; M_{Pl}$, respectively. 
 From this figure we can see that now the deviations from GR decays rapidly when away from the throat only in the black hole direction, that is, only for $x > x_H^+$, and near the white hole horizon the quantum effects 
 become very large again. 
 In fact, near the two horizons now we find that $R(x_H^{+}) \lesssim 10^{-25},\; R(x_H^{-}) \simeq 10^{10}$, and $\left|\Delta {\cal{K}}(x_H^{+})\right| \lesssim 10^{-12}$ and $\left|\Delta {\cal{K}}(x_H^{-})\right|\simeq  0.05$.  
 Thus, in the current case  the quantum gravitational effects  are negligible only at the black hole horizon but still very large at the white hole horizon. This is due to the fact that the throat is now very close to the white hole 
  {horizon.}
 
 
 The above examples show clearly that, depending on the values of the three free parameters ${\cal{C}}, \; {\cal{D}}, \; x_0$ (or $ {\cal{D}}, \; \lambda_1,\; \lambda_2$),  quantum gravitational effects can be
 large, even for the cases in which the black/white hole masses are of order of solar masses. In particular, near the two horizons $x = x_H^{\pm}$, we find
  \bqn
\lb{eq3.25horizon}
 R &=& \begin{cases}
-\frac{x_{0}^6}{\mathcal{D} ^2 \left(2 \mathcal{D}  \left(\sqrt{\mathcal{D} ^2-x_{0}^2}-\mathcal{D} \right)+x_{0}^2\right) {\cal{R}}_H^+}, & x  =  x_H^+,\cr
\frac{x_{0}^6}{\mathcal{D} ^2 \left(2 \mathcal{D}  \left(\mathcal{D} +\sqrt{\mathcal{D} ^2-x_{0}^2}\right)-x_{0}^2\right)  {\cal{R}}_H^-}, & x  =  x_H^-,\cr
\end{cases}
\eqn
where ${\cal{R}}_H^{\pm} \equiv \left(\left(\mathcal{D} \pm \sqrt{\mathcal{D} ^2-x_{0}^2}\right)^6+\mathcal{C} ^6\right)^{2/3}$. Thus, for ${\cal{D}} \gg |x_0|$, we have
  \bqn
\lb{eq3.25horizon}
 R &\simeq& \begin{cases}
\frac{4x_{0}^2}{\left[(2{\cal{D}})^6 + \mathcal{C}^6\right]^{2/3}}, & x  =  x_H^+,\cr
\frac{x_{0}^6}{4\mathcal{D}^4\mathcal{C}^4}, & x  =  x_H^-.\cr
\end{cases}
\eqn
Therefore, to have the effects negligibly small near the two horizons, we must require
 \bq
  \lb{eq3.25e}
  {\cal{C}}  \gtrsim |x_0|, \quad   {\cal{D}} \gg |x_0|.
  \eq

On the other hand, as $x \rightarrow \pm \infty$, we find that
\bqn
\lb{eq3.25c}
 R &\simeq& \begin{cases}
-\frac{x_{0}^2}{4 x^4}+\frac{\mathcal{D}  x_{0}^2}{2 x^5}+{\cal{O}}\left(\epsilon^6\right), & x  \rightarrow  \infty,\cr
-\frac{x_{0}^6}{4 x^4 \mathcal{C} ^4}-\frac{\mathcal{D}  x_{0}^6}{2 x^5 \mathcal{C} ^4}+{\cal{O}}\left(\epsilon^6\right), & x  \rightarrow  -\infty,\cr
\end{cases}
\eqn
and 
\bqn
\lb{eq3.25d}
\Delta {\cal{K}}  &\simeq&  \begin{cases}
-\frac{4 x_{0}^2}{3 M_{BH} x}+{\cal{O}}\left(\epsilon^2\right), &  x  \rightarrow   \infty,\cr
 +\frac{4   \mathcal{C} ^2}{3 M_{WH} x }  +{\cal{O}}\left(\epsilon^2\right), &  x  \rightarrow   - \infty,\cr
 \end{cases}
\eqn
where $M_{BH}$ and $M_{WH}$ are given by Eq.(\ref{eq3.22}). Then, we have   $|\Delta {\cal{K}}_+/\Delta {\cal{K}}_-|=1 + {\cal{O}}\left(\epsilon^2\right)$, as $|x| \rightarrow \infty$. 
That is,  whether $M_{WH} \gg M_{BH}$ or not, $|\Delta {\cal{K}}_+|$ will always have the same asymptotic magnitude as $|\Delta {\cal{K}}_-|$, and both of them approach their GR limits as
${\cal{O}}(1/|x|)$.

Therefore, in the present case we find the following:

\begin{itemize}

\item The throat is always located in the region between the black and white hole horizons, $ x_H^- \le x_m \le x_H^+$, and each of the three energy conditions, WEC, DEC, and SEC, is satisfied at the throat only
in the case where the condition (\ref{eq3.14-3a}) holds. In this case the quantum gravitational effects are always large at the  {black hole horizon $x = x_H^+$.  This is expected, 
as at the throat the quantum effects need to be strong, in order to  keep the throat open, and when the condition (\ref{eq3.14-3a}) is satisfied, the black hole horizon always coincides with the throat,
$x_m = x_H^+$.}

\item Even  the condition (\ref{eq3.14-3a}) does not hold, and the throat is far from both of the white and black hole horizons, that is, $\left|x_m\right| \ll \left|x_H^{\pm}\right|$, the quantum gravitational effects can be still large
  at the two horizons, including the cases in which both of the white and black hole masses  are large, $M_{BH}, \; M_{WH} \gg 10^{6}\; M_{Pl}$. Only in the case where the conditions (\ref{eq3.25e}) hold, 
  can the effects become negligible at the two horizons.

\item  {In general,  none of the three energy conditions is satisfied in the neighborhoods of  the white and black hole horizons, $x = x_{H}^{\pm}$, except precisely at these two surfaces.  }
 However, the surface gravity at the black (white) hole horizon is always positive (negative), as now the condition $\rho + p_r + 2p_{\theta} > 0$ is still satisfied in the most part of the spacetime \cite{Komar59}, as can be seen 
 from Figs. \ref{fig2rho} and  \ref{fig2rhoB}.  So, the trapped region ($x_H^{-} < x < x_H^{+}$) is still attractive to observers outside of it.
 
 \item In the two asymptotically flat regions $x \rightarrow \pm \infty$, for which the geometrical radius becomes infinitely large, $b(\pm \infty) = \infty$, none of the
three energy conditions  is satisfied.   

\item The black and white hole masses read off from these two asymptotically flat regions are given by Eq.(\ref{eq3.22}), which are always  positive, no matter the condition (\ref{eq3.14-3a}) is satisfied or not. 
Again, this is because the relativistic  {Komar mass density $\rho + p_r + 2p_{\theta}$ is still positive in a large part of the spacetime. As a result, the total masses of the spacetime read off at the two asymptotically 
flat region are  positive. }

\end{itemize}

 It should be noted that the absence of spacetime singularities in this case does not contradict to the Hawking-Penrose singularity theorems \cite{HE73}, as now none of the three energy conditions is satisfied in
 the two  asymptotically flat regions, including the case in which the condition (\ref{eq3.14-3a}) holds,  { as shown in the above explicitly. }

 \subsubsection{$|x_m|>x_{H}^{+}$}
 
  Now, let us turn to consider the case $|x_m| >x_{H}^{+}$,  which implies that 
   \bq
 \lb{eq3.8a}
  \beta <  1 + \frac{\left(\alpha - 1\right)^2}{2\alpha}.
 \eq
In this case, since the throat is located in the region where $a(x) > 0$,  then   at the throat we have $\rho = \rho^+$ and $p_r = p_r^+$. Hence, from Eq.(\ref{eq3.5}) we find that
 the effective energy density $\rho$ and pressures $p_r$ and $p_{\theta}$ at the throat are given by
  \bqn
 \lb{eq3.8}
  \rho  &= & \frac{\mathcal{C}  (12 \mathcal{D} -5 \mathcal{C} )-5 x_{0}^2}{2^{2/3} \mathcal{C} ^2 \left(x_{0}^2+\mathcal{C} ^2\right)},\nb\\
  p_r  &= & -\frac{1}{2^{2/3} \mathcal{C} ^2},\nb\\
  p_{\theta}  &= & \frac{{\left(x_{0}^2+\mathcal{C} ^2\right)^3} -{4 \mathcal{D}  x_{0}^2 \mathcal{C} ^3}}{2^{2/3} \mathcal{C} ^2{\left(x_{0}^2+\mathcal{C} ^2\right)^3}}.
 \eqn
 From these expressions, we find that in the 3D parameter space,   WEC is satisfied when,
\bqn
\lb{eq3.14}
\beta   \ge  1 + \frac{\left(\alpha -1\right)^2}{2\alpha}, \quad {\text{and}} \quad \beta  > \frac{1}{2} \alpha.  
 \eqn
Clearly, these conditions contradict to the condition $|x_m| >x_{H}^{+}$, as it can be seen from Eq.(\ref{eq3.8a}). Therefore, in the current case WEC is always violated at the throat.
 In addition, for $\rho, \; p_r$ and $p_{\theta}$ given by  Eq.(\ref{eq3.8}), we also find that   neither DEC nor SEC   is  satisfied, after the conditions (\ref{eq3.8a}) are taken into account. 
Therefore, in the current case, {\it none of the three energy conditions is satisfied at the throat.}

On the other hand, following the analyses provided in the last subsection,  it can be also shown that in the current case the following is true:
 (i) All the three energy conditions are not satisfied generically  in the regions near the 
black hole and white hole horizons in the whole 3D phase space. But, the surface gravity at the black (white) hole horizon can be still  positive (negative), as the relativistic Komar mass density can be still positive
over a large region of the spacetime, so that its integration over the 3D spatial space can be positive, $\int_V{\left(\rho + p_r + 2p_{\theta}\right) dV} > 0$. 
(ii) In the two asymptotically flat regions $x \rightarrow \pm \infty$,  none of the
three energy conditions  is satisfied for any given values of ${\cal{C}}, \; {\cal{D}}$ and $x_0$, as longer as the condition (\ref{eq3.6b})   holds, which is resulted from the condition $\Delta > 0$.
 (iii)  The black/white hole masses are also given by Eq.(\ref{eq3.22}), which are all positive in the current case, too. 
 (iv) The quantum effects are mainly concentrated near the throat.   Since now the throat is always located either outside the black hole horizon ($ x_m > x_H^+$) or outside the white hole horizon 
 ($ x_m < x_H^-$), the quantum effects can be large near the two horizons, even for the cases where the white/black hole masses are of order of  solar masses.

 It should be  noted that the above analysis  { is not valid for   the limit cases  $x_0 \rightarrow 0$ and ${\cal{C}} \rightarrow 0$. So, in } the following, let us consider these particular
 cases, separately.

\subsubsection{$\; x_0 = 0, \; {\cal{C}} \not= 0$}

 {If we assume that } $\lambda_2 \not= 0$, from Eq.(\ref{eq2.3}) we can see that this corresponds to the limit $\sqrt{n} \rightarrow \infty$. However, to keep ${\cal{D}} >0$ and finite, we 
 must require $D/\sqrt{n} \rightarrow \text{finite}$ and $CD > 0$. Then, we find that $\Delta = {\cal{D}}^2$, and from Eq.(\ref{eq2.6}) we find $X = |x|$, and 
  \bqn
\lb{eq3.26}
Y =x +|x| = \begin{cases}
2x, & x \ge 0,\cr
0, & x < 0.\cr
\end{cases}
\eqn
 Hence, Eq.(\ref{eq2.5}) implies $a(x) = 0$ and $b(x) = \infty$ for $x \le 0$, that is, the metric becomes singular for $x \le 0$. 
 However, since $b(0) = \infty$, it is clear that now $x = 0$
 already represents the spatial infinity. Therefore, in this case we only need to consider  the region $x \in (0, \infty)$ [cf. Fig. \ref{fig1a-c}(b)]. Then, we find that 
 \bqn
\lb{eq3.27}
X = x, \;\; Y = 2x, \;\; Z = 4 \left(x^6 + \hat{\cal{C}}^6\right)^{1/3},\; (x \ge 0), ~~~~~~
\eqn
where $\hat{\cal{C}} \equiv {\cal{C}}/2$,  and 
 \bqn
\lb{eq3.28}
a(x) &=& \frac{x^3\left(x -\mathcal{D}\right)}{4\left(x^6 + {\hat{\cal{C}}}^6\right)^{2/3}}, \nb\\
  b(x) &=& \frac{2}{x}\left(x^6 + {\hat{\cal{C}}}^6\right)^{1/3}.
\eqn
 Clearly, $a(x) = 0$ leads to two roots,
 \bq
 \lb{eq3.29}
 x^{-}_H= 0, \quad x^{+}_H = \mathcal{D},
 \eq
 while the minimum of $b(x)$ now is located at $x_{m} \equiv {\hat{\cal{C}}}$, so we have
 \bq
 \lb{eq3.30}
b(x) = \begin{cases}
\infty, & x = 0,  \cr
2^{4/3}{\hat{\cal{C}}}, &  x = {\hat{\cal{C}}},\cr
\infty, & x = \infty.\cr
\end{cases}
 \eq
  It is interesting to note that  the outer (black hole) horizon located at $x = x^{+}_H$ can be  smaller than the throat $ x = x_{m}$, that is, ${\hat{\mathcal{C}}} > {\mathcal{D}}$. In addition,
    the spacetime becomes antitrapped at $x^-_H = 0$. Since $b(x= 0) = \infty$, this antitrapped point now also corresponds to the spatial infinity at the other side (the white hole
    side) of the throat. 
  
To study the solutions further, in the following let us consider the cases ${\mathcal{D}} \ge {\hat{\mathcal{C}}}$ and $ {\mathcal{D}} < {\hat{\mathcal{C}}}$, separately.

   \begin{figure}[h!]
\includegraphics[height=4.8cm]{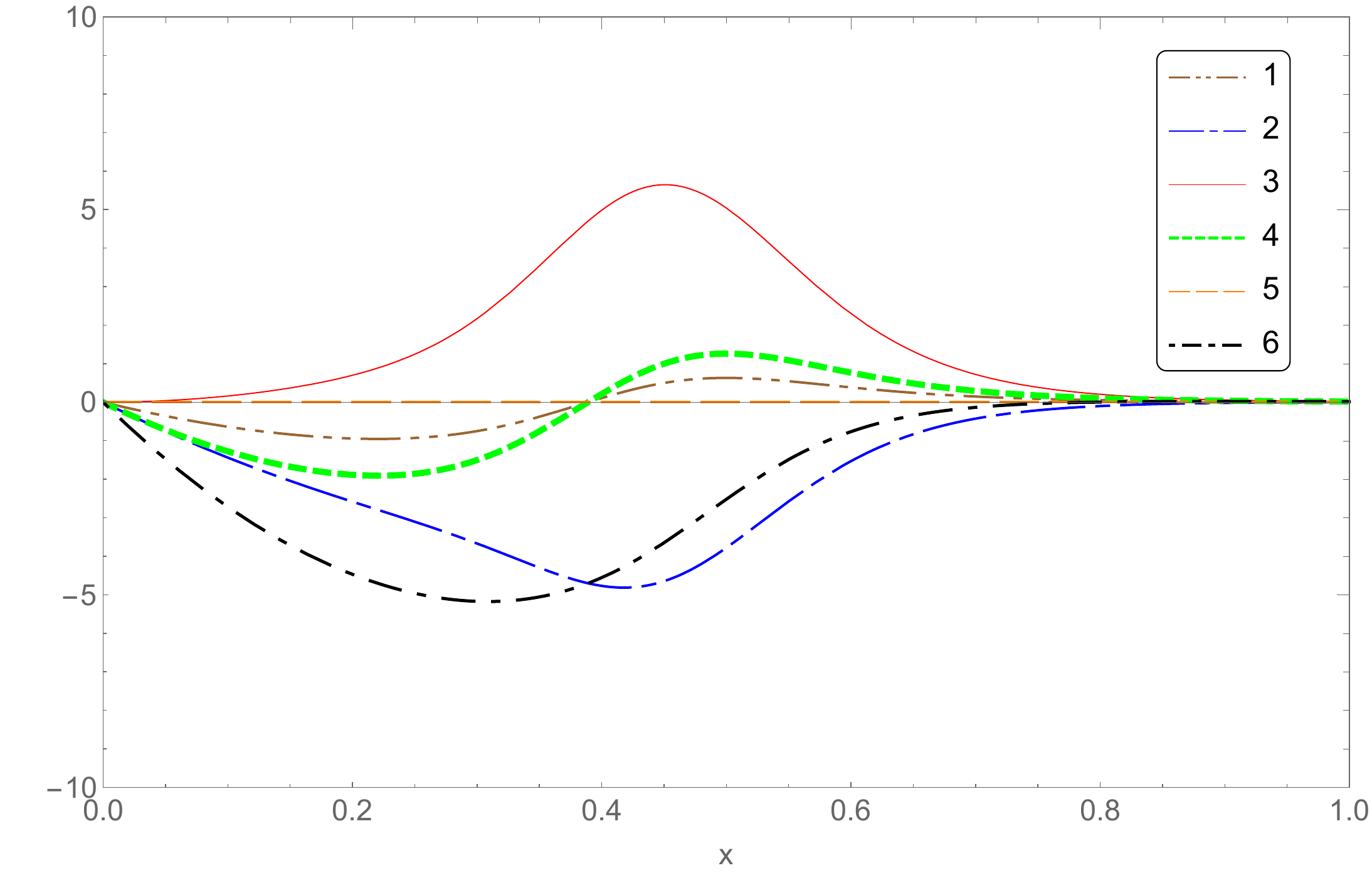}
\caption{Case $\Delta > 0, \; {\cal{D}} > 0,\; x_0 = 0,\; {\cal{C}} \not =0$: The  physical quantities, $\rho$, $(\rho + p_r)$,  $(\rho - p_r)$, $(\rho + p_{\theta})$, $(\rho - p_{\theta})$, and  $(\rho + p_r + 2p_{\theta})$, represented, respectively, by Curves 1 - 6,  vs $x$ in the neighborhood of the throat.
 All curves are plotted with $\mathcal{C}=1,\; \mathcal{D}=1$, for which
the throat is at $x_{m} = 0.5$, and the outer horizon is at $x^{+}_H = 1$.} 
\lb{fig5}
\end{figure}

{\it (Case III.3.1)  ${\mathcal{D}} \ge {\hat{\mathcal{C}}}$:}  In this case the throat locates always inside the black hole horizon, so in the region $x < x_H^+$ we always have 
$a (x) < 0$, and the corresponding   effective energy density and pressures are given by
 \bqn
\lb{eq3.31a}
\rho(x) &=& \frac{\mathcal{C} ^6 \left[64 \mathcal{D}  x^6+\mathcal{C} ^6 (2 x-\mathcal{D} )\right]x}{2^{13}\left(x^6+\hat{\mathcal{C}} ^6\right)^{8/3}}, \nb\\
 p_r(x) &=& -\frac{\mathcal{C} ^6 \left(\mathcal{D}  \mathcal{C} ^6-640 x^7+704 \mathcal{D}  x^6\right)x}{2^{13}\left(x^6+\hat{\mathcal{C}} ^6\right)^{8/3}}, \nb\\
  p_{\theta}(x) &=& \frac{\mathcal{C} ^6 \left[64 \mathcal{D}  x^6+\mathcal{C} ^6 (2 x-\mathcal{D} )\right]x}{2^{13}\left(x^6+\hat{\mathcal{C}} ^6\right)^{8/3}}.
\eqn
 In particular, at the throat $(x =\hat{\mathcal{C}})$, we have
 \bqn
 \lb{eq3.33a}
 \rho =p_{\theta}=\frac{1}{2^{2/3} \mathcal{C} ^2},\;\;
  p_r  = \frac{5 \mathcal{C} -12 \mathcal{D}  }{2^{2/3} \mathcal{C} ^3},
  \eqn
from which we find that   the WEC, SEC, and DEC are satisfied in the domain,
\bqn
\lb{eq3.34-1}
2\mathcal{D} \le {\mathcal{C}} \le  3\mathcal{D}.
 \eqn
Combining Eq.(\ref{eq3.34-1}) with ${\mathcal{C}}/2 \le  \mathcal{D}$, we have ${\mathcal{C}}/2 =  \mathcal{D}$, which implies that the effective energy-momentum tensor satisfies all the three energy conditions at the throat only when the location of the throat and location of the black hole horizon coincide. 
 
  In Fig. \ref{fig5} we plot the physical quantities $\rho, \; \rho \pm p_r,\;  \rho \pm p_{\theta}$ and $ \rho + p_r + 2  \pm p_{\theta}$ in the neighborhood of the throat.

\begin{widetext} 

 \begin{figure}[h!]
 \begin{tabular}{cc}
 \includegraphics[height=4.8cm]{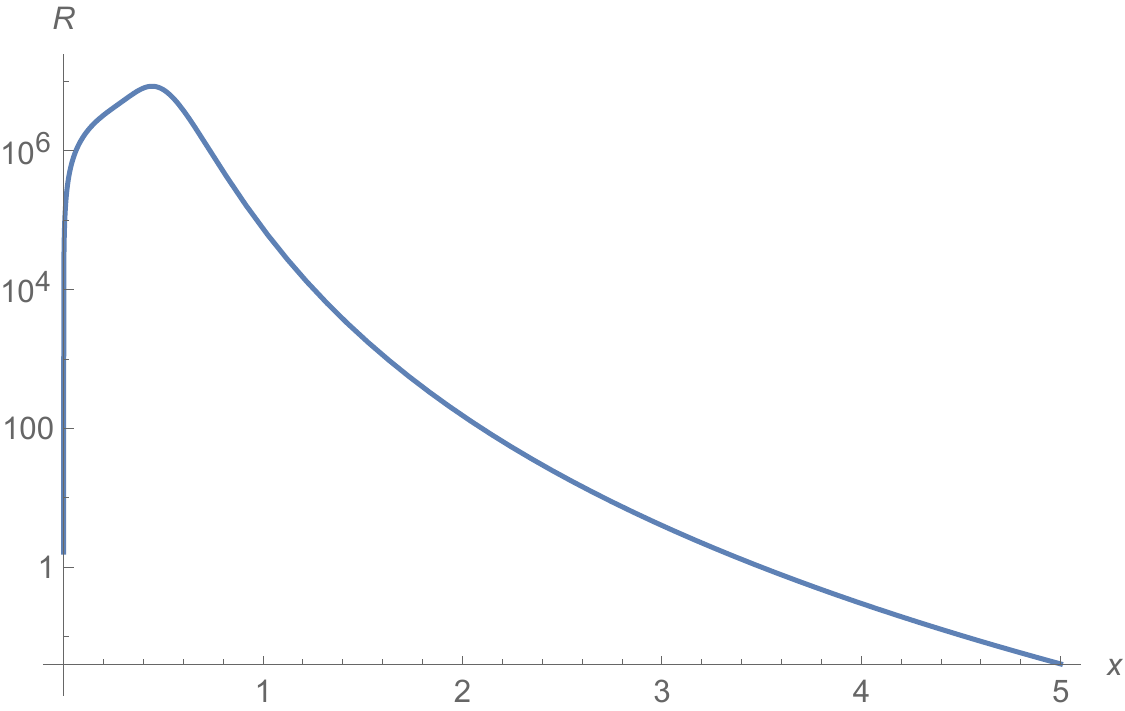}&
\includegraphics[height=4.8cm]{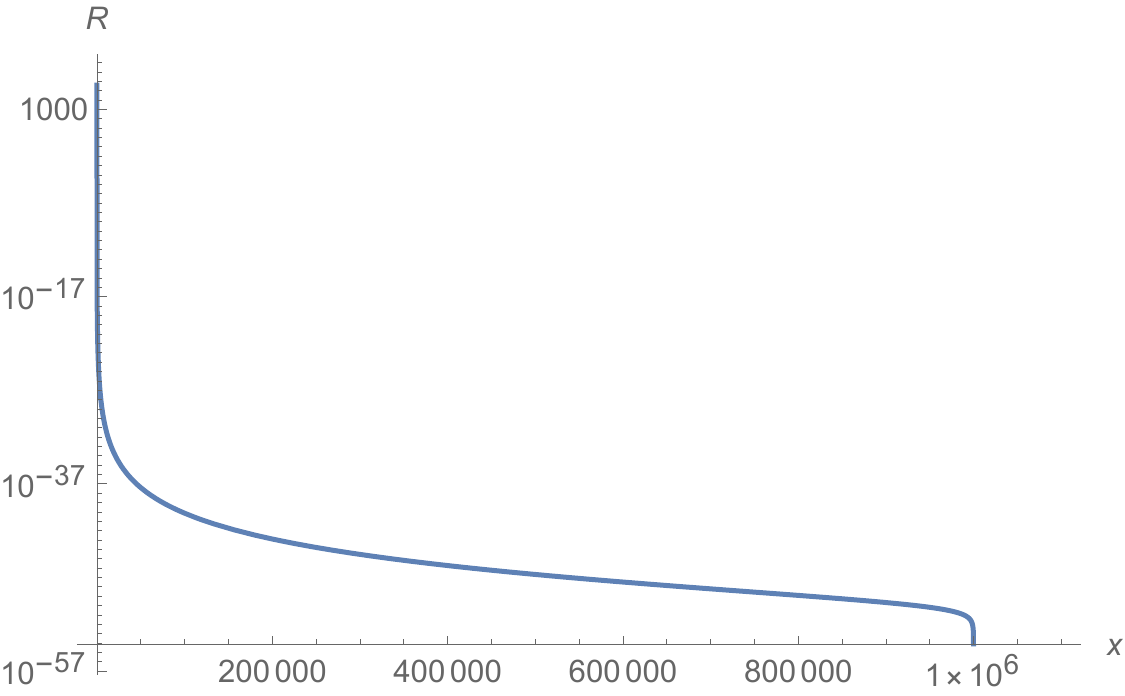}\\
(a) & (b)  \\[6pt]
\includegraphics[height=4.8cm]{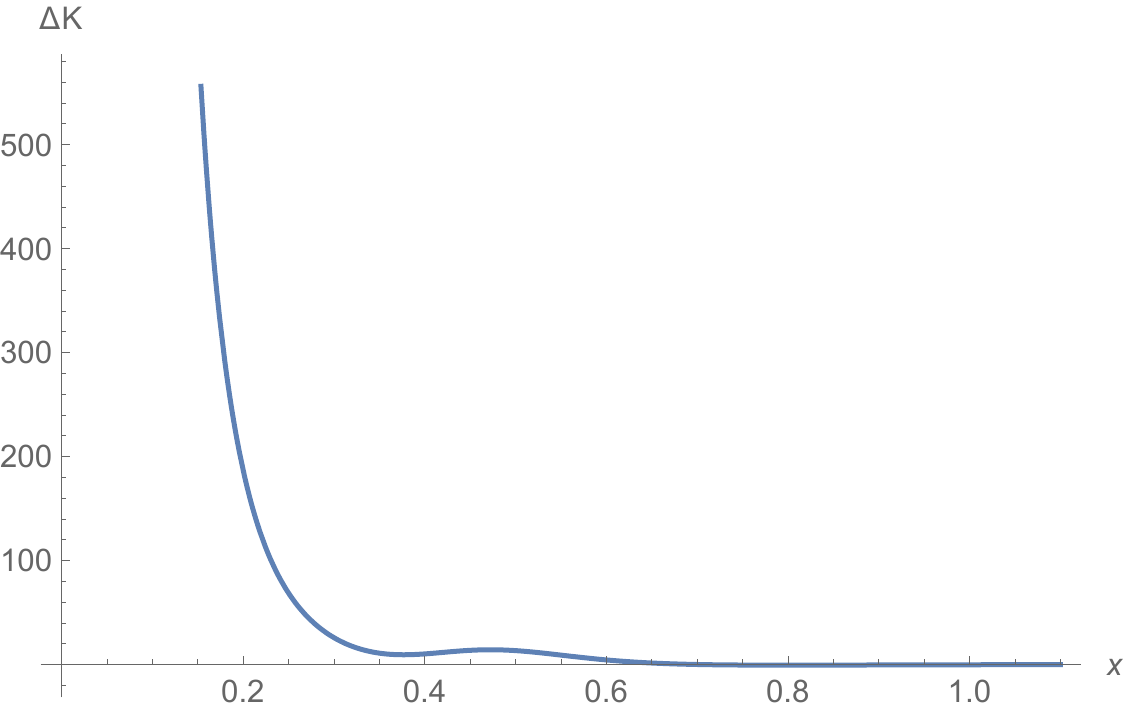}&
\includegraphics[height=4.8cm]{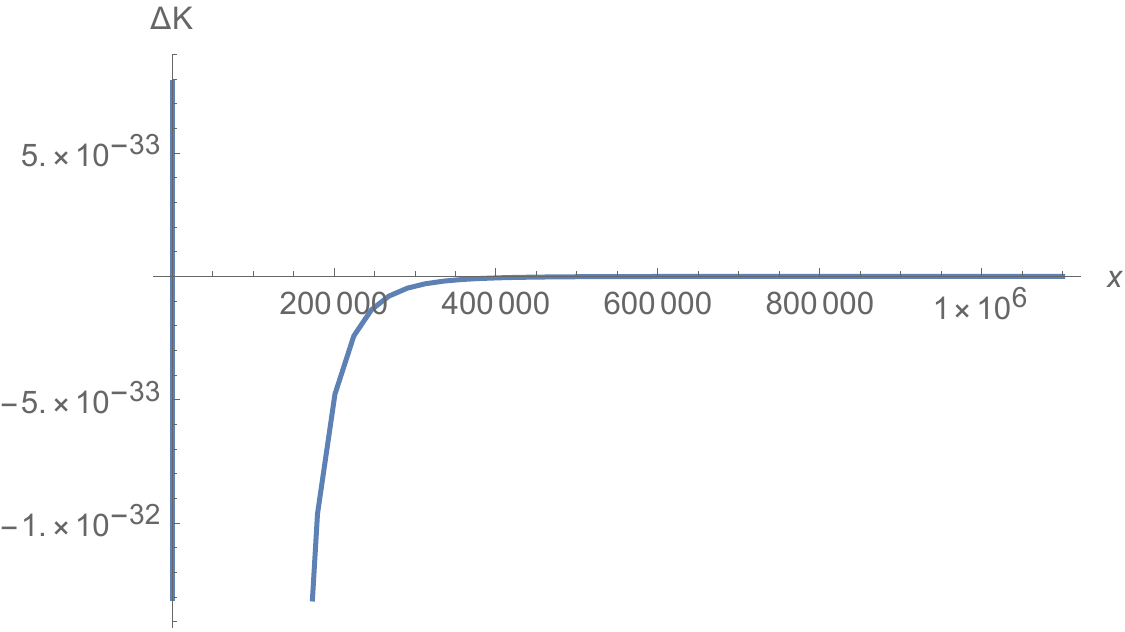}\\
(c) & (d)  \\[6pt]
\end{tabular}
\caption{Case $\Delta > 0, \; {\cal{D}} > 0,\; x_0 = 0,\; {\cal{C}} \not =0$: The physical quantities $R$ and $\Delta  {\cal{K}}$.   Here we choose $\mathcal{C}=1$, $\mathcal{D}=10^6$, so that $M_{BH}=10^6\; M_{Pl},\; x^+_H =  \mathcal{D}=10^6,\; x_m =\hat{\cal{C}}=1/2$.} 
\label{fig6-1}
\end{figure}

\end{widetext} 

 In addition, as $x \rightarrow 0 $ (or $b(x) \rightarrow \infty$), we find that 
\bqn
 \lb{eq3.33b}
 \rho &=&  p_{\theta} =-\frac{8 \mathcal{D}  x}{\mathcal{C} ^4}+\frac{16 x^2}{\mathcal{C} ^4}+\mathcal{O}\left(x^3\right),\nb\\
  p_r  &=& -\frac{8  \mathcal{D}  x}{\mathcal{C} ^4}+\mathcal{O}\left(x^3\right),
  \eqn
 from which we find that   the WEC, SEC, and DEC are satisfied  only at $x=0$.

On the other hand, outside of the black hole horizon ($x > x_H^+$),  we always have 
$a (x) >  0$, and the corresponding   effective energy density and pressures are given by
 \bqn
\lb{eq3.31}
\rho(x) &=& \frac{\mathcal{C} ^6 \left(\mathcal{D}  \mathcal{C} ^6-640 x^7+704 \mathcal{D}  x^6\right)x}{2^{13}\left(x^6+\hat{\mathcal{C}} ^6\right)^{8/3}}, \nb\\
 p_r(x) &=& -\frac{\mathcal{C} ^6 \left[64 \mathcal{D}  x^6+\mathcal{C} ^6 (2 x-\mathcal{D} )\right]x}{2^{13}\left(x^6+\hat{\mathcal{C}} ^6\right)^{8/3}}, \nb\\
  p_{\theta}(x) &=& \frac{\mathcal{C} ^6 \left[64 \mathcal{D}  x^6+\mathcal{C} ^6 (2 x-\mathcal{D} )\right]x}{2^{13}\left(x^6+\hat{\mathcal{C}} ^6\right)^{8/3}}.
\eqn
In particular,  at the black hole horizon $(x^+_H =\mathcal{D})$, we have
 \bqn
 \lb{eq3.32}
 \rho &=& -p_r  =  p_{\theta} =  \frac{8 \mathcal{D} ^2 \mathcal{C} ^6}{\left(64 \mathcal{D} ^6+\mathcal{C} ^6\right)^{5/3}},  
  \eqn
 so all the three energy conditions, WEC, SEC, and DEC, are satisfied at the black hole horizon. The surface gravity now is given by,
\begin{equation}  
\lb{eq3.32a}
\kappa_{BH} \equiv \frac{1}{2}a'(x ={\cal{D}}) =\frac{2 \mathcal{D} ^3}{\left(64 \mathcal{D} ^6+\mathcal{C} ^6\right)^{2/3}},
\end{equation}
which is always positive, as now we have  $\mathcal{D}>0$.

 At the spatial infinity $x \rightarrow \infty$,  we find
  \bqn
 \lb{eq3.34}
  \rho  &\approx&-\frac{5 \mathcal{C} ^6}{64 x^8}+\frac{11 \mathcal{D}  \mathcal{C} ^6}{128 x^9}+{\cal{O}}\left(\epsilon^{10}\right),\nb\\
     \rho + p_r &\approx & -\frac{5 \mathcal{C} ^6}{64 x^8}+\frac{5 \mathcal{D}  \mathcal{C} ^6}{64 x^9}+{\cal{O}}\left(\epsilon^{10}\right),\nb\\
  \rho + p_{\theta}   &\approx & -\frac{5 \mathcal{C} ^6}{64 x^8}+\frac{3 \mathcal{D}  \mathcal{C} ^6}{32 x^9}+{\cal{O}}\left(\epsilon^{10}\right),\nb\\
   \rho +  p_r  + 2 p_{\theta}   &\approx & -\frac{5 \mathcal{C} ^6}{64 x^8}+\frac{3 \mathcal{D}  \mathcal{C} ^6}{32 x^9}+{\cal{O}}\left(\epsilon^{10}\right),
 \eqn
from which we can see that none of the three energy conditions is satisfied. In addition, we also have
\bqn
\lb{eq3.34b}
a(x) &=& \begin{cases}
 \frac{1}{4}\left(1-\frac{2\mathcal{D} }{b}\right) +  {\cal{O}}\left(\epsilon^2\right), & x  \rightarrow  \infty,\cr
-\frac{4 \mathcal{D}  x^3}{\mathcal{C} ^4}+\frac{4 x^4}{\mathcal{C} ^4}+{\cal{O}}\left(x^6\right), & x  \rightarrow  0,\cr
\end{cases}\nb\\
b(x) &\simeq&  \begin{cases}
2x, & x  \rightarrow  \infty,\cr
\frac{\mathcal{C} ^2}{2 x}+\frac{32 x^5}{3 \mathcal{C} ^4}+{\cal{O}}\left(x^6\right), & x  \rightarrow  0.\cr
\end{cases}
\eqn
 {Therefore,} the mass of the black  hole is given by 
\bqn
\lb{eq3.34c}
M_{BH} &=& \mathcal{D}.
\eqn

 \begin{figure}[h!]
\includegraphics[height=4.8cm]{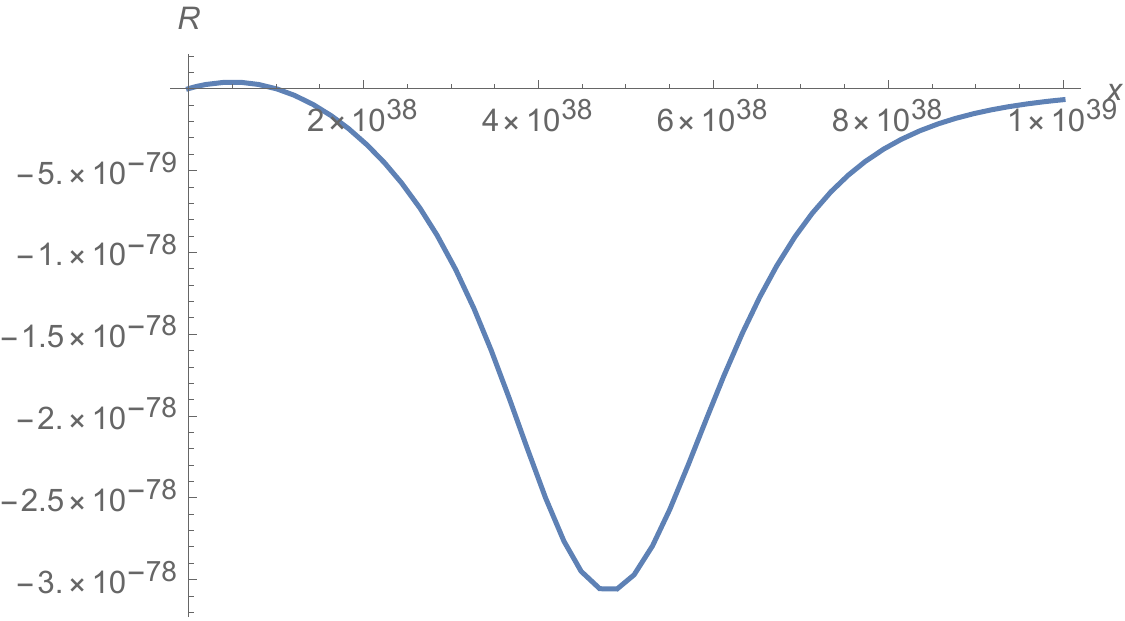}\\
(a)\\
\includegraphics[height=4.8cm]{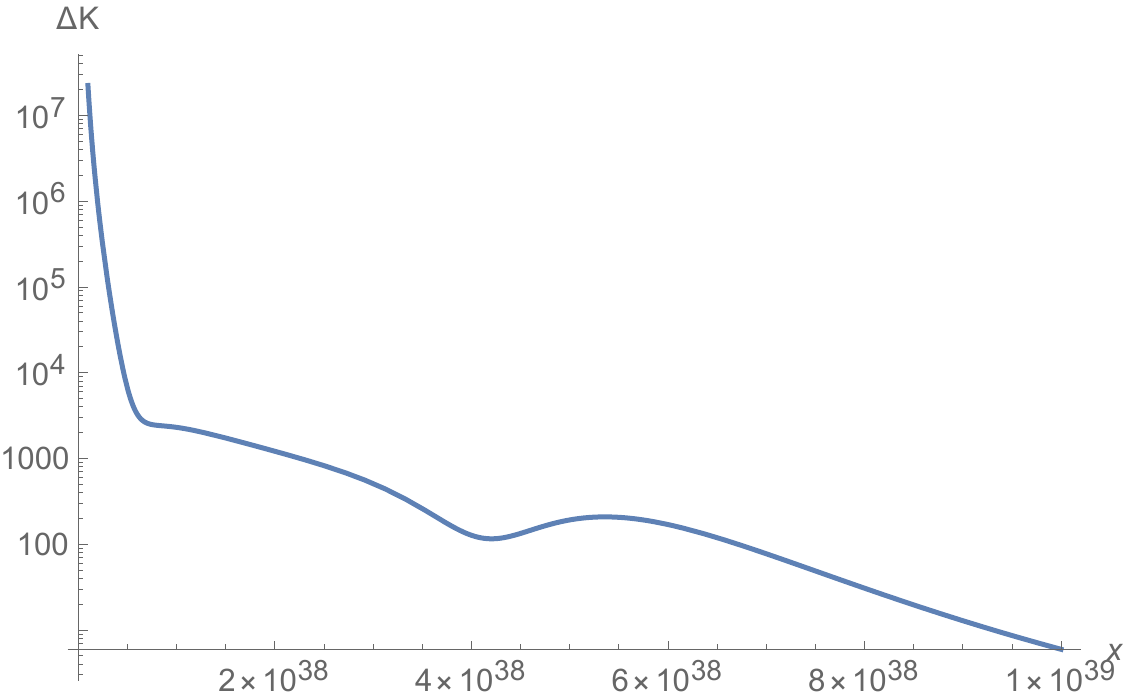}\\
 (b)  \\
\caption{Case $\Delta > 0,\; {\cal{D}} > 0, \; x_0 =0,\;x_m > x_H^+,\;  $: 
The quantities $R$ and $\Delta  {\cal{K}}$ vs $x$. 
 Here we choose $\mathcal{C}=10^{39},\;  \mathcal{D}= 10^{38}$,  for which the outer horizon is located at 
$x_H^{+} = 10^{38}$, and the throat  is at $x_{m}=5 \times 10^{38}$, while the black  hole mass is
$M_{BH}= 10^{38}\; M_{Pl}$. } 
\label{fig11}
\end{figure}

To study the quantum gravitational effects further, in Fig. \ref{fig6-1} we plot $R$ and $\Delta  {\cal{K}}$ in the region that covers the throat and the horizon, from which it can be seen that the deviation from GR quickly becomes vanishingly small around the horizon.  In addition, as $x \rightarrow  \infty$, we find that
\bqn
\lb{eq3.34d}
 R &\simeq& -\frac{20 \mathcal{C} ^6}{b^8}+{\cal{O}}\left(\epsilon^{9}\right),\nb\\
\Delta {\cal{K}}  &\simeq&  \frac{32  \mathcal{C} ^6}{3 M_{BH} b^5}+{\cal{O}}\left(\epsilon^{6}\right),
\eqn
 from which we can see that quantum corrections are decaying rapidly when $x \rightarrow  \infty$. 
 
 When $x \rightarrow 0\; (b \rightarrow \infty)$, we have
\bqn
\lb{eq3.34d-2}
 R &\simeq& \frac{8 \mathcal{D}  }{\mathcal{C} ^2 b}+ {\cal{O}}\left(b^{-2}\right),\nb\\
 {\cal{K}}  &\simeq& \frac{240  \mathcal{D}^2 }{\mathcal{C} ^4 b^2}+ {\cal{O}}\left(b^{-3}\right),
\eqn
which decays much less slowly than that in the Schwarzschild case,  ${\cal{K}}^{GR} \rightarrow b^{-6}$. \footnote{In \cite{BMM19} a different conclusion was derived, as the authors implicitly assumed that 
$x_0{\cal{C}} \not= 0$. Therefore, our conclusion  in this case does not essentially contradict to the one obtained in \cite{BMM19}.}   It is even slower than that of the loop quantum black hole solution found by Ashtekar, Olmedo and Singh
\cite{AOS18a,AOS18b}, in which  $R \rightarrow b^{-2}$ and ${\cal{K}} \rightarrow b^{-4}$ \cite{AO20}.

 {\it (Case III.3.2)  ${\mathcal{D}} < {\hat{\mathcal{C}}}$: } In this case the throat locates always outside the black hole horizon, so in the region $x > x_H^+$ we always have 
$a (x) > 0$, and the corresponding   effective energy density and pressures are given by Eq.(\ref{eq3.31}).
 In particular, at the throat $(x =\hat{\mathcal{C}})$, we have
 \bqn
 \lb{eq3.34e}
  \rho = \frac{6 \mathcal{D} -5 \mathcal{\hat C} }{2^{8/3}  \mathcal{\hat C} ^3 },\;\;
  p_r  =-p_{\theta} =  -\frac{1}{2^{8/3} \mathcal{\hat C} ^2}, 
  \eqn
from which we find that   the WEC, SEC, and DEC are satisfied in the domain,
\bqn
\lb{eq3.34f}
0<{\mathcal{C}}/2 <  \mathcal{D}. 
 \eqn
Combining Eq.(\ref{eq3.34f}) with ${\mathcal{D}} < {\hat{\mathcal{C}}}$, we find that in this case, all the energy conditions are  violated at the throat.

 { In addition, as $x \rightarrow 0 $ (or $b(x) \rightarrow \infty$), we still have Eq.(\ref{eq3.33b}),  from which we find that   the WEC, SEC, and DEC are satisfied  only at $x=0$.  At the spatial infinity $x \rightarrow \infty$,  we still have Eq.(\ref{eq3.34}), from which we can see that none of the three energy conditions is satisfied. In addition, we also have Eq.(\ref{eq3.34b}), thus  the mass of the black  hole is given by Eq.(\ref{eq3.34c}).}

 {For the quantum gravitational effects, we still have   Eqs.(\ref{eq3.34d}) and (\ref{eq3.34d-2}). In Fig. \ref{fig11} we plot $R$ and $\Delta  {\cal{K}}$ in the region that covers the throat and the horizon, from which it can be seen that the deviation from GR is still large around the horizon even for solar mass black holes, due to the fact that $\mathcal{C}$ is very large in this case and thus makes $\Delta  {\cal{K}}$ large around horizon which can be seen from  Eq.(\ref{eq3.34d}).}

 \subsubsection{$\; \mathcal{C} = 0,\; x_0 \not= 0$}

 {If we assume that } $\lambda_1 \not= 0$, from Eq.(\ref{eq2.3}) we can see that this corresponds to the limit $C \rightarrow0$. However, to keep ${\cal{D}} >0$ and finite, we 
 must require $DC  \rightarrow$ finite and positive. Thus,  we have 
 \bqn
\lb{eq3.35}
a(x) = \frac{\left(x^2 - \Delta\right)X}{\left(X + {\cal{D}}\right)Y^2}, \quad
  b(x)= Y.
\eqn
Clearly, $a(x) = 0$ leads to two real roots,
 \bq
 \lb{eq3.36}
 x^{\pm}_H = \pm \sqrt{\Delta},  
 \eq
 while $b(x)$ is a monotonically increasing function with $b(x = -\infty) =0$ [cf. Fig. \ref{fig1a-c}(c)].  
 Therefore, in contrast to the above cases, now the spacetime is not asymptotically flat
 as $x \rightarrow -\infty$, but rather it represents the center of the spacetime, at which a spacetime curvature singularity appears, as to be shown below. Therefore,  in the current case the spacetime
 represents a black hole with two horizons located at $x = \pm \sqrt{\Delta}$. This is quite similar to the charged Reissner-Nordstr{\"o}m (RN) solution.

In the trapped region,  $x_H^-<x<x_H^+$,   the effective energy density and pressures are given by
\begin{widetext}
 \bqn
\lb{eq3.37a}
\rho(x) &=& \frac{x_{0}^2 Y^3}{X^2 \left(Y^6\right)^{8/3}} \Bigg(\Big[1024 x^{10}-512 \mathcal{D}  x^9+2560 x^8 x_{0}^2-1024 \mathcal{D}  x^7 x_{0}^2+2240 x^6 x_{0}^4-672 \mathcal{D}  x^5 x_{0}^4+800 x^4 x_{0}^6\nb\\
 &&-160 \mathcal{D}  x^3 x_{0}^6+100 x^2 x_{0}^8-10 \mathcal{D}  x x_{0}^8+2 x_{0}^{10}\Big]X+1024 x^{11}-512 \mathcal{D}  x^{10}+3072 x^9 x_{0}^2-1280 \mathcal{D}  x^8 x_{0}^2\nb\\
 &&+3392 x^7 x_{0}^4-1120 \mathcal{D}  x^6 x_{0}^4+1664 x^5 x_{0}^6-400 \mathcal{D}  x^4 x_{0}^6+340 x^3 x_{0}^8-50 \mathcal{D}  x^2 x_{0}^8+20 x x_{0}^{10}-\mathcal{D}  x_{0}^{10}
 \Bigg), \nb\\
 p_r(x) &=& -\frac{\mathcal{D}  x_{0}^2 Y}{X^2 \left(Y^6\right)^{2/3}}, \nb\\
  p_{\theta}(x) &=& \frac{x_{0}^2 Y^2 }{2 X^{3} \left(Y^6\right)^{8/3}}\Bigg(\Big[4096 x^{12}-4096 \mathcal{D}  x^{11}+13312 x^{10} x_{0}^2-10752 \mathcal{D}  x^9 x_{0}^2+16384 x^8 x_{0}^4-10240 \mathcal{D}  x^7 x_{0}^4\nb\\
  &&+9408 x^6 x_{0}^6-4256 \mathcal{D}  x^5 x_{0}^6+2480 x^4 x_{0}^8-720 \mathcal{D}  x^3 x_{0}^8+244 x^2 x_{0}^{10}-34 \mathcal{D}  x x_{0}^{10}+4 x_{0}^{12}\Big]X+4096 x^{13}\nb\\
  &&-4096 \mathcal{D}  x^{12}+15360 x^{11} x_{0}^2-12800 \mathcal{D}  x^{10} x_{0}^2+22528 x^9 x_{0}^4-15104 \mathcal{D}  x^8 x_{0}^4+16192 x^7 x_{0}^6-8288 \mathcal{D}  x^6 x_{0}^6\nb\\
  &&+5808 x^5 x_{0}^8-2080 \mathcal{D}  x^4 x_{0}^8+924 x^3 x_{0}^{10}-194 \mathcal{D}  x^2 x_{0}^{10}+44 x x_{0}^{12}-3 \mathcal{D}  x_{0}^{12}
  \Bigg).
\eqn

On the other hand, in the region  $x<x_H^-$ or $x>x_H^+$,  the effective energy density and pressures are given by
 \bqn
\lb{eq3.37}
\rho(x) &=& \frac{\mathcal{D}  x_{0}^2 Y}{X^2 \left(Y^6\right)^{2/3}}, \nb\\
 p_r(x) &=& -\frac{x_{0}^2 Y^3}{X^2 \left(Y^6\right)^{8/3}} \Bigg(\Big[1024 x^{10}-512 \mathcal{D}  x^9+2560 x^8 x_{0}^2-1024 \mathcal{D}  x^7 x_{0}^2+2240 x^6 x_{0}^4-672 \mathcal{D}  x^5 x_{0}^4+800 x^4 x_{0}^6\nb\\
 &&-160 \mathcal{D}  x^3 x_{0}^6+100 x^2 x_{0}^8-10 \mathcal{D}  x x_{0}^8+2 x_{0}^{10}\Big]X+1024 x^{11}-512 \mathcal{D}  x^{10}+3072 x^9 x_{0}^2-1280 \mathcal{D}  x^8 x_{0}^2\nb\\
 &&+3392 x^7 x_{0}^4-1120 \mathcal{D}  x^6 x_{0}^4+1664 x^5 x_{0}^6-400 \mathcal{D}  x^4 x_{0}^6+340 x^3 x_{0}^8-50 \mathcal{D}  x^2 x_{0}^8+20 x x_{0}^{10}-\mathcal{D}  x_{0}^{10}
 \Bigg), \nb\\
  p_{\theta}(x) &=& \frac{x_{0}^2 Y^2 }{2 X^{3} \left(Y^6\right)^{8/3}}\Bigg(\Big[4096 x^{12}-4096 \mathcal{D}  x^{11}+13312 x^{10} x_{0}^2-10752 \mathcal{D}  x^9 x_{0}^2+16384 x^8 x_{0}^4-10240 \mathcal{D}  x^7 x_{0}^4\nb\\
  &&+9408 x^6 x_{0}^6-4256 \mathcal{D}  x^5 x_{0}^6+2480 x^4 x_{0}^8-720 \mathcal{D}  x^3 x_{0}^8+244 x^2 x_{0}^{10}-34 \mathcal{D}  x x_{0}^{10}+4 x_{0}^{12}\Big]X+4096 x^{13}\nb\\
  &&-4096 \mathcal{D}  x^{12}+15360 x^{11} x_{0}^2-12800 \mathcal{D}  x^{10} x_{0}^2+22528 x^9 x_{0}^4-15104 \mathcal{D}  x^8 x_{0}^4+16192 x^7 x_{0}^6-8288 \mathcal{D}  x^6 x_{0}^6\nb\\
  &&+5808 x^5 x_{0}^8-2080 \mathcal{D}  x^4 x_{0}^8+924 x^3 x_{0}^{10}-194 \mathcal{D}  x^2 x_{0}^{10}+44 x x_{0}^{12}-3 \mathcal{D}  x_{0}^{12}
  \Bigg).
\eqn
 
\end{widetext}

 \begin{figure}[ht]
\includegraphics[height=4.8cm]{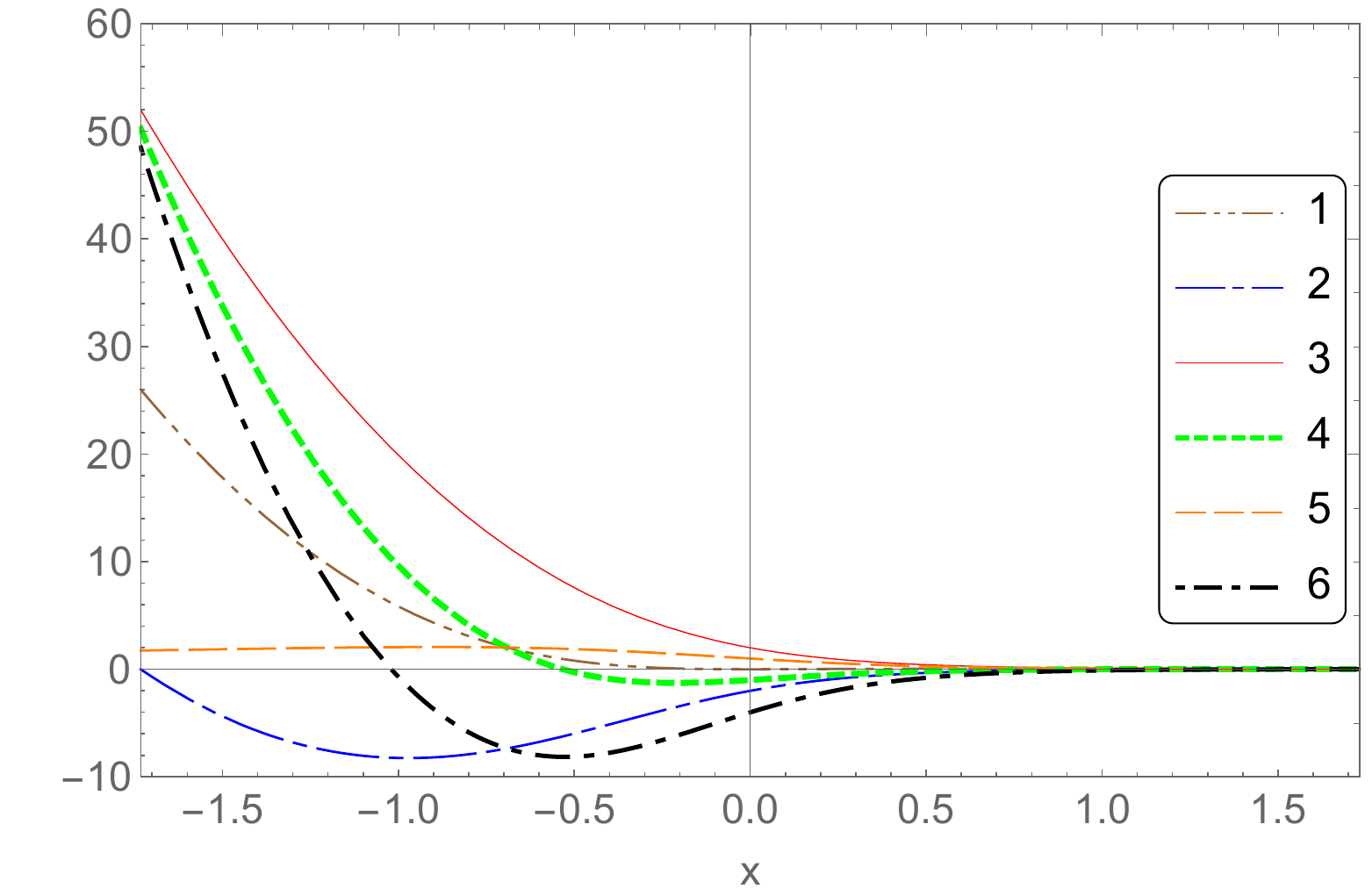}\\
(a)  \\
\includegraphics[height=4.8cm]{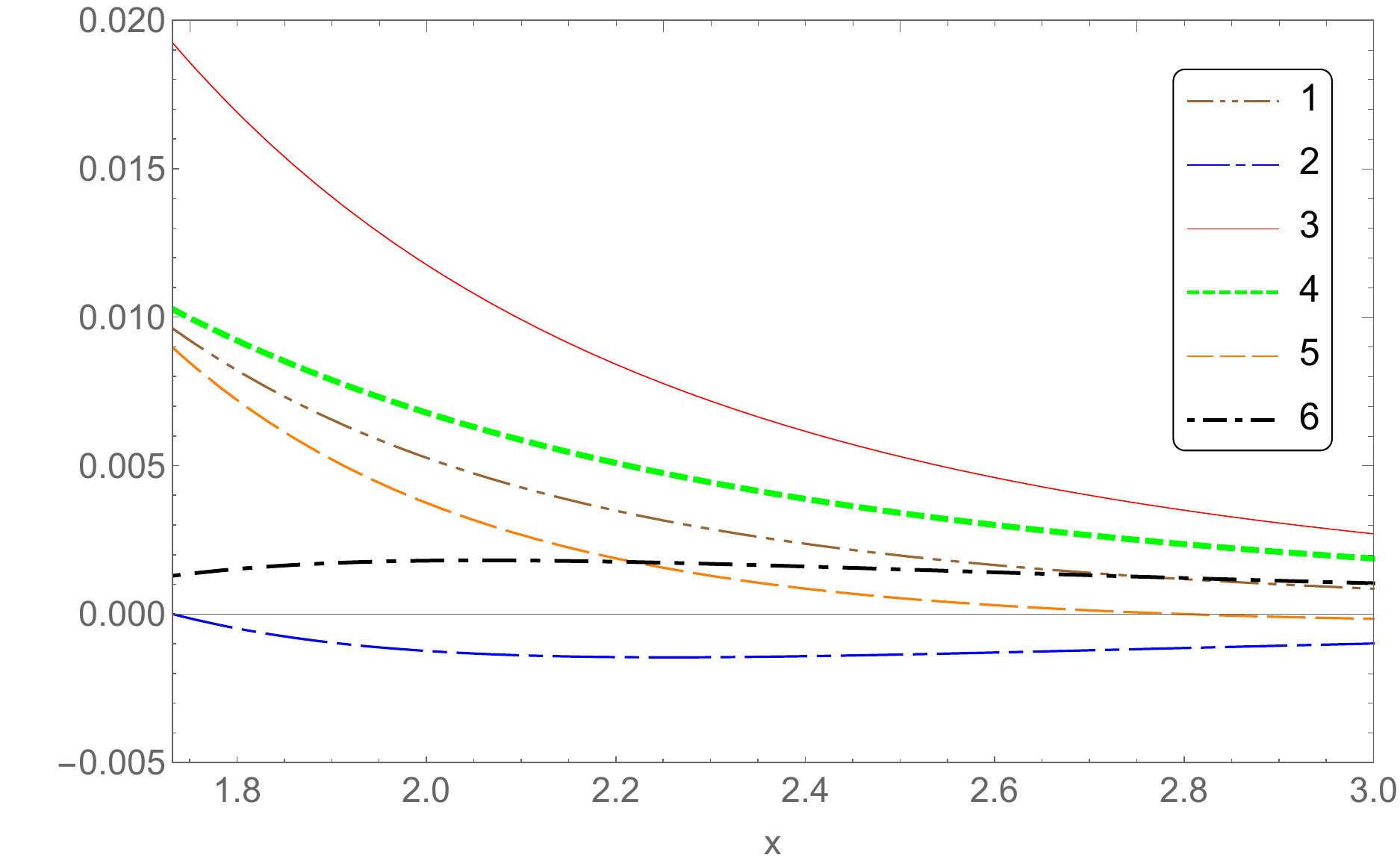}\\
(b)\\
\includegraphics[height=4.8cm]{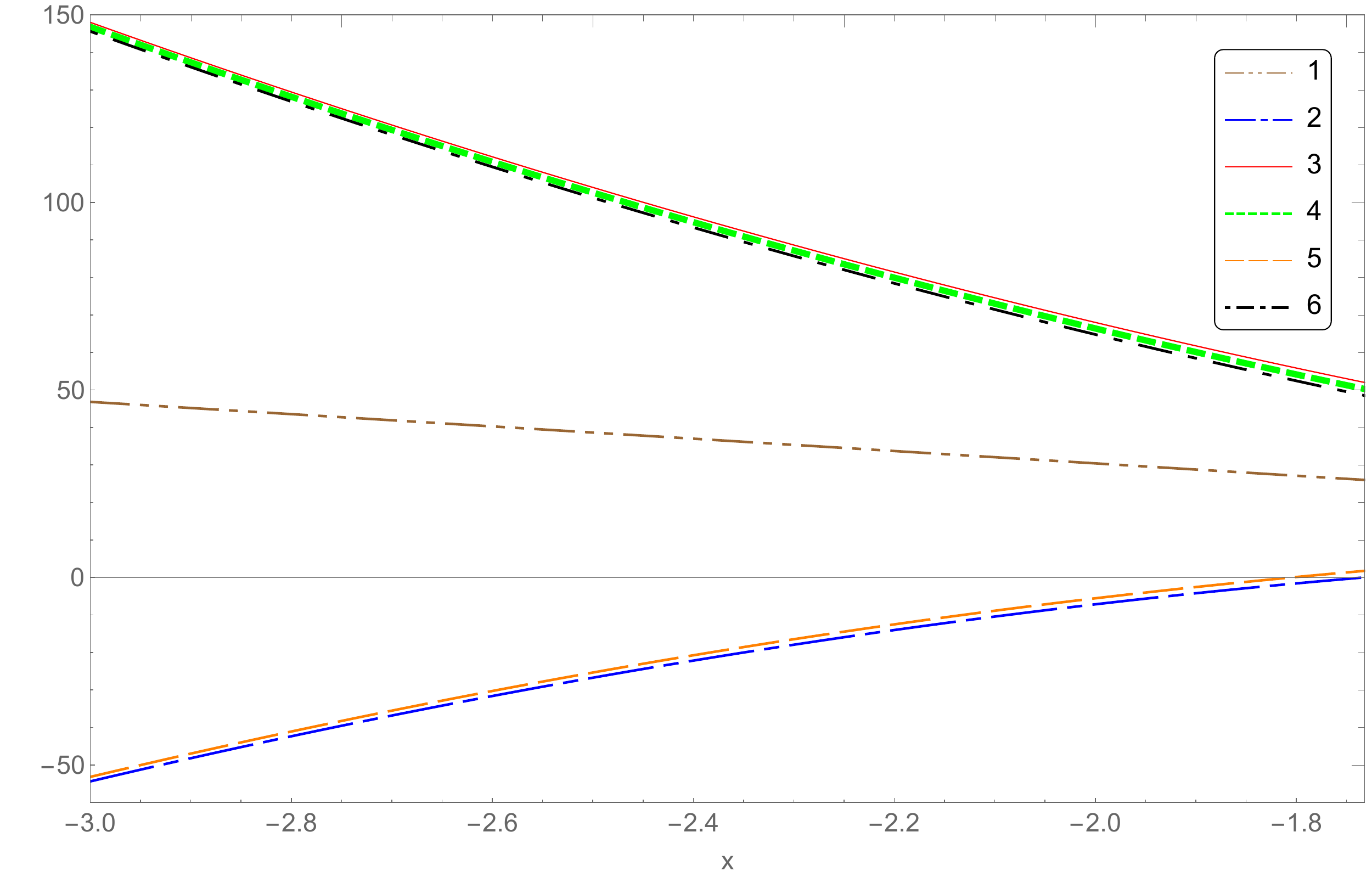}\\
(c)   \\
\caption{Case $\Delta > 0, \; {\cal{D}} > 0,\; x_0 \not= 0,\; {\cal{C}} =0$: The  physical quantities, $\rho$, $(\rho + p_r)$,  $(\rho - p_r)$, $(\rho + p_{\theta})$, $(\rho - p_{\theta})$, and  $(\rho + p_r + 2p_{\theta})$, represented, respectively, by Curves 1 - 6,  vs $x$: (a) between the white and black horizons, $ x_H^{-} \le x \le x_{H}^{+}$; (b) outside  the black  horizon, $ x \ge x_H^{+} = \sqrt{3}$; (c):  outside  the white  horizon, $ x \le x_H^{-} = -\sqrt{3}$. All curves are plotted with $x_{0}=1,\; \mathcal{D}=2$, for which the 
two horizons are located respectively at $x^{\pm}_H =\pm \sqrt{\Delta}=\pm \sqrt{3}$. } 
\lb{fig7}
\end{figure}

   In Fig. \ref{fig7} we plot the physical quantities $\rho, \; \rho \pm p_r,\;  \rho \pm p_{\theta}$, and $ \rho + p_r + 2  \pm p_{\theta}$ in the neighborhood of the two horizons, from which we can see that all these quantities become unbounded as
   $x \rightarrow -\infty$ (or $b(x) \rightarrow 0$).

In particular,   at the horizon $(x =\sqrt{\Delta})$, we have
 \bqn
 \lb{eq3.38}
 \rho &=& -p_r  =  \frac{\left(\sqrt{\Delta}+\mathcal{D} \right) x_{0}^2}{\mathcal{D}  Z^{2}}, \nb\\
 p_{\theta} &=&  \frac{x_{0}^4}{2 \mathcal{D} ^2 Z^{2}},  
  \eqn
  so all the three energy conditions, WEC, SEC, and DEC, are satisfied in the domain
 \bq
 \lb{eq3.38a}
 \left|x_{0}\right| < \mathcal{D}, \; \left(x_{0} \not= 0\right).
 \eq
The surface gravity at this horizon is given by,
\bqn
\lb{eq3.38b}
\kappa_{BH} &\equiv& \frac{1}{2}a'(x =\sqrt{\Delta}) \nb\\
&=& \frac{Y^2}{2Z^5}\Bigg(\Big[32 \mathcal{D} ^6-x_{0}^6+18 \mathcal{D} ^2 x_{0}^4-48 \mathcal{D} ^4 x_{0}^2
\Big]\sqrt{\Delta}\nb\\
&&+32 \mathcal{D} ^7-6 \mathcal{D}  x_{0}^6+38 \mathcal{D} ^3 x_{0}^4-64 \mathcal{D} ^5 x_{0}^2
\Bigg),
\eqn
which is always positive, provided that the conditions (\ref{eq3.38a}) hold.

On the other hand,  at the  horizon $x =-\sqrt{\Delta}$, we have
 \bqn
 \lb{eq3.38c}
 \rho &=& -p_r  =  \frac{Y}{\mathcal{D}  x_{0}^8} \Bigg(16 \mathcal{D} ^4 \left(\mathcal{D}+\sqrt{\Delta }\right)+x_{0}^4 \left(5 \mathcal{D} +\sqrt{\Delta }\right)\nb\\
 &&~~~~~~-4 \mathcal{D} ^2 x_{0}^2 \left(5 \mathcal{D} +3 \sqrt{\Delta }\right)\Bigg), \nb\\
 p_{\theta} &=& \frac{x_{0}^4}{2 \mathcal{D} ^2 Y^{2}},  
  \eqn
 so all the three energy conditions, WEC, SEC, and DEC, are satisfied in the domain given by Eq.(\ref{eq3.38a}). The surface gravity at this horizon is given by,
\bqn 
\lb{eq3.38d}
\kappa_{BH} &&\equiv \frac{1}{2}a'(x = - \sqrt{\Delta}) \nb\\
&&= -\frac{Y^2}{2Z^5}\Bigg(\Big[32 \mathcal{D} ^6-x_{0}^6+18 \mathcal{D} ^2 x_{0}^4-48 \mathcal{D} ^4 x_{0}^2
\Big]\sqrt{\Delta}\nb\\
&&-32 \mathcal{D} ^7+6 \mathcal{D}  x_{0}^6-38 \mathcal{D} ^3 x_{0}^4+64 \mathcal{D} ^5 x_{0}^2
\Bigg),
\eqn
 which is always negative  when the conditions (\ref{eq3.38a}) hold.

As $x \rightarrow \pm \infty$, we find that 
\bqn
\lb{eq3.39}
\rho(x) &=& 
\begin{cases}
\frac{\mathcal{D}  x_{0}^2}{8 x^5}+ {\cal{O}}\left(\epsilon^6\right), & x \rightarrow \infty,\cr
-\frac{8 \mathcal{D}  x}{x_{0}^4}+{\cal{O}}\left(\epsilon\right), &  x \rightarrow - \infty, \cr
\end{cases}\nb\\
p_r(x) &=&
\begin{cases}
-\frac{x_{0}^2}{4 x^4}+\frac{\mathcal{D}  x_{0}^2}{8 x^5}+ {\cal{O}}\left(\epsilon^6\right), &  x \rightarrow \infty,\cr
-\frac{16 x^2}{x_{0}^4}-\frac{8 \mathcal{D}  x}{x_{0}^4}-\frac{4}{x_{0}^2}+{\cal{O}}\left(\epsilon\right), &  x \rightarrow - \infty,\cr
\end{cases}\nb\\
p_{\theta}(x) &=&
\begin{cases}
\frac{x_{0}^2}{4 x^4}-\frac{\mathcal{D}  x_{0}^2}{4 x^5}, &  x \rightarrow \infty,\cr
\frac{16 x^2}{x_{0}^4}+\frac{8 \mathcal{D}  x}{x_{0}^4}+\frac{4}{x_{0}^2}+{\cal{O}}\left(\epsilon\right), &  x \rightarrow - \infty,\cr
\end{cases}\nb\\
\eqn
from which we can show that none of the three energy conditions, WEC, SEC, and DEC,  is satisfied at spatial infinity $x = \infty$ as well as at the center $b(x = -\infty) = 0$. In addition, we also have
\bqn
\lb{eq3.40}
a(x) &=& \begin{cases}
 \frac{1}{4}\left(1-\frac{2\mathcal{D} }{b}\right) +  {\cal{O}}\left(\epsilon^2\right), & x  \rightarrow  \infty,\cr
 \frac{4 x^4}{x_{0}^4}+\frac{4 \mathcal{D}  x^3}{x_{0}^4}+\frac{6 x^2}{x_{0}^2}+\frac{4 \mathcal{D}  x}{x_{0}^2}\nb\\
 +\frac{7}{4}+\frac{\mathcal{D} }{4 x}+{\cal{O}}\left(\epsilon^2\right), & x  \rightarrow  -\infty,\cr
\end{cases}\nb\\
b(x) &\simeq&  \begin{cases}
2x, & x  \rightarrow  \infty,\cr
-\frac{x_{0}^2}{2 x}+\frac{x_{0}^4}{8 x^3}+{\cal{O}}\left(\epsilon^4\right), & x  \rightarrow  -\infty.\cr
\end{cases}
\eqn
 {Thus,} the mass of the black hole is given by 
\bqn
\lb{eq3.41}
M_{BH} &=& \mathcal{D}.
\eqn

However, at $x = - \infty$ we have $b(-\infty) = 0$, and  the physical quantities, $\rho, \; p_r$ and $p_{\theta}$, all become unbounded,  so a spacetime curvature singularity
appears at $x = - \infty$, and the solution has a RN-like structure, i.e., two horizons, one is  inner and the other is outer, located, respectively, at
$x = \pm \sqrt{\Delta}$.  The spacetime singularity located at $b(-\infty) = 0$ is timelike.

\begin{widetext} 

 \begin{figure}[h!]
 \begin{tabular}{cc}
 \includegraphics[height=4.8cm]{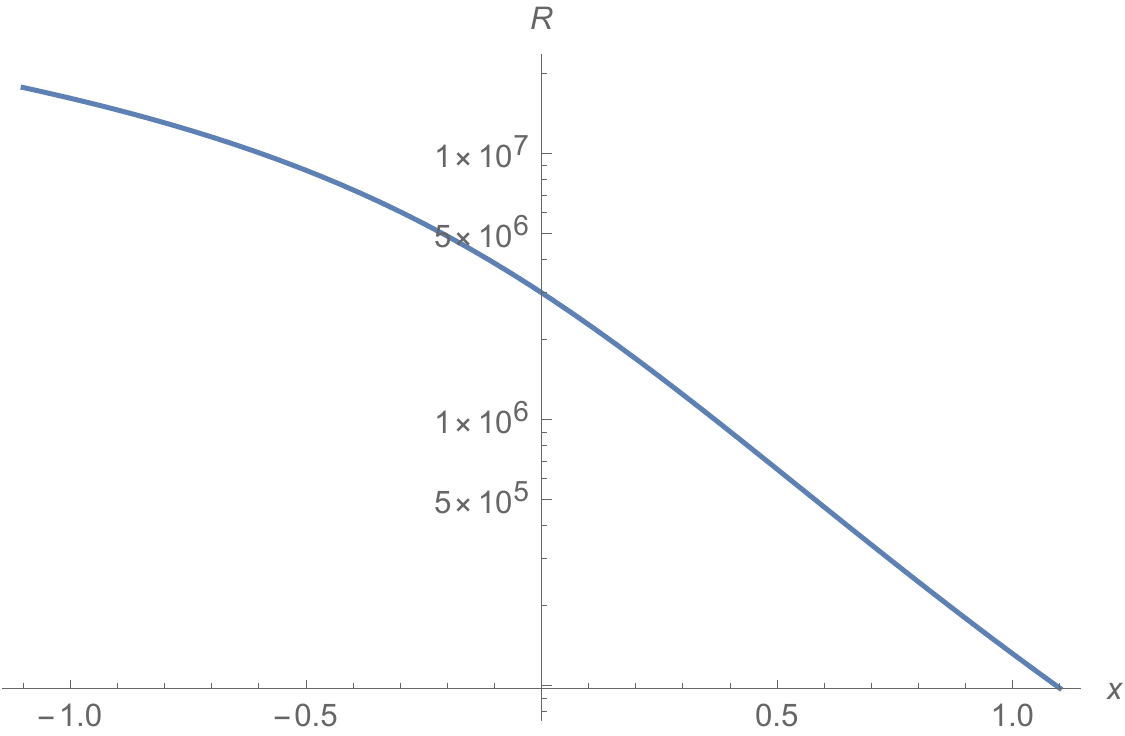}&
\includegraphics[height=4.8cm]{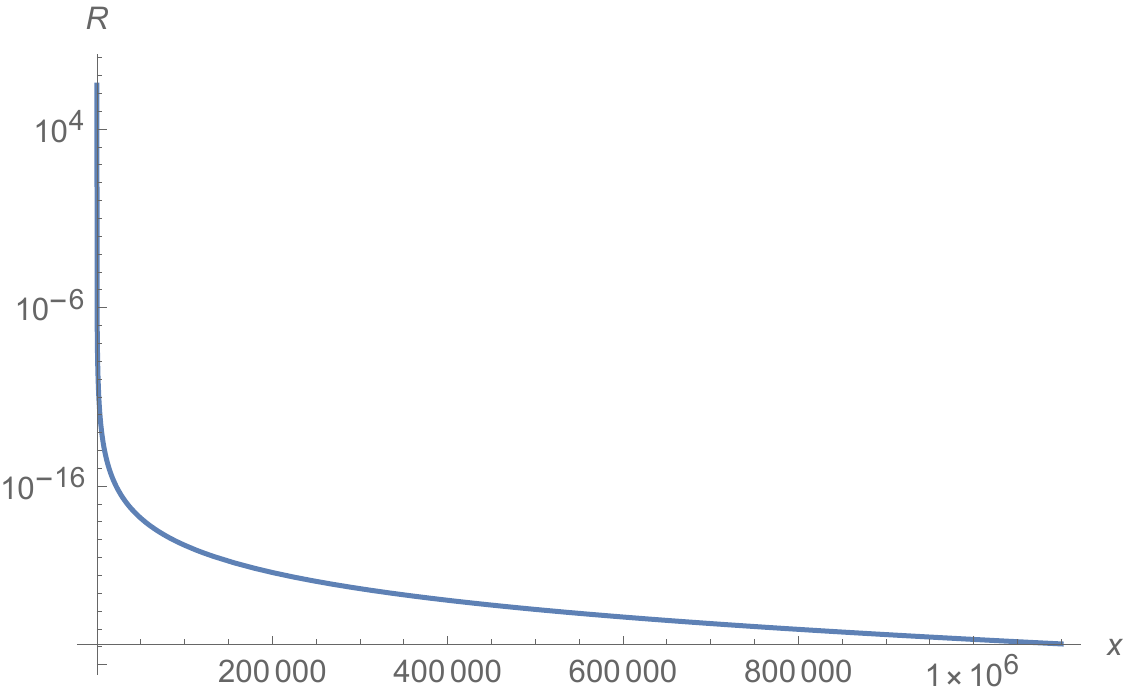}\\
		(a) & (b)  \\[6pt]
\includegraphics[height=4.8cm]{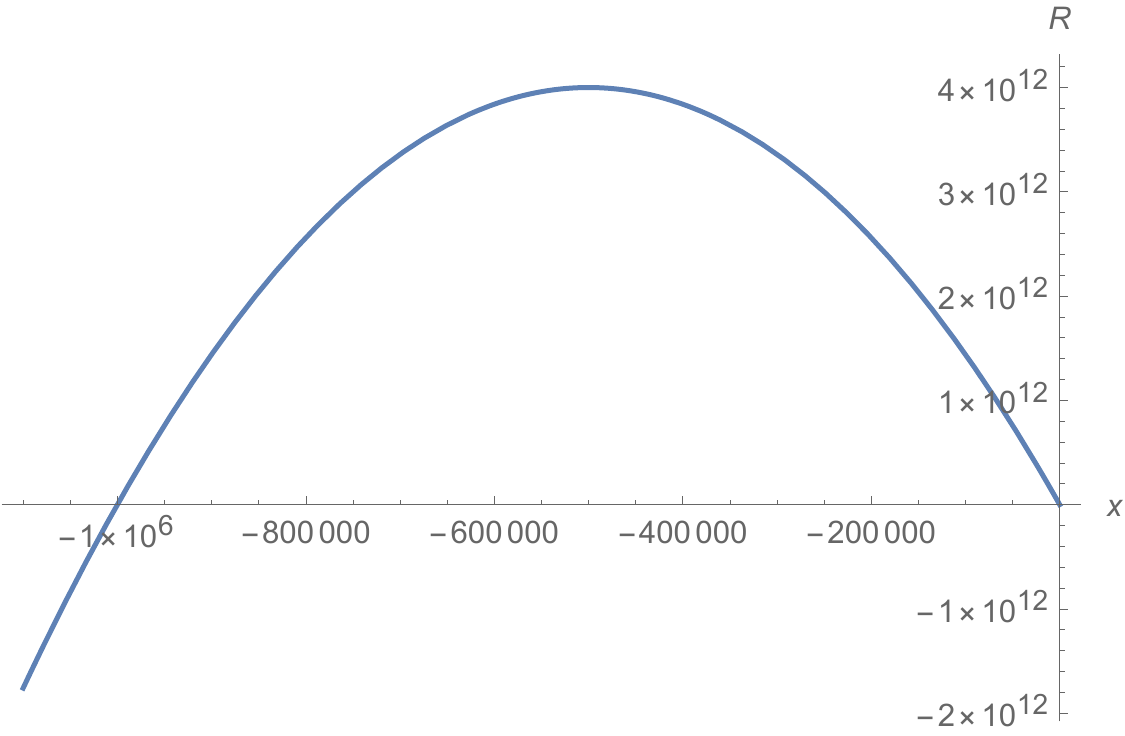}&
\includegraphics[height=4.8cm]{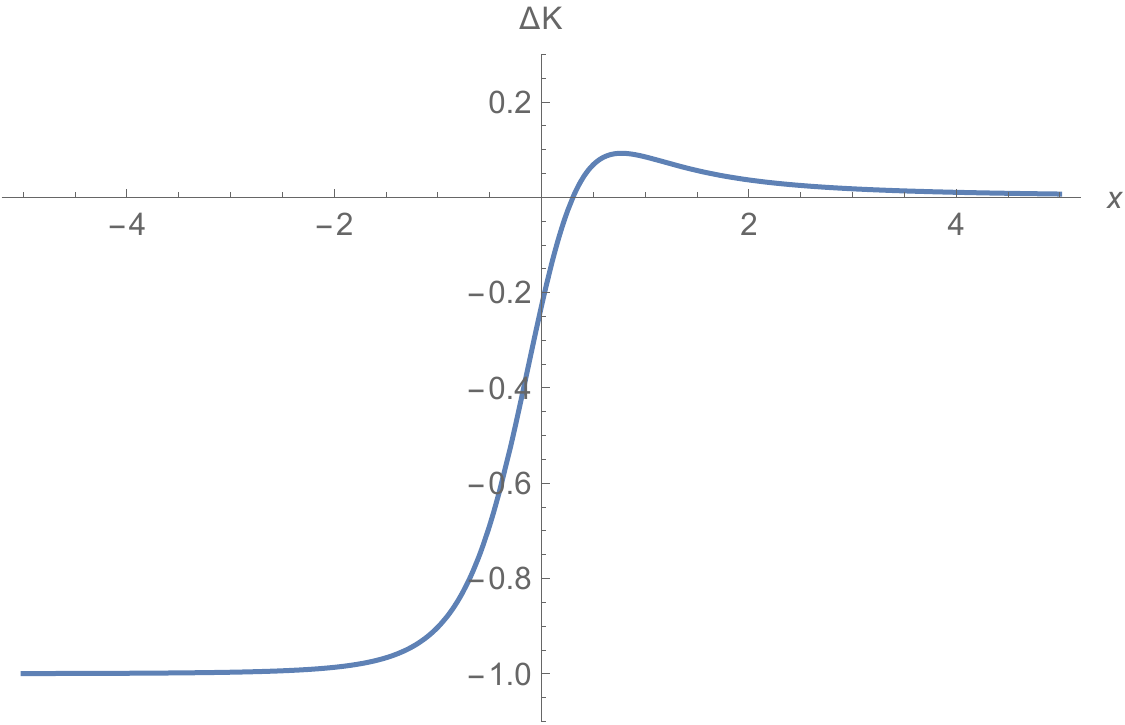}\\
		(c) & (d)  \\[6pt]
\includegraphics[height=4.8cm]{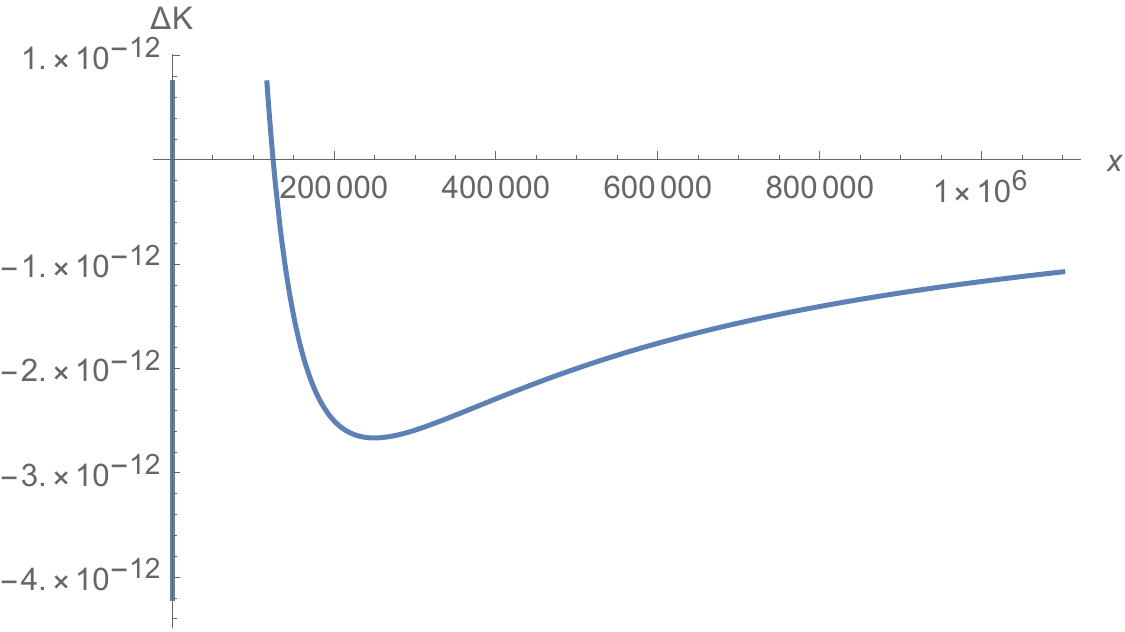}&
\includegraphics[height=4.8cm]{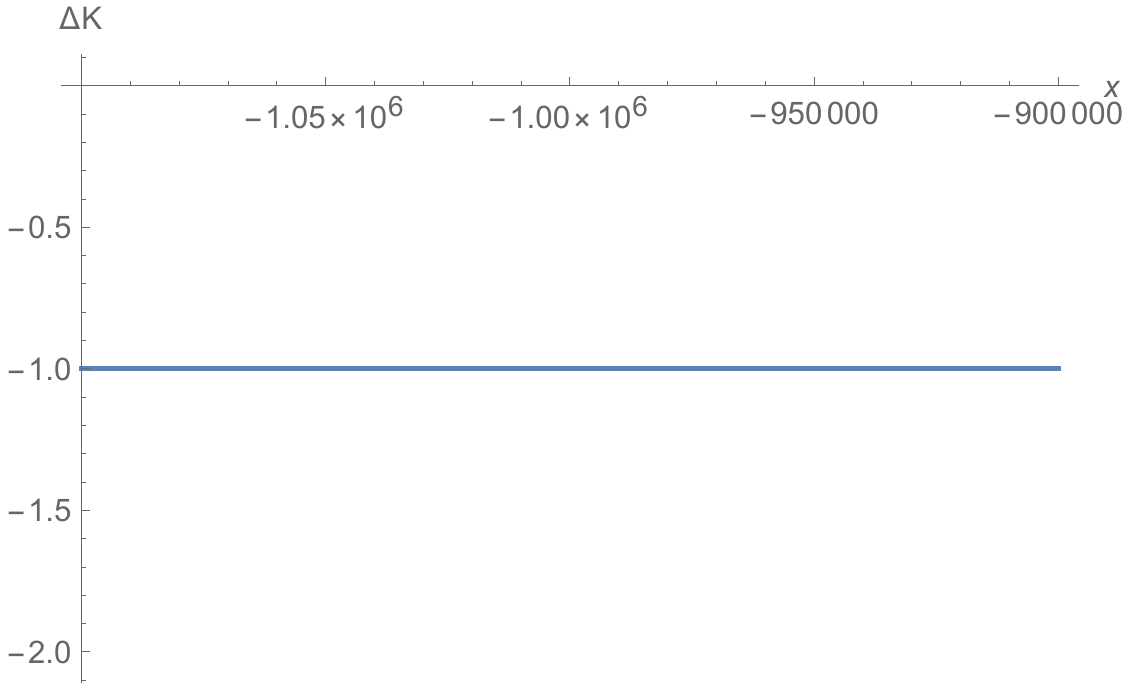}\\
		(e)  & (f) \\ [6pt]
	\end{tabular}
\caption{Case $\Delta > 0, \; {\cal{D}} > 0,\; x_0 \not= 0,\; {\cal{C}} =0$: The physical quantities $R$ and $\Delta  {\cal{K}}$ vs $x$.   Here we choose $x_{0}=1$, $\mathcal{D}=10^6$, so that $M_{BH}=10^6\;
M_{Pl}$, and  the horizons are located at 
$x = \pm \mathcal{D}=\pm 10^6$.}
\label{fig7-1}
\end{figure}

\end{widetext}

On the other hand,    in Fig. \ref{fig7-1} we plot $R$ and $\Delta  {\cal{K}}$ in the region that covers the throat and the horizons, from which it can be seen that the deviation from GR quickly becomes vanishing small around the outer  horizon, but around the inner horizon, $R$ deviates from GR significantly. In fact, as $x \rightarrow \pm \infty$, we find that
\bqn
\lb{eq3.41a}
 R &\simeq& \begin{cases}
-\frac{x_{0}^2}{4 x^4}+\frac{\mathcal{D}  x_{0}^2}{2 x^5}+{\cal{O}}\left(\epsilon^6\right), & x  \rightarrow  \infty,\cr
-\frac{16 x^2}{x_{0}^4}-\frac{16 \mathcal{D}  x}{x_{0}^4}-\frac{4}{x_{0}^2}+{\cal{O}}\left(\epsilon\right), & x  \rightarrow  -\infty,\cr
\end{cases}
\eqn
and
\bqn
\lb{eq3.41b}
 \mathcal{K} &\simeq& \begin{cases}
\frac{3 \mathcal{D} ^2}{4 x^6}+{\cal{O}}\left(\epsilon^7\right), & x  \rightarrow  \infty,\cr
\frac{2816 x^4}{x_{0}^8}+\frac{3072 \mathcal{D}  x^3}{x_{0}^8}+\frac{64 x^2 \left(15 \mathcal{D}^2+22 x_{0}^2\right)}{x_{0}^8}\\
+\frac{640 \mathcal{D}  x}{x_{0}^6}+\frac{16 \left(11 x_{0}^2-8  \mathcal{D} ^2\right)}{x_{0}^6}+{\cal{O}}\left(\epsilon\right), & x  \rightarrow  -\infty,\cr
\end{cases}
\eqn
 from which we can see that,  as $x \rightarrow -\infty$, both of the Ricci and  Kretschmann scalars   become unbounded,  and
  a spacetime singularity appears at $b(x = -\infty) = 0$. 
  
  It is interesting to note that  $\Delta {\cal{K}} $ is bounded and  approaches a nonzero constant $-1$,
  as $x \rightarrow -\infty$. In fact, we have
 \bqn
\lb{eq3.41c}
\Delta {\cal{K}}  &\simeq&  \begin{cases}
-\frac{4 x_{0}^2}{3 M_{BH} x}+{\cal{O}}\left(\epsilon^2\right), \;  x  \rightarrow   \infty, \cr
 -1+\frac{11 x_{0}^4}{12 M_{BH}^2 x^2}+{\cal{O}}\left(\epsilon^3\right), & x  \rightarrow  -\infty,\cr
\end{cases}
\eqn
where in writing the above expressions we had set ${\cal{K}}^{GR} = 48M_{BH}/b^6$ over the whole region $x \in(-\infty, \infty)$. Thus, near the singular point $b(x = -\infty) = 0$,
the   Kretschmann scalar of the quantum black hole diverges much more slowly than that of the Schwarzschild black hole. This can be seen from Eqs.(\ref{eq3.40}) and (\ref{eq3.41b}),
from which we find that ${\cal{K}} \propto b^{-4}$ as $x \rightarrow -\infty$.

\subsubsection{$\;x_{0}= \mathcal{C} = 0$}
  
 Since $\lambda_1 \lambda_2  \not= 0$, from Eq.(\ref{eq2.3}) we can see that this corresponds to the limits
  $C \rightarrow 0$ and $\sqrt{n} \rightarrow \infty$. However, to keep ${\cal{D}} >0$, at these limits, we must require $DC/\sqrt{n}  \rightarrow$ finite and positive. 
 Then, we find that $\Delta = {\cal{D}}^2$, and from Eq.(\ref{eq2.6}) we find $X = |x|$, and 
  \bqn
\lb{eq3.42}
Y =x +|x| = \begin{cases}
2x, & x \ge 0,\cr
0, & x < 0.\cr
\end{cases}
\eqn
  Therefore, the spacetime must be restricted to the region $x \ge 0$, in which we have 
 \bqn
\lb{eq3.43}
a(x) &=& 
\frac{x-\mathcal{D}}{4x}=\frac{1}{4}\left(1-\frac{2 \mathcal{D}}{b}\right), \nb\\
  b(x) &=&  (x+\sqrt{x^2})=2x,
\eqn
and
 \bqn
\lb{eq3.44}
\rho(x) = p_r=  p_{\theta}(x)=0.
\eqn
In fact, this is  precisely the Schwarzschild solution, and will take its standard form, by setting $r = 2x$ and rescaling $t$,
\bq
\lb{eq3.45}
ds^2 = \left(1 - \frac{2m}{r}\right) dt^2 +  \left(1 - \frac{2m}{r}\right)^{-1}dr^2 + r^2d\Omega^2,
\eq
where $m \equiv {\cal{D}}$. This case can be also considered as the limit of $\lambda_{1, 2} \rightarrow 0$, for which the GR limit is obtained. Therefore, the results are consistent with
the effective theory of quantum black holes, as the singularities are always avoided exactly because of the replacement (\ref{eq2.12}). When  $\lambda_{1, 2} \rightarrow 0$, 
the classical limits are recovered.

 \subsection{$\mathcal{D}<0$}
 
 In this case, similar to the last one, let us consider  $x_0 \mathcal{C}  \not= 0$ and $ x_0 \mathcal{C}  = 0$, separately.

  \subsubsection{$\; x_0 \mathcal{C}  \not= 0$}

Then, since $\Delta = {\cal{D}}^2 - x_0^2 >0$, we must have 
 \bq
 \lb{eq3.46}
  {\cal{D}} < - \left|x_0\right|.
  \eq
Thus, from Eq.(\ref{eq2.5}) we find that 
 \bqn
\lb{eq3.47}
{a}(x) = \frac{\left(X + \left|{\cal{D}}\right|\right)XY^2}{Z^2},\quad {b}(x) =  \frac{Z}{Y},
\eqn
where $X, \; Y$, and $Z$ are given by Eq.(\ref{eq2.6}). From the above expressions, it can be shown that   there are two 
 asymptotically flat regions, corresponding to $x \rightarrow \pm \infty$,
 respectively. They are still connected by a throat located at $x_{m}$ given by Eq.(\ref{eq3.7}) [cf. Fig. \ref{fig1a-c}(a)]. But since $a(x) \not =0$ for any given $x$, horizons, either WHs or BHs, do not exist.

 At the throat, the effective energy density $\rho$ and pressures $p_r$ and $p_{\theta}$ are still given by Eq.(\ref{eq3.8}). Then, it can be easily shown that none of the three energy conditions
   can be satisfied in the current case, because condition Eq.(\ref{eq3.14}) is always violated for $\mathcal{D}<0$.

 At the spatial  infinities $x \rightarrow \pm \infty$, we find that the expression of $\rho, p_r, p_{\theta}$ are still given by Eq.(\ref{eq3.20}), from which we can see that {\it none of the three energy conditions is satisfied either}.  The asymptotic expressions of $a(x)$ and $b(x)$ are still given by Eq.(\ref{eq3.21}), and the total masses  measured  at $x \rightarrow \pm \infty$ are 
\bqn
\lb{eq3.47a}
M_{+} &=& \mathcal{D},\quad
 M_{-} =   \frac{\mathcal{D} \mathcal{C} ^2}{x_{0}^2},
\eqn
 but since we now have $\mathcal{D} < 0$, they are all negative.  Note that from now on, we use $M_{\pm}$ to denote the total masses of the spacetimes measured at
 $x = \pm \infty$, when no horizons (either BHs or WHs) exist, while reserve $M_{BH/WH}$ to denote  the black (white) hole masses.

It can be shown that  in the present case the deviation from GR decays rapidly when away from the throat from both directions of it only for some particular choice of the free parameters. 
In particular,  as $x \rightarrow \pm \infty$, we find that the asymptotic expressions of $R(x)$ and $\Delta \mathcal{K}(x)$ are still given by Eq.(\ref{eq3.25c}) and Eq.(\ref{eq3.25d}), with $M_{BH} (M_{WH})$ being replaced by $M_{+} (M_{-})$. Therefore, we still have $|\Delta {\cal{K}}_+/\Delta {\cal{K}}_-|=1 + {\cal{O}}\left(\epsilon^2\right)$, as $|x| \rightarrow \infty$. 
That is,  whether $M_{-} \gg M_{+}$ or not, $|\Delta {\cal{K}}_+|$ will always have the same asymptotic magnitude as $|\Delta {\cal{K}}_-|$, and both of them approach their GR limits as
${\cal{O}}(1/|x|)$.

 \subsubsection{ $ x_0 = 0, \;\; {\cal{C}} \not= 0$}

 In this case $a(x)$ and $b(x)$ are still given by Eq.(\ref{eq3.28}), but since $\mathcal{D}<0$, $a(x) = 0$ is possible only when   
  \bq
 \lb{eq3.48}
 x_H= 0,
 \eq
where $b(x = 0) = \infty$. Therefore, in the current case  there is no black/white hole horizon either, while the minimum of $b(x)$ now is still located at $x_{m} \equiv {\hat{\cal{C}}}$ [cf. Fig. \ref{fig1a-c}(b)].
 On the other hand,  in this case  the effective energy density and pressures are still given by Eq.(\ref{eq3.31}), which are all become zero as $x \rightarrow 0$.

 At the throat $(x =\hat{\mathcal{C}})$, $\rho, p_r, p_{\theta}$ are given by Eq.(\ref{eq3.34e}), but since now we  have $\mathcal{D} < 0$, none of the three energy conditions is satisfied at the throat.

    At the spatial infinity $x \rightarrow \infty$, on the other hand, we have the same expressions as given by Eq.(\ref{eq3.34}), from which we can see that none of the three energy conditions is satisfied. The asymptotic behavior of $a(x)$ and $b(x)$ are still given by Eq.(\ref{eq3.34b}). Therefore, the total mass at $x \rightarrow \infty$ is given by 
\bqn
\lb{eq3.48a}
M_{+} &=& \mathcal{D}<0.
\eqn

On the other hand,  to study the quantum gravitational effects further,   we consider the physical quantities $R$ and $\Delta  {\cal{K}}$ and  find  that the deviation from GR also
quickly becomes vanishingly small as $x \rightarrow  \infty$ for some particular choice of the free parameters.
In particular, as $x \rightarrow  \infty$, we find that  the asymptotic expressions of $R(x)$ and $\Delta \mathcal{K}(x)$ are still given by Eq.(\ref{eq3.34d}), with $M_{BH}$ being replaced by $M_{+}$.

 \begin{widetext}

\begin{table}
\caption{\label{tab:table1}%
The main properties of the solutions given by Eqs.(\ref{eq2.4})-(\ref{eq2.3b}) with $\Delta > 0$ in various cases, where
bhH $\equiv$ black hole horizon, whH $\equiv$ white hole horizon,  ECs $\equiv$ energy conditions, 
SAF $\equiv$ spacetime is asymptotical flat,   SCS $\equiv$ spacetime curvature singularity, and Sch.S $\equiv$ Schwarzschild solution.  
In addition, ``$\checkmark$" means yes,  ``$\times$" means no, while  ``N/A"  means not applicable. }
\begin{tabular}{|l|c|c|c|c|c|c|c|c|}  \hline
{\bf {\mbox{\hspace{.2cm}}} Properties {\mbox{\hspace{.2cm}}}} 
& \multicolumn{8}{c|}{\bf  $\Delta>0$}\\ \cline{2-9}
 & \multicolumn{4}{c|}{$\mathcal{D}>0$} & \multicolumn{4}{c|}{$\mathcal{D}<0$}   \\ \cline{2-9}
 & $\mathcal{C} \not= 0,$ & $\mathcal{C} \not= 0,$ & $\mathcal{C}  =0,$ & $\mathcal{C} = x_0=0$ & $\mathcal{C} \not= 0,$ &
 $\mathcal{C} \not= 0,$ & $\mathcal{C} = 0,$ & $\mathcal{C} = x_0=0$  \\ 
  & $ x_0 \not= 0$ & $ x_0 = 0$ & $ x_0 \not= 0$ & (Sch.S) & $ x_0 \not= 0$ &
 $ x_0 = 0$ & $x_0 \not= 0$ & (Sch.S)  \\ \hline  
bhH exists?  & \checkmark  & \checkmark  & \checkmark  & \checkmark  & $\times$   &  $\times$  &  $\times$  & $\times$    \\ \hline  
ECs at bhH     & Eq.(\ref{eq3.6b})   & \checkmark  &  Eq.(\ref{eq3.38a})  & \checkmark  &  N/A &  N/A & N/A  &   N/A  \\ \hline
whH exists?  &  \checkmark &  $\times$  & \checkmark  & $\times$   &  $\times$ &  $\times$  & $\times$   &  $\times$   \\ \hline
ECs at whH     & Eq.(\ref{eq3.6b})  &  N/A & Eq.(\ref{eq3.38a})  &  N/A &  N/A &  N/A & N/A  &  N/A  \\ \hline
Throat exists? &  \checkmark &  \checkmark &  $\times$  & $\times$   & \checkmark  & \checkmark  & $\times$   &  $\times$   \\ \hline
ECs at throat     & Eq.(\ref{eq3.14-3a})  & $\mathcal{C}=2\mathcal{D}$  &  N/A &  N/A & $\times$  &  $\times$  & N/A  &   N/A  \\ \hline
ECs at $x = \infty$     & $\times$  &  $\times$ &  $\times$  & \checkmark  &  $\times$ & $\times$  &  $\times$  &  \checkmark   \\ \hline
Mass at $x = \infty$  & $\mathcal{D}$  &  $\mathcal{D}$ &  $\mathcal{D}$ &  $\mathcal{D}$ &  $\mathcal{D}$ &  $\mathcal{D}$ & $\mathcal{D}$  &   $\mathcal{D}$  \\ \hline 
ECs at $x = -\infty$    & $\times$  & N/A ($x \ge 0$)   & $\times$  &  N/A ($x \ge 0$) & $\times$  &  N/A ($x \ge 0$)   &  $\times$  &   N/A ($x \ge 0$)   \\ \hline
Mass at $x = -\infty$  & $\frac{\mathcal{D} \mathcal{C} ^2}{x_{0}^2}$  &  SAF at $x=0$ & SCS  & SCS at $x = 0$  &  $\frac{\mathcal{D} \mathcal{C} ^2}{x_{0}^2}$ &  SAF at $x=0$ & SCS  &   SCS at $x = 0$  \\ \hline     
\end{tabular}
\lb{Table1}
\end{table}

\end{widetext}

  \subsubsection{ $\; x_0 \not= 0,\; \mathcal{C} = 0$} 
  
 From Eq.(\ref{eq2.5}) we find that 
 \bqn
\lb{eq3.49}
{a}(x) = \frac{\left(X + \left|{\cal{D}}\right|\right)X}{Y^2},\quad {b}(x) =  Y,
\eqn
where $X, \; Y$, and $Z$ are given by Eq.(\ref{eq2.6}). Clearly, $a(x) = 0$ has no real roots,  thus  no horizons exist,  while $b(x)$ is still a monotonically increasing function with $b(x = -\infty) =0$ [cf. Fig. \ref{fig1a-c}(c)].

On the other hand,  in this case  the effective energy density and pressures are still given by Eq.(\ref{eq3.37}).  In particular, at the spatial  infinities $x \rightarrow \pm \infty$, they stall take the forms   
of Eq.(\ref{eq3.39}), from which we find none of  the three energy conditions, WEC, SEC, and DEC, is satisfied. In addition, the asymptotic behaviors of $a(x)$ and $b(x)$ are  given by Eq.(\ref{eq3.40}). Therefore, the total mass at $x = \infty$   is still given by 
Eq.(\ref{eq3.41}), which is always negative.
 
However, at $x = - \infty$ we have $b(-\infty) = 0$, and  the physical quantities, $\rho, \; p_r$ and $p_{\theta}$, all become unbounded,  so a spacetime curvature singularity appears at $x = - \infty$.

In addition,  from $R$ and $\Delta  {\cal{K}}$ we find that  the deviation from GR quickly becomes vanishingly small as  $x \rightarrow + \infty$, but as  $x \rightarrow - \infty$, $R$ deviates from GR significantly, as a spacetime curvature singularity now appears at $x = -\infty$, at which we have $b(x= -\infty) = 0$.

\subsubsection{ $\;x_{0}= \mathcal{C} = 0$}

In this case,  the solution is  precisely the Schwarzschild solution with negative mass, and will take its standard form, by setting $r = 2x$ and rescaling $t$,
\bq
\lb{eq3.49b}
ds^2 = \left(1 - \frac{2m}{r}\right) dt^2 +  \left(1 - \frac{2m}{r}\right)^{-1}dr^2 + r^2d\Omega^2,
\eq
where $m \equiv {\cal{D}}<0$.

This completes the analysis of the solutions in the case $\Delta > 0$. In Table \ref{Table1}, we summarize the main properties of these solutions.

\section{Spacetimes with $\Delta = 0$}
\label{Section.IV}
 \renewcommand{\theequation}{4.\arabic{equation}}\setcounter{equation}{0}

From Eq.(\ref{eq2.3b}) we find that this case corresponds to 
\bq
\lb{eq3.50}
\left|\lambda_2\right| = \frac{3}{2} \left|CD\right|, \quad {\mbox{or}} \quad \left|{\cal{D}}\right| = \left|x_0\right|.
\eq
Then, from Eqs.(\ref{eq2.5}) and (\ref{eq2.6}) we obtain
\bqn
\lb{eq3.51}
{a}(x) &=& \frac{x^2XY^2}{\left(X + {\cal{D}}\right)Z^2},\quad {b}(x) =  \frac{Z}{Y},
\eqn
where 
\bqn
\lb{eq3.52}
X &\equiv& \sqrt{x^2 + {\cal{D}}^2}, \quad Y \equiv x + X, \nb\\
Z &\equiv& \left(Y^6 + {\cal{C}}^6\right)^{1/3}.
\eqn

To study these solutions further, in the following  let us consider the three possibilities, ${\cal{D}} > 0$, ${\cal{D}} = 0$
 and ${\cal{D}} < 0$, separately. 
 
\subsection{ ${\cal{D}} > 0$}

In this subcase, there are still two possibilities, $\mathcal{C} \not= 0$ and  $\mathcal{C} = 0$.

  \subsubsection{ $\; \mathcal{C} \not= 0$}

In this case, since we also have  ${\cal{D}} > 0$, we find that
\bq
\lb{eq3.53}
b(x) = \begin{cases}
\infty, & x = \infty,\cr
2^{1/3}{\cal{C}}, & x = x_{m},\cr
\infty, & x = - \infty,\cr
\end{cases}
\eq
where $x_{m} \equiv ({\cal{C}}^2 - {\cal{D}}^2)/(2{\cal{C}})$ [cf. Fig. \ref{fig1a-c}(a)]. 

On the other hand, $a(x) = 0$ leads to $x_H^{\pm} = 0$, which is a double root. This is similar to the charged RN solution in the extreme case $|e| = m$.  
At the horizon, we have
\bq
\lb{eq3.54}
b(0) = \frac{\left({\cal{C}}^6 + {\cal{D}}^6\right)^{1/3}}{\left|{\cal{D}}\right|},
\eq
and  
   \bqn
 \lb{eq3.55}
  \rho  &= & - p_r = 2  p_{\theta} =  \frac{\mathcal{D} ^2}{\left(\mathcal{D} ^6+\mathcal{C} ^6\right)^{2/3}},
 \eqn
 from which we find that all the WEC, SEC, and DEC are satisfied.
 In addition, the surface gravity at the horizon is,
\bqn 
\lb{eq3.55a}
\kappa_{BH} &&\equiv \frac{1}{2}a'(x =0)= 0,
\eqn
as in the extremal case of the RN solution.

 At the throat, the effective energy density $\rho$ and pressures $p_r$ and $p_{\theta}$ are given by
  \bqn
 \lb{eq3.56}
  \rho  &= & \frac{-5 \mathcal{D} ^2+12 \mathcal{D}  \mathcal{C} -5 \mathcal{C} ^2}{2^{2/3} \mathcal{C} ^2 \left(\mathcal{D} ^2+\mathcal{C} ^2\right)},\quad
  p_r  = -\frac{1}{2^{2/3} \mathcal{C} ^2},\nb\\
  p_{\theta}  &= & \frac{\left(\mathcal{D} ^2+\mathcal{C} ^2\right)^3-4 \mathcal{D} ^3 \mathcal{C} ^3}{2^{2/3} \left(\mathcal{D} ^2+\mathcal{C} ^2\right)^3},
 \eqn
from which we find that WEC, SEC, and DEC are satisfied only when
 \bq
 \lb{eq3.56a}
 \mathcal{D} =\mathcal{C}.
 \eq
 Then,  from the expression $x_{m} = ({\cal{C}}^2 - {\cal{D}}^2)/(2{\cal{C}})$, we can see when $\mathcal{D} =\mathcal{C}$ we also have  $x_{m}=0$, i.e., the black hole horizon
 now coincides with the throat.

 \begin{figure}[h!]
\includegraphics[height=4.8cm]{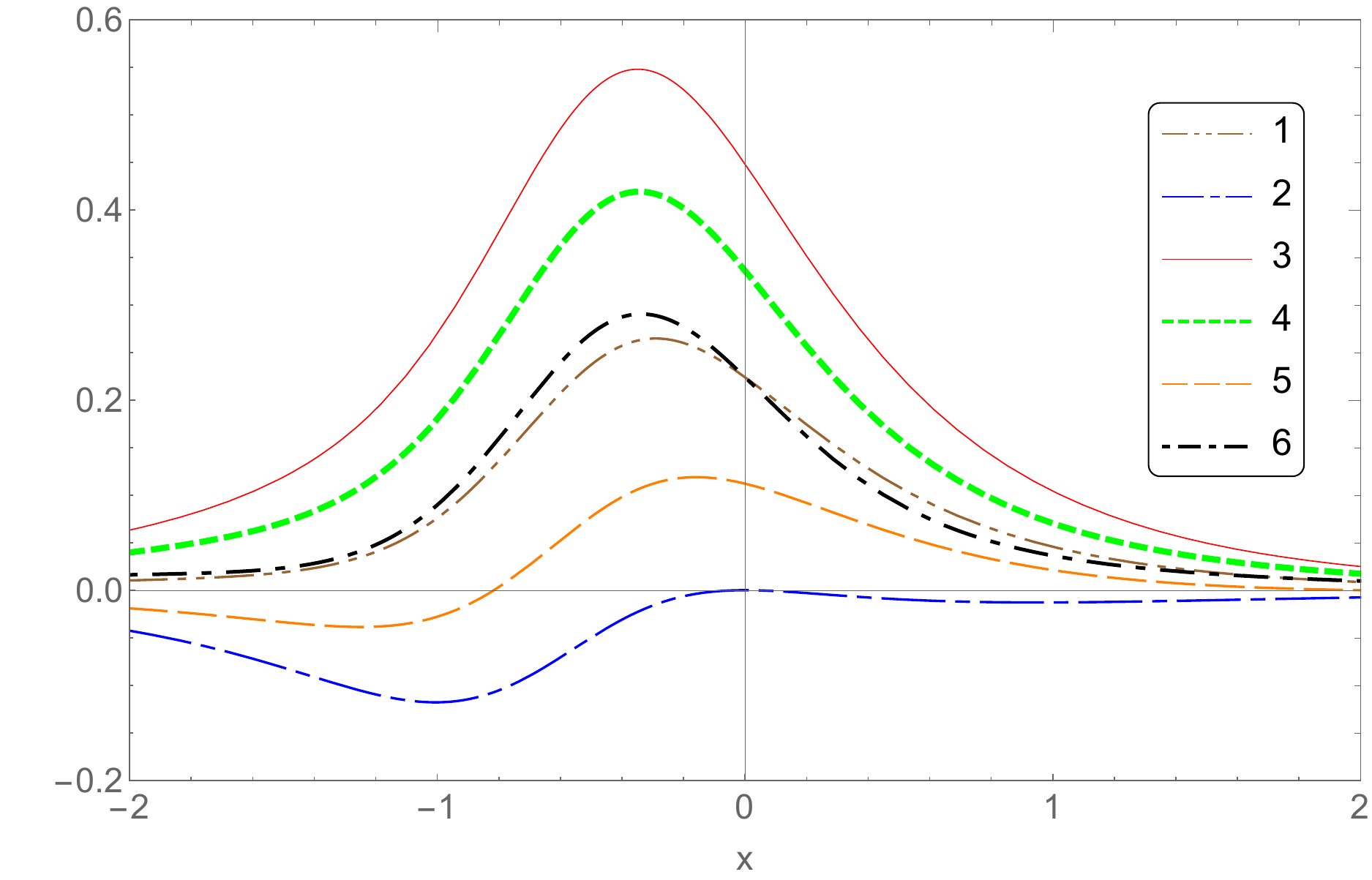}
\caption{Case $\Delta = 0, \; {\cal{D}} > 0,\; {\cal{C}} \not=0$: The  physical quantities, $\rho$, $(\rho + p_r)$,  $(\rho - p_r)$, $(\rho + p_{\theta})$, $(\rho - p_{\theta})$, and  $(\rho + p_r + 2p_{\theta})$, represented, respectively, by Curves 1 - 6,  vs $x$ 
 in the neighborhood of the throat. All graphs are plotted with $\mathcal{C}=1.5,\; \mathcal{D}=2$, for which
the throat is at $x_{m} \approx -0.437$, and horizons are at $x_H^{\pm} =0$.} 
\lb{fig12}
\end{figure}

In Fig. \ref{fig12} we plot out the quantities $\rho,\; \rho \pm p_r, \; \rho \pm p_{\theta}$ and  $\rho +p_r +2p_{\theta} $ vs $x$ in the neighborhood of the throat for 
 $\mathcal{C}=1.5,\; \mathcal{D}=2$. With these choices, 
the throat is located at $x_{m} \approx -0.437$, and the  horizon is  at $x_H^{\pm} =0$.
From these curves we can see clearly that  the three energy conditions, WEC, SEC, and DEC, are satisfied only at the horizon.

At the spatial  infinities $x \rightarrow \pm \infty$, we find that 
\bqn
\lb{eq3.57}
\rho(x) &=& 
\begin{cases}
\frac{\mathcal{D} ^3}{8 x^5}+{\cal{O}}\left(\epsilon^6\right), & x \rightarrow \infty,\cr
-\frac{\mathcal{D} ^7}{8 x^5 \mathcal{C} ^4}+ {\cal{O}}\left(\epsilon^6\right), &  x \rightarrow - \infty,\cr
\end{cases}\nb\\
p_r(x) &=&
\begin{cases}
-\frac{\mathcal{D} ^2}{4 x^4}+\frac{\mathcal{D} ^3}{8 x^5}+ {\cal{O}}\left(\epsilon^6\right), &  x \rightarrow \infty,\cr
-\frac{\mathcal{D} ^6}{4 x^4 \mathcal{C} ^4}-\frac{\mathcal{D} ^7}{8 x^5 \mathcal{C} ^4}+ {\cal{O}}\left(\epsilon^6\right), &  x \rightarrow - \infty,\cr
\end{cases}\nb\\
p_{\theta}(x) &=&
\begin{cases}
\frac{\mathcal{D} ^2}{4 x^4}-\frac{\mathcal{D} ^3}{4 x^5}+{\cal{O}}\left(\epsilon^6\right), &  x \rightarrow \infty,\cr
\frac{\mathcal{D} ^6}{4 x^4 \mathcal{C} ^4}+\frac{\mathcal{D} ^7}{4 x^5 \mathcal{C} ^4}+{\cal{O}}\left(\epsilon^6\right), &  x \rightarrow - \infty,\cr
\end{cases}
\eqn
and 
\bqn
\lb{eq3.58}
a(x) &=& \begin{cases}
 \frac{1}{4}\left(1-\frac{2\mathcal{D} }{b}\right) +  {\cal{O}}\left(\epsilon^2\right), & x  \rightarrow  \infty,\cr
 \frac{\mathcal{D}^4}{4 \mathcal{C} ^4}\left(1-\frac{\left(2 \mathcal{C} ^2/\mathcal{D}\right)}{ b}\right) +  {\cal{O}}\left(\epsilon^2\right), & x  \rightarrow  -\infty,\cr
\end{cases}\nb\\
b(x) &\simeq&  \begin{cases}
2x+  {\cal{O}}\left(\epsilon\right), & x  \rightarrow  \infty,\cr
-2 \left(\mathcal{C} ^2 /\mathcal{D}^{2}\right) x+  {\cal{O}}\left(\epsilon\right), & x  \rightarrow  -\infty,\cr
\end{cases}
\eqn
where $\epsilon \equiv 1/x$.
Therefore, the masses of the black and white holes are given, respectively, by 
\bqn
\lb{eq3.59}
M_{BH} &=& \mathcal{D},\quad
 M_{WH} =   \frac{\mathcal{C} ^2}{\mathcal{D} }.
\eqn

  \begin{figure}[h!]
\includegraphics[height=4.5cm]{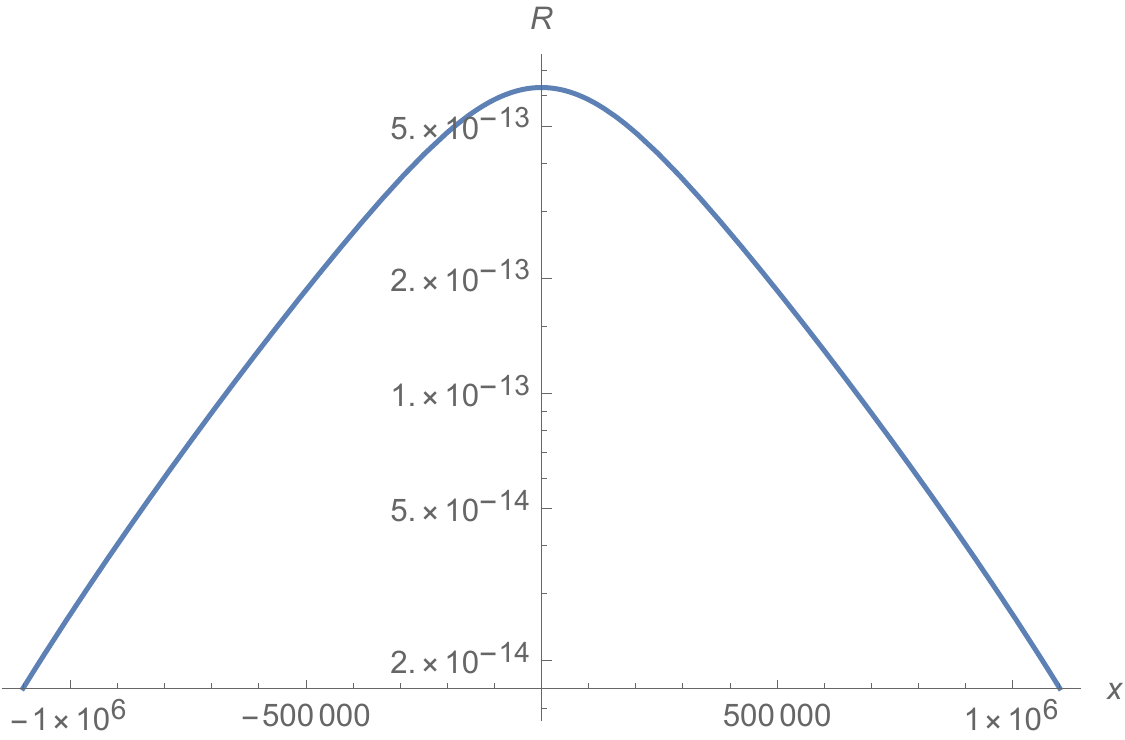}\\
{(a)}\\
\vspace{.5cm}
\includegraphics[height=4.5cm]{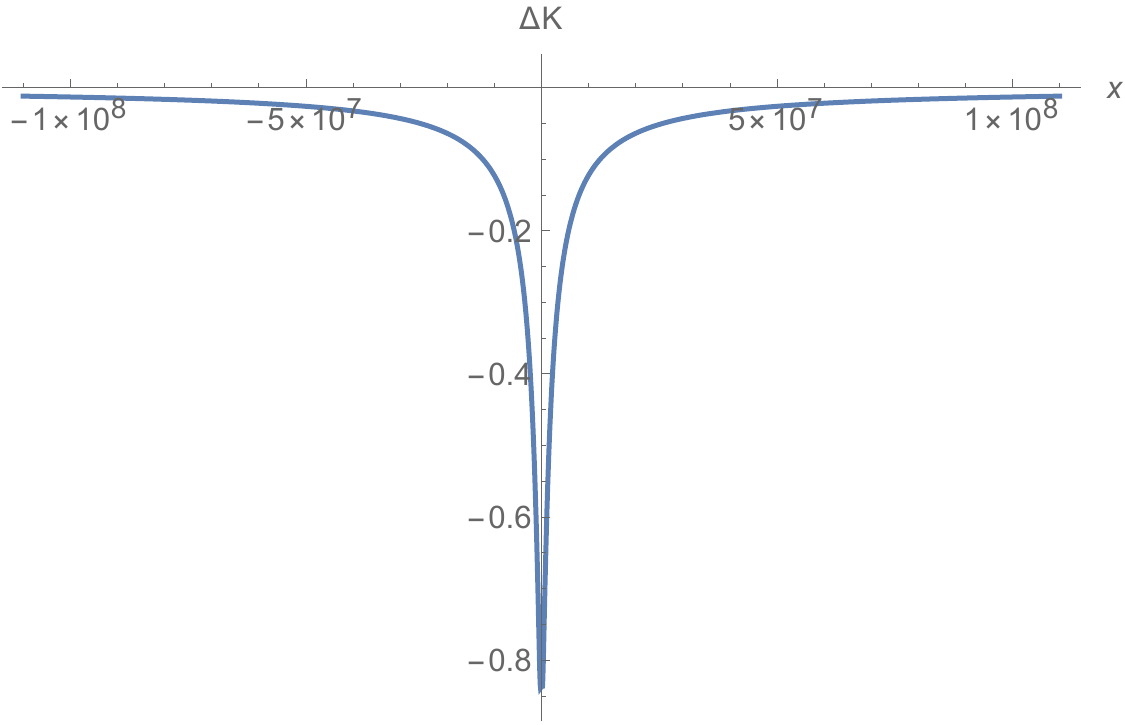}\\
{(b)}\\
\vspace{.5cm}
\caption{Case $\Delta = 0, \; {\cal{D}} > 0,\; {\cal{C}} \not=0$: $R$ and $\Delta  {\cal{K}}$ vs $x$.  Here we choose $\mathcal{C}= x_0 =10^6$,  for which the horizon and the throat are all located at 
$x_H^{\pm} = x_m =0$. } 
\label{fig12-1}
\end{figure}

On the other hand, from Eq.(\ref{eq3.57}) we find that  in the limit $x \rightarrow \infty$ we have
  \bqn
 \lb{eq3.61}
  \rho  &\approx&\frac{\mathcal{D} ^3}{8 x^5}+{\cal{O}}\left(\epsilon^6\right),\nb\\
     \rho + p_r &\approx & -\frac{\mathcal{D} ^2}{4 x^4}+\frac{\mathcal{D} ^3}{4 x^5}+{\cal{O}}\left(\epsilon^6\right),\nb\\
  \rho + p_{\theta}   &\approx & \frac{\mathcal{D} ^2}{4 x^4}-\frac{\mathcal{D} ^3}{8 x^5}+{\cal{O}}\left(\epsilon^6\right),\nb\\
   \rho +  p_r  + 2 p_{\theta}   &\approx & \frac{\mathcal{D} ^2}{4 x^4}-\frac{\mathcal{D} ^3}{4 x^5}+{\cal{O}}\left(\epsilon^6\right),
 \eqn
  while in the limit $x \rightarrow -\infty$, we have
   \bqn
 \lb{eq3.62}
  \rho  &\approx&-\frac{\mathcal{D} ^7}{8 x^5 \mathcal{C} ^4}+{\cal{O}}\left(\epsilon^6\right),\nb\\
     \rho + p_r &\approx & -\frac{\mathcal{D} ^6}{4 x^4 \mathcal{C} ^4}-\frac{\mathcal{D} ^7}{4 x^5 \mathcal{C} ^4}+{\cal{O}}\left(\epsilon^6\right),\nb\\
  \rho + p_{\theta}   &\approx & \frac{\mathcal{D} ^6}{4 x^4 \mathcal{C} ^4}+\frac{\mathcal{D} ^7}{8 x^5 \mathcal{C} ^4}+{\cal{O}}\left(\epsilon^6\right),\nb\\
   \rho +  p_r  + 2 p_{\theta}   &\approx &  \frac{\mathcal{D} ^6}{4 x^4 \mathcal{C} ^4}+\frac{\mathcal{D} ^7}{4 x^5 \mathcal{C} ^4}+{\cal{O}}\left(\epsilon^6\right).
 \eqn
 Therefore, {\it none of the three energy conditions is satisfied at both $x = -\infty$ and $x = \infty$}.

 In Fig. \ref{fig12-1}, we plot $R$ and $\Delta  {\cal{K}}$ for solar mass black/white holes in the region that covers the throat, with $\mathcal{C}= \mathcal{D} = x_0 =10^6$,  for which the horizon and the throat are all located at 
$x_H^{\pm} = x_m =0$. In this case, it can be seen   that the deviations from GR decay rapidly when away from the throat from both directions,  and the quantum gravitational  effects are mainly concentrated  in the neighborhood of it. 

In addition, as $x \rightarrow \pm \infty$, we find that
\bqn
\lb{eq3.62a}
 R &\simeq& \begin{cases}
-\frac{\mathcal{D}^2}{4 x^4}+\frac{\mathcal{D}^3}{2 x^5}+{\cal{O}}\left(\epsilon^6\right), & x  \rightarrow  \infty,\cr
-\frac{\mathcal{D}^6}{4 x^4 \mathcal{C} ^4}-\frac{\mathcal{D}^7}{2 x^5 \mathcal{C} ^4}+{\cal{O}}\left(\epsilon^6\right), & x  \rightarrow  -\infty,\cr
\end{cases}
\eqn
and 
\bqn
\lb{eq3.62b}
\Delta {\cal{K}}  &\simeq&  \begin{cases}
-\frac{4 M_{BH}}{3  x}+{\cal{O}}\left(\epsilon^2\right), &  x  \rightarrow   \infty,\cr
 +\frac{4   \mathcal{C} ^2}{3 M_{WH} x }  +{\cal{O}}\left(\epsilon^2\right), &  x  \rightarrow   - \infty,\cr
 \end{cases}
\eqn
where $M_{BH}$ and $M_{WH}$ are given by Eq.(\ref{eq3.59}).

  \subsubsection{$\mathcal{C} = 0$}

  In this case,  we have 
 \bqn
\lb{eq3.63}
a(x) &=& \frac{x^2X}{\left(X + {\cal{D}}\right)Y^2}, \quad
  b(x) = Y.
\eqn
Then,  $a(x) = 0$ leads to $x = 0$, which is a double root, as mentioned above.
 The geometric radius  $b(x)$ is a monotonically increasing function with $b(x = -\infty) =0$ [cf. Fig. \ref{fig1a-c}(c)]. 

 \begin{figure}[ht]
\includegraphics[height=4.8cm]{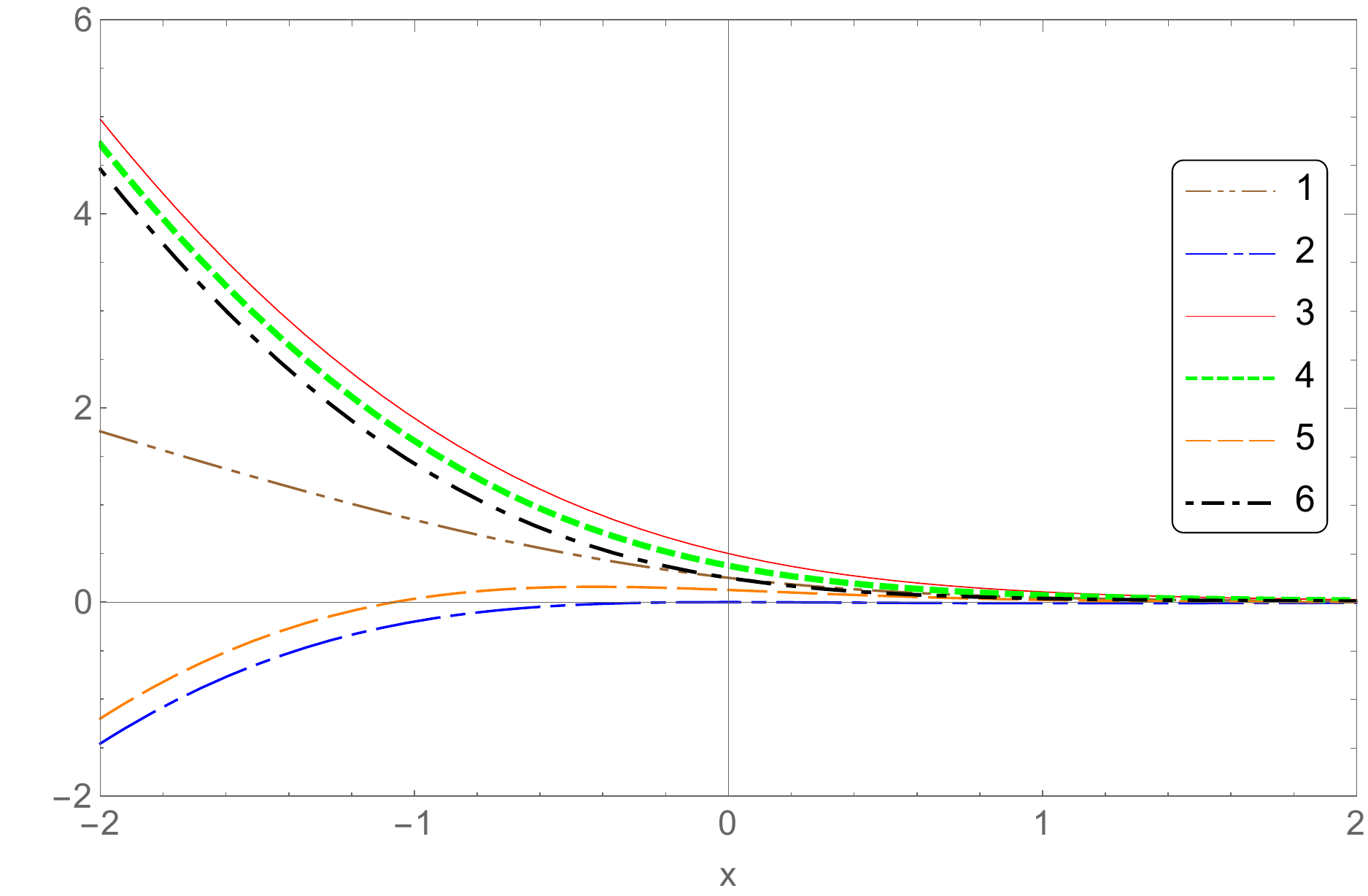}
\caption{Case   $\Delta = 0, \; {\cal{D}} >0,\; {\cal{C}} = 0$: The  physical quantities, $\rho$, $(\rho + p_r)$,  $(\rho - p_r)$, $(\rho + p_{\theta})$, $(\rho - p_{\theta})$, and  $(\rho + p_r + 2p_{\theta})$, represented, respectively, by Curves 1 - 6,  vs $x$
  in the neighborhood of the  horizon $x_H^{\pm} = 0$. All graphs are plotted with $\mathcal{D}=2$.} 
\lb{fig14}
\end{figure}

   In Fig. \ref{fig14} we plot the physical quantities $\rho, \; \rho \pm p_r,\;  \rho \pm p_{\theta}$ and $ \rho + p_r + 2  \pm p_{\theta}$ in the neighborhood of the  horizon $x_H = 0$,
   at which,   we have
 \bqn
 \lb{eq3.64}
 \rho &=& -p_r  =  2p_{\theta} = \frac{1}{\mathcal{D} ^2}, 
  \eqn
  so all the three energy conditions, WEC, SEC, and DEC, are satisfied.
  In addition, the surface gravity at this horizon also vanishes.
 
 At the spatial  infinities $x \rightarrow \pm \infty$, we find that 
\bqn
\lb{eq3.65}
\rho(x) &=& 
\begin{cases}
\frac{\mathcal{D} ^3}{8 x^5}+{\cal{O}}\left(\epsilon^6\right), & x \rightarrow \infty,\cr
-\frac{8 x}{\mathcal{D} ^3}+{\cal{O}}\left(\epsilon\right), &  x \rightarrow - \infty, \cr
\end{cases}\nb\\
p_r(x) &=&
\begin{cases}
-\frac{\mathcal{D} ^2}{4 x^4}+\frac{\mathcal{D} ^3}{8 x^5}+\left(\epsilon^6\right), &  x \rightarrow \infty,\cr
-\frac{16 x^2}{\mathcal{D} ^4}-\frac{8 r}{\mathcal{D} ^3}-\frac{4}{\mathcal{D} ^2}+{\cal{O}}\left(\epsilon\right), &  x \rightarrow - \infty,\cr
\end{cases}\nb\\
p_{\theta}(x) &=&
\begin{cases}
\frac{\mathcal{D} ^2}{4 x^4}-\frac{\mathcal{D} ^3}{4 x^5}+{\cal{O}}\left(\epsilon^6\right), &  x \rightarrow \infty,\cr
\frac{16 x^2}{\mathcal{D} ^4}+\frac{8 x}{\mathcal{D} ^3}+\frac{4}{\mathcal{D} ^2}+{\cal{O}}\left(\epsilon\right), &  x \rightarrow - \infty,\cr
\end{cases}
\eqn
from which we can see that none of the three energy conditions, WEC, SEC, and DEC,  is satisfied at the spatial infinities. In addition, we also have
\bqn
\lb{eq3.66}
a(x) &=& \begin{cases}
 \frac{1}{4}\left(1-\frac{2\mathcal{D} }{b}\right) +  {\cal{O}}\left(\epsilon^2\right), & x  \rightarrow  \infty,\cr
 \frac{4 x^4}{\mathcal{D}^4}+\frac{4   x^3}{\mathcal{D}^3}+\frac{6 x^2}{\mathcal{D}^2}+\frac{4   x}{\mathcal{D}}\nb\\
 +\frac{7}{4}+\frac{\mathcal{D} }{4 x}+{\cal{O}}\left(\epsilon^2\right), & x  \rightarrow  -\infty,\cr
\end{cases}\nb\\
b(x) &\simeq&  \begin{cases}
2x+{\cal{O}}\left(\epsilon\right), & x  \rightarrow  \infty,\cr
-\frac{\mathcal{D}^2}{2 x}+\frac{\mathcal{D}^4}{8 x^3}+{\cal{O}}\left(\epsilon^4\right), & x  \rightarrow  -\infty.\cr
\end{cases}
\eqn
Therefore, the mass of the black hole is given by 
\bqn
\lb{eq3.67}
M_{BH} &=& \mathcal{D}.
\eqn
However, at $x = - \infty$ we have $b(-\infty) = 0$, and  the physical quantities, $\rho, \; p_r$ and $p_{\theta}$, all become unbounded,  so a spacetime curvature singularity
appears at $x = - \infty$.

 \begin{figure}[h!]
 \includegraphics[height=4.8cm]{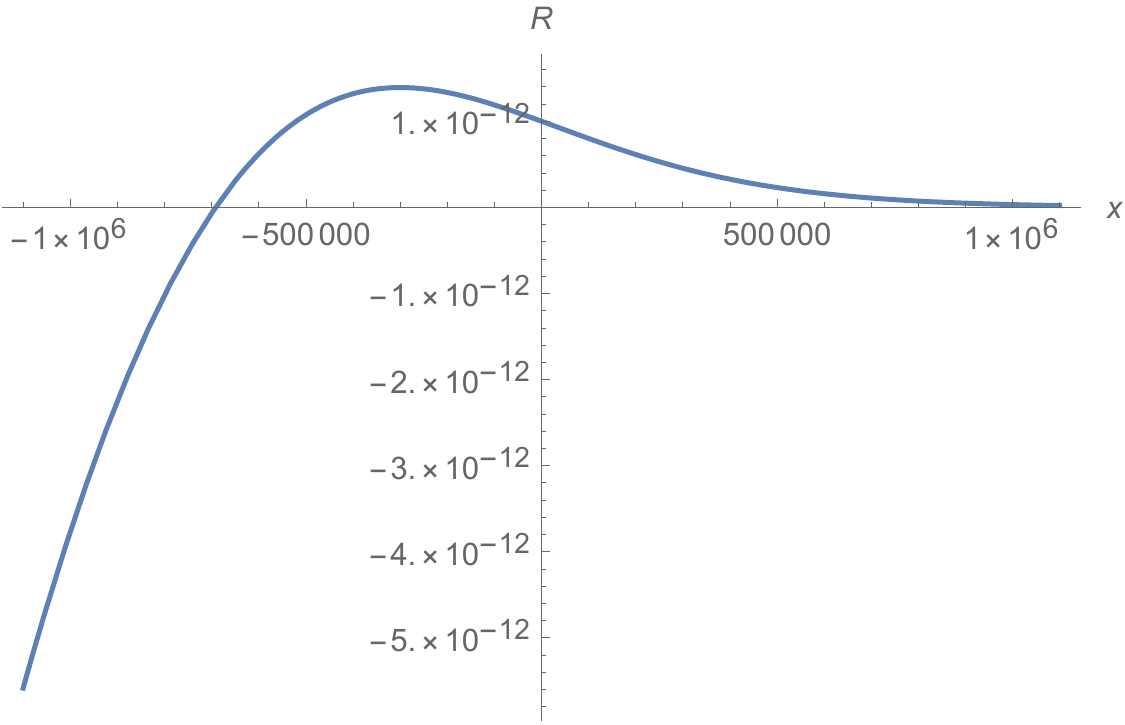}\\
 {(a)}\\
\vspace{.5cm}
\includegraphics[height=4.8cm]{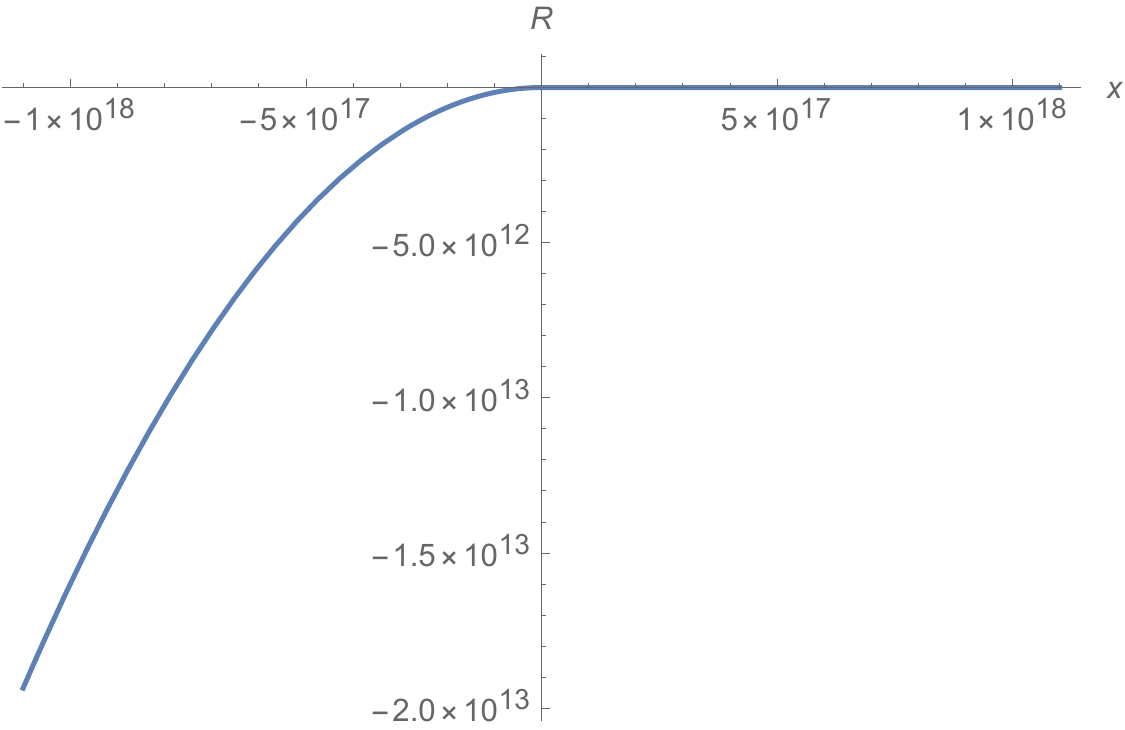}\\
{(b)}\\
\vspace{.5cm}
\includegraphics[height=4.8cm]{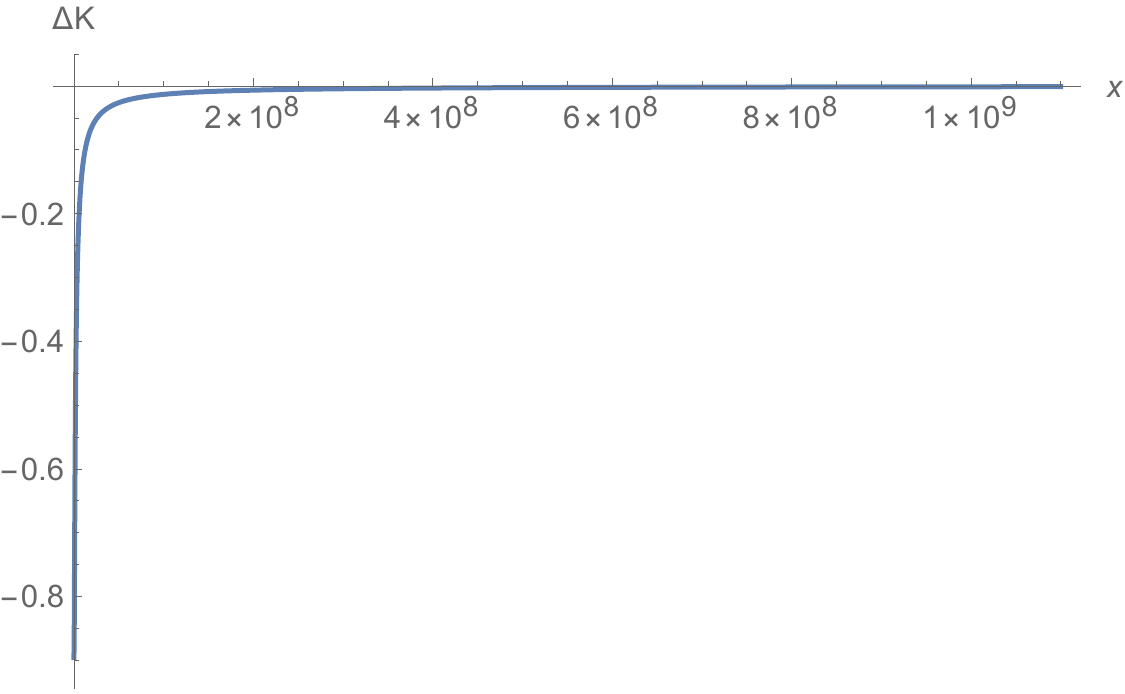}\\
{(c)}\\
\vspace{.5cm}
\caption{Case $\Delta = 0, \; {\cal{D}} >0,\; {\cal{C}} = 0$: $R$ and $\Delta  {\cal{K}}$ vs $x$.   Here we choose $x_{0}=10^6$, $\mathcal{D}=10^6$, so that $M_{BH}=10^6\; M_{Pl}$. Note that the horizon is located at 
$x_H^{\pm} =0$, and the spacetime is singular at $b(x= -\infty) = 0$.} 
\label{fig14-1}
\end{figure}

To study the quantum gravitational effects further, in Fig. \ref{fig14-1} we plot $R$ and $\Delta  {\cal{K}}$, from which it can be seen that the deviation from GR quickly becomes vanishingly small as $x \rightarrow \infty$. However,
 as $x \rightarrow -\infty$, $R$ diverges, as now the spacetime is singular at $b(x = -\infty) = 0$.
In fact, as $x \rightarrow \pm \infty$, we find that
\bqn
\lb{eq3.67a}
 R &\simeq& \begin{cases}
-\frac{\mathcal{D} ^2}{4 x^4}+\frac{\mathcal{D}^3}{2 x^5}+{\cal{O}}\left(\epsilon^6\right), & x  \rightarrow  \infty,\cr
-\frac{16 x^2}{\mathcal{D} ^4}-\frac{16   x}{\mathcal{D}^3}-\frac{4}{\mathcal{D} ^2}+{\cal{O}}\left(\epsilon\right), & x  \rightarrow  -\infty,\cr
\end{cases}
\eqn
\bqn
\lb{eq3.67b}
 \mathcal{K} &\simeq& \begin{cases}
\frac{3 \mathcal{D} ^2}{4 x^6}-\frac{\mathcal{D} ^3}{x^7}+{\cal{O}}\left(\epsilon^8\right), & x  \rightarrow  \infty,\cr
\frac{2816 x^4}{\mathcal{D} ^8}+\frac{3072 x^3}{\mathcal{D} ^7}+\frac{2368 x^2}{\mathcal{D} ^6}\\
+\frac{640 x}{\mathcal{D} ^5}+\frac{48}{\mathcal{D} ^4}+{\cal{O}}\left(\epsilon\right), & x  \rightarrow  -\infty,\cr
\end{cases}
\eqn
and 
\bqn
\lb{eq3.67c}
\Delta {\cal{K}}  &\simeq&  
 \begin{cases}
-\frac{4  M_{BH}}{3 x}+{\cal{O}}\left(\epsilon^2\right), &  x  \rightarrow   \infty, \cr 
-1+\frac{11 \mathcal{D} ^4}{12 M_{BH}^2 x^2}+{\cal{O}}\left(\epsilon^3\right), & x  \rightarrow  - \infty.
 \end{cases}
\eqn

\subsection{ ${\cal{D}} = 0$}

In this case, since $\left|{\cal{D}}\right| = \left|x_0\right|$, we also have $x_0=0$. Then, from Eq.(\ref{eq2.3}), this corresponds to the limit  $n \rightarrow \infty$. Again, to study the solutions further, we consider the two
cases $ \mathcal{C} \not= 0$ and  $\mathcal{C} = 0$, separately.

  \subsubsection{$\; \mathcal{C} \not= 0$}
  
From Eq.(\ref{eq2.6}) we find $X = |x|$, and 
  \bqn
\lb{eq3.67-1}
Y =x +|x| = \begin{cases}
2x, & x \ge 0,\cr
0, & x < 0.\cr
\end{cases}
\eqn
 Thus, from Eq.(\ref{eq2.5}) we find $a(x) = 0$ and $b(x) = \infty$ for $x \le 0$, that is, the metric becomes singular for $x \le 0$. 
 However, since $b(0) = \infty$, it is clear that now $x = 0$
 already represents the spatial infinity. Therefore, in this case we only need to consider  the region $x \in (0, \infty)$ [cf. Fig. \ref{fig1a-c}(b)].
In this case we have 
\bqn
\lb{eq3.67-2}
{a}(x) = \frac{x^2 Y^2}{Z^2},\quad {b}(x) =  \frac{Z}{Y}.
\eqn
Clearly, $a(x) = 0$ leads to a double root,
 $x_H^{\pm}= 0$,
 while the minimum of $b(x)$ now is located at $x_{m} \equiv {\hat{\cal{C}}}=\mathcal{C}/2$, so we have
 \bq
 \lb{eq3.67-4}
b(x) = \begin{cases}
\infty, & x = 0,  \cr
2^{4/3}{\hat{\cal{C}}}, &  x = {\hat{\cal{C}}},\cr
\infty, & x = \infty.\cr
\end{cases}
 \eq
 
The spacetime becomes antitrapped at $x = 0$. Since $b(x= 0) = \infty$, this antitrapped point now also corresponds to the spatial infinity at the other side of the throat, located at $x_{m}
  =  {\hat{\cal{C}}}$.   

On the other hand,   the effective energy density and pressures are now given by
 \bqn
\lb{eq3.67-5}
\rho(x) &=& -\frac{5120 x^8 \mathcal{C} ^6}{\left(64 x^6+\mathcal{C} ^6\right)^{8/3}}, \nb\\
 p_r(x) &=& -\frac{16 x^2 \mathcal{C} ^{12}}{\left(64 x^6+\mathcal{C} ^6\right)^{8/3}}, \nb\\
  p_{\theta}(x) &=& \frac{16 x^2 \mathcal{C} ^{12}}{\left(64 x^6+\mathcal{C} ^6\right)^{8/3}}, 
\eqn
which all become zero as $x \rightarrow 0$. They are also vanishing as $x \rightarrow \infty$. 

 At the throat $(x =\hat{\mathcal{C}})$, we have
 \bqn
 \lb{eq3.67-6}
 \rho = -\frac{5}{2^{8/3} \mathcal{\hat C} ^2},\;\;
  p_r  =-p_{\theta} =  -\frac{1}{2^{8/3} \mathcal{\hat C} ^2}, 
  \eqn
 so we find that none of the WEC, SEC, and DEC is satisfied.

 At the spatial infinity $x \rightarrow \infty$, on the other hand, we find
  \bqn
 \lb{eq3.67-7}
  \rho  &\approx&-\frac{5 \mathcal{C} ^6}{64 x^8}+{\cal{O}}\left(\epsilon^{9}\right),\nb\\
     \rho + p_r &\approx & -\frac{5 \mathcal{C} ^6}{64 x^8}+{\cal{O}}\left(\epsilon^{9}\right),\nb\\
  \rho + p_{\theta}   &\approx & -\frac{5 \mathcal{C} ^6}{64 x^8}+{\cal{O}}\left(\epsilon^{9}\right),\nb\\
   \rho +  p_r  + 2 p_{\theta}   &\approx & -\frac{5 \mathcal{C} ^6}{64 x^8}+{\cal{O}}\left(\epsilon^{9}\right),
 \eqn
 while as $x \rightarrow 0$ (or $b(x) \rightarrow \infty$), we find that
   \bqn
 \lb{eq3.67-8}
  \rho  &\approx& -\frac{5120 x^8}{\mathcal{C} ^{10}}+{\cal{O}}\left(x^{11}\right),\nb\\
     \rho + p_r &\approx & -\frac{16 x^2}{\mathcal{C} ^4}-\frac{7168 x^8}{3 \mathcal{C} ^{10}}+{\cal{O}}\left(x^{11}\right),\nb\\
  \rho + p_{\theta}   &\approx &\frac{16 x^2}{\mathcal{C} ^4}-\frac{23552 x^8}{3 \mathcal{C} ^{10}}+{\cal{O}}\left(x^{11}\right),\nb\\
   \rho +  p_r  + 2 p_{\theta}   &\approx & \frac{16 x^2}{\mathcal{C} ^4}-\frac{23552 x^8}{3 \mathcal{C} ^{10}}+{\cal{O}}\left(x^{11}\right),
 \eqn
 from which we can see that none of the three energy conditions is satisfied. 

In addition, we also have
\bqn
\lb{eq3.67-8}
a(x) &=& \begin{cases}
 \frac{1}{4}\left(1-\frac{2\mathcal{C}^6 }{3 b^6}\right) +  {\cal{O}}\left(\epsilon^7\right), & x  \rightarrow  \infty,\cr
\frac{4 x^4}{\mathcal{C} ^4}+{\cal{O}}\left(x^6\right), & x  \rightarrow  0,\cr
\end{cases}\nb\\
b(x) &\simeq&  \begin{cases}
2x+{\cal{O}}\left(\epsilon\right), & x  \rightarrow  \infty,\cr
\frac{\mathcal{C} ^2}{2 x}+\frac{32 x^5}{3 \mathcal{C} ^4}+{\cal{O}}\left(x^6\right). & x  \rightarrow  0.\cr
\end{cases}
\eqn
Thus,  the space-time is asymptotically flat as $x  \rightarrow  \infty$, with a black/hole mass  given by
\bq
\lb{eq3.67-9}
M_{BH/WH} = 0.
\eq

 \begin{figure}[h!]
 \includegraphics[height=4.8cm]{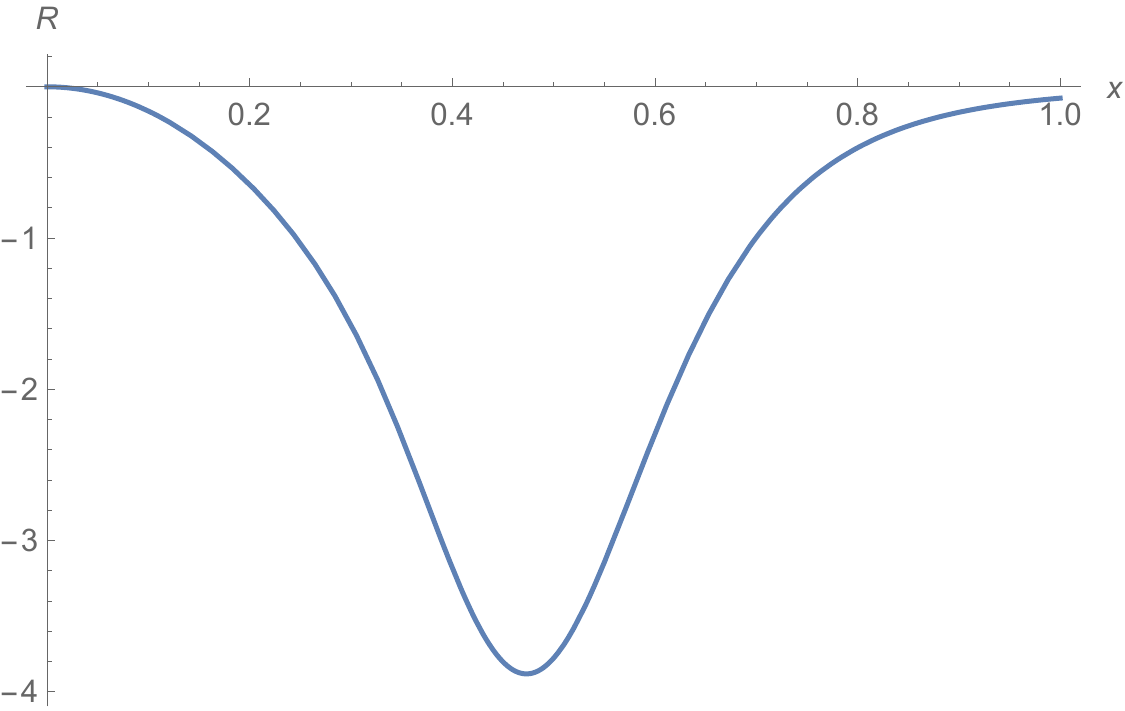}\\
 {(a)}\\
\vspace{.5cm}
 \includegraphics[height=4.8cm]{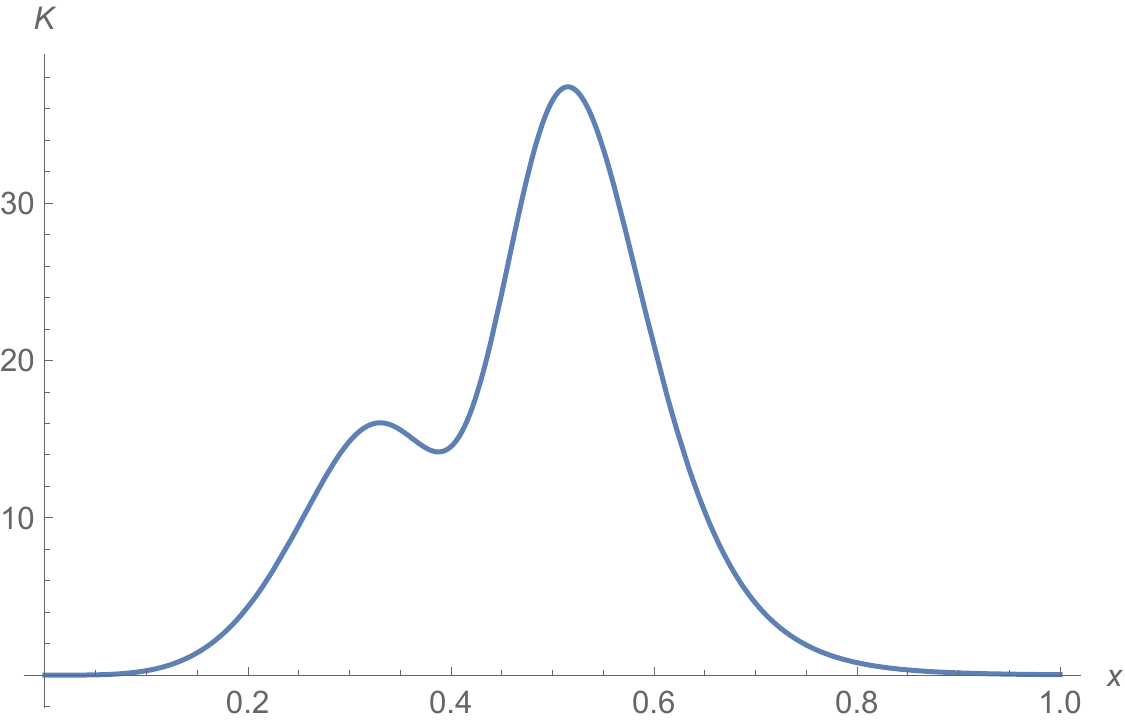}\\
 {(b)}\\
\vspace{.5cm}
\caption{Case $\Delta = 0, \; {\cal{D}} =0,\; {\cal{C}} \not= 0$: $R$ and $ {\cal{K}}$ vs $x$. Here we choose $\mathcal{C}=1$. Note that now the throat is at $x=\hat{\cal{C}}=1/2$.} 
\label{fig16-1}
\end{figure}

On the other hand,  to study the quantum gravitational effects, in Fig. \ref{fig16-1} we plot $R$ and $ {\cal{K}}$ in the region that covers the throat, and in the asymptotical regions
$x \rightarrow 0$ and $x \rightarrow \infty$,  from which it can be seen that the deviation from GR is mainly in the region near the throat, and quickly becomes vanishingly small as $x \rightarrow \infty$ or  $x \rightarrow 0$.

The spacetime is also asymptotically flat as $x \rightarrow 0\; (b(0) = \infty$). In fact, 
we find
\bqn
\lb{eq3.67-8a}
 R &\simeq& \begin{cases}
 -\frac{5 \mathcal{C} ^6}{64 x^8}+{\cal{O}}\left(\epsilon^{9}\right), & x \rightarrow \infty, \cr
-\frac{16 x^2}{\mathcal{C} ^4}+{\cal{O}}\left(x^4\right), & x \rightarrow 0, \cr
\end{cases}\nb\\
 {\cal{K}} &\simeq& \begin{cases}
\frac{127 \mathcal{C} ^{12}}{4096 x^{16}}+O\left(\epsilon^{19}\right), & x \rightarrow \infty, \cr
\frac{2816 x^4}{\mathcal{C} ^8}+O\left(x^6\right), & x \rightarrow 0. \cr
\end{cases} 
\eqn

\subsubsection{$\; \mathcal{C} = 0$} 

From Eq.(\ref{eq2.6}) we find  
  \bqn
\lb{eq3.67-9}
Y =x +|x| = \begin{cases}
2x, & x \ge 0,\cr
0, & x < 0.\cr
\end{cases}
\eqn
  Therefore, the spacetime must be restricted to the region $x \ge 0$, in which we have 
 \bqn
\lb{eq3.67-10}
a(x) &=& 
\frac{1}{4}, \quad
  b(x) =  (x+\sqrt{x^2})=2x,
\eqn
and
 \bqn
\lb{eq3.67-11}
\rho(x) = p_r=  p_{\theta}(x)=0.
\eqn
In fact, this is  precisely the Minkowski solution, and will take its standard form, by setting $r = 2x$ and rescaling $t$.
 
\subsection{ ${\cal{D}} < 0$}

Similar to the last subcase, now we also need to consider the cases  $\mathcal{C} \not= 0$ and $ \mathcal{C} = 0$ separately.

 \begin{widetext}
 
 \begin{table}
\caption{\label{tab:table1}%
The main properties of the solutions given by Eqs.(\ref{eq2.4})-(\ref{eq2.3b}) with $\Delta = 0$, for which we have $x_{H}^{\pm} = 0$, and the white and black  hole horizons 
always  coincide. Here
bhH $\equiv$ black hole horizon, whH $\equiv$ white hole horizon,  ECs $\equiv$ energy conditions, 
SAF $\equiv$ spacetime is asymptotical flat,   and SCS $\equiv$ spacetime curvature singularity.  
In addition, ``$\checkmark$" means yes,  ``$\times$" means no, while  ``N/A"  means not applicable.  }
\begin{tabular}{|l|c|c|c|c|c|c|c|c|c|c|c|c|}  \hline
\multicolumn{1}{|c|}{\bf  Properties}
& \multicolumn{6}{c|}{\bf  $\Delta=0$}\\ \cline{2-7}
 & \multicolumn{2}{c|}{$\mathcal{D}>0$} & \multicolumn{2}{c|}{$\mathcal{D}=0$}& \multicolumn{2}{c|}{$\mathcal{D}<0$}   \\ \cline{2-7}
 & $\mathcal{C} \not= 0$  & $\mathcal{C} = 0$  & $\mathcal{C} \not= 0$  & $\mathcal{C} = 0$  & $\mathcal{C} \not= 0$  & $\mathcal{C} = 0$     \\ \hline  
bhH/whH exists?  & \checkmark  &   \checkmark  &  \checkmark  & (Minkowski)  &  $\times$  &    $\times$  \\ \hline  
ECs at bhH/whH   & \checkmark   &   \checkmark   & $\times$  &  N/A & N/A   &  N/A    \\ \hline
Throat exists? &  \checkmark & $\times$  & \checkmark  &   N/A   &\checkmark  &   $\times$   \\ \hline
ECs at throat   & Eq.(\ref{eq3.56a})  &  N/A & $\times$   & N/A   &   $\times$ &  N/A    \\ \hline
ECs at $x = \infty$    &   $\times$  &  $\times$   & $\times$ &  N/A  & $\times$  &    $\times$ \\ \hline
Mass at $x = \infty$  & $\mathcal{D} $  & $\mathcal{D} $  & $0$  &  0   & $\mathcal{D} $ &   $\mathcal{D} $  \\ \hline 
ECs at $x = -\infty$    &   $\times$  & $\times$  &   N/A($x\ge 0$)  &  N/A  & $\times$  &  $\times$   \\ \hline
Mass at $x = -\infty$  & $\frac{\mathcal{C} ^2}{\mathcal{D} }$  &  SCS($b(-\infty)=0$) &  SAF($x=0$) &  N/A  & $\frac{\mathcal{C} ^2}{\mathcal{D} }$ &   SCS($b(-\infty)=0$)  \\ \hline     
\end{tabular}
\lb{Table2}
\end{table}

\end{widetext}

  \subsubsection{$\; \mathcal{C} \not= 0$}

When  ${\cal{D}} < 0$, we find that
\bq
\lb{eq3.67-13}
b(x) = \begin{cases}
\infty, & x = \infty,\cr
2^{1/3}{\cal{C}}, & x = x_{m},\cr
\infty, & x = - \infty,\cr
\end{cases}
\eq
where $x_{m} \equiv ({\cal{C}}^2 - {\cal{D}}^2)/(2{\cal{C}})$ [cf. Fig. \ref{fig1a-c}(a)]. 
On the other hand, $a(x) = 0$ has  no real roots, thus in the current case no black/white hole horizons exist.

 But, as shown by Eq.(\ref{eq3.67-13}), a  throat still exists at $x = x_{m}$, at which 
 the effective energy density $\rho$ and pressures $p_r$ and $p_{\theta}$ are still given by Eq.(\ref{eq3.56}), from which we find that none of the three energy conditions is satisfied at this point.

 At the spatial  infinities $x \rightarrow \pm \infty$, we find that the effective energy density $\rho$ and pressures $p_r$ and $p_{\theta}$ are still given by Eqs.(\ref{eq3.57}), (\ref{eq3.61}), and(\ref{eq3.62}),  from which we can see that {\it none of the three energy conditions is satisfied   at both $x = -\infty$ and $x = \infty$}.
 In addition,  the asymptotic expression of $a(x)$ and $b(x)$ are still given by Eq.(\ref{eq3.58}). Therefore, the total mass at $x \rightarrow \infty$ is given by 
\bqn
\lb{eq3.67-14a}
M_{+} &=& \mathcal{D}<0,
\eqn
while the total mass at $x \rightarrow -\infty$ is given by 
\bqn
\lb{eq3.67-14}
 M_{-} =   \frac{\mathcal{C} ^2}{\mathcal{D} }< 0.
\eqn

It can be shown that in the present case the quantum gravitational effects are also concentrated in the region near the throat, and are vanishing rapidly when away from it in each side of the throat.

  \subsubsection{$\; \mathcal{C} = 0$}
  
   In this case,  we have 
 \bqn
\lb{eq3.67-16}
a(x) &=& \frac{\left(X + \left|{\cal{D}}\right|\right)X}{Y^2}, \quad
  b(x) = Y.
\eqn
Thus, $a(x) = 0$ has no real roots, and $b(x)$ becomes a monotonically increasing function with $b(-\infty) = 0$ and $b(\infty) = \infty$ [cf. Fig. \ref{fig1a-c}(c)]. Therefore, in this case a throat does not exist.

 At the spatial  infinities $x \rightarrow \pm \infty$, we find that the effective energy density $\rho$ and pressures $p_r$ and $p_{\theta}$ are still given by Eq.(\ref{eq3.65}), from which we find that none of the three energy conditions, WEC, SEC and DEC,  is satisfied at the spatial infinity. In addition, the asymptotic expressions of $a(x)$ and $b(x)$ are still given by Eq.(\ref{eq3.66}). 
Therefore, the total mass at $x \rightarrow \infty$ is given by 
\bqn
\lb{eq3.67-17}
M_{+} &=& \mathcal{D}<0.
\eqn

However, at $x = - \infty$ we have $b(-\infty) = 0$, and  the physical quantities, $\rho, \; p_r$ and $p_{\theta}$, all become unbounded,  so a spacetime curvature singularity
appears at $x = - \infty$. Since no horizon exists, such a singularity is also naked.

 This completes our analysis for the case $\Delta = 0$, and the main properties of these solutions are summarized in Table \ref{Table2}.

\section{Spacetimes with $\Delta < 0$}
\label{Section.V}
 \renewcommand{\theequation}{5.\arabic{equation}}\setcounter{equation}{0}

In this case we have 
\bqn
\lb{eq3.68}
{a}(x) = \frac{\left(x^2 + \left|\Delta\right|\right)XY^2}{\left(X + {\cal{D}}\right)Z^2},\quad {b}(x) =  \frac{Z}{Y},
\eqn
where $X,\; Y,\; Z$ are given by Eq.(\ref{eq2.6}), while $\Delta$  is given by Eq.(\ref{eq2.3b}), from which we find $\Delta < 0$ implies 
\bqn
\lb{eq3.69}
\left|{\cal{D}}\right| <  \left|x_0\right|.  
\eqn
Then, we find that 
\bq
\lb{eq3.70}
b(x) = \begin{cases}
\infty, & x = \infty,\cr
2^{1/3}{\cal{C}}, & x = x_{m},\cr
\infty, & x = - \infty,\cr
\end{cases}
\eq
where $x_{m} \equiv ({\cal{C}}^2 - x_0^2)/(2{\cal{C}})$ [cf. Fig. \ref{fig1a-c}(a)].

To study the solutions further, as what we did in the last case, let us consider the solutions with $ {\cal{D}} > 0,  \; {\cal{D}} = 0$ and $ {\cal{D}} < 0$, separately. 

\subsection{$ {\cal{D}} > 0$}

Then, we find $a(x)$ is nonzero for any $x  \in (-\infty, \infty)$, and in particular we have
\bq
\lb{eq3.71}
a(x) = \begin{cases}
\frac{1}{4}, & x = \infty,\cr
\frac{x_0^4}{4{\cal{C}}^2}, & x = - \infty.\cr
\end{cases}
\eq
Thus, in the current case horizons do not exist. 
But, a throat does exist, as shown by Eq.(\ref{eq3.70}). At the throat, the effective energy density $\rho$ and pressures $p_r$ and $p_{\theta}$ are still given by Eq.(\ref{eq3.8}), from which we find that 
WEC, SEC and DEC are still  satisfied, provided that 
\bq
\lb{eq3.72}
|x_{0}| \leq \sqrt{\mathcal{C}  (2 \mathcal{D} -\mathcal{C} )}\; , \;0<\mathcal{C} \leq 2 \mathcal{D}.
\eq
In addition,   we also have the constraint $\left|{\cal{D}}\right| <  \left|x_0\right| $, as now we are considering the case $\Delta < 0$.

 At the spatial  infinities $x \rightarrow \pm \infty$, we find that the effective energy density $\rho$ and pressures $p_r$ and $p_{\theta}$ can be also written in the forms of  Eq.(\ref{eq3.20}), from which we can see that {\it none of the three energy conditions is satisfied   at both $x = -\infty$ and $x = \infty$}.
 
The asymptotic expressions of $a(x)$ and $b(x)$ are still given by Eq.(\ref{eq3.21}). Therefore, the total mass at $x \rightarrow \infty$ is given by 
\bqn
\lb{eq3.73}
M_{+} &=& \mathcal{D},
\eqn
while the total mass at $x \rightarrow -\infty$ is given by 
\bqn
\lb{eq3.74}
 M_{-} =   \frac{\mathcal{D} \mathcal{C} ^2}{x_0^2 }.
\eqn

It can be shown that the quantum gravitational effects are concentrated in the region near the throat, and are rapidly vanishing as away from the throat in each side of it only by proper choice of the free parameters involved
in the solutions,  as in the corresponding case $\Delta > 0, \; {\cal{D}} > 0$ and $x_0{\cal{C}} \not= 0$. 

Although no horizons exist in the present case, the corresponding solution is very interesting on its own rights: {\it it represents a wormhole spacetime, in which all the three energy conditions, WEC, SEC, and
DEC, are satisfied in the neighborhood of the throat, provided that Eq.(\ref{eq3.72}) holds, while none of them is satisfied at the asymptotically flat regions (spatial infinities) $x \rightarrow \pm \infty$}.

 It should be also noted that the above analysis does not cover the limit cases  $x_0 \rightarrow 0$ and ${\cal{C}} \rightarrow 0$. However, since now
 $\left|{\cal{D}}\right| <  \left|x_0\right|$, the cases  $x_0 = 0, \; {\cal{C}} \not= 0$ and $x_{0}= \mathcal{C} = 0$ do not exist. So, only the limiting case,
 $\mathcal{C} = 0,\; x_0 \not= 0$, exists.

  {\bf $\bullet \; \mathcal{C} = 0,\; x_0 \not= 0$:}  In this case, we have 
 \bqn
\lb{eq3.76}
a(x) &=& \frac{\left(x^2 + |\Delta|\right)X}{\left(X + {\cal{D}}\right)Y^2}, \quad
  b(x) = Y.
\eqn
Clearly, $a(x) = 0$ does not have real solutions, while $b(x)$ is a monotonically increasing function with $b(x = -\infty) =0$, as shown in Fig. \ref{fig1a-c}(c).
  
 At the spatial  infinities $x \rightarrow \pm \infty$, we find that the effective energy density $\rho$ and pressures $p_r$ and $p_{\theta}$ are still given by Eq.(\ref{eq3.39}), from which we can see that {\it none of the three energy conditions is satisfied   at both $x = -\infty$ and $x = \infty$}.

 The asymptotic expression of $a(x)$ and $b(x)$ are still given by Eq.(\ref{eq3.40}). Therefore, the total mass at $x \rightarrow \infty$ is given by 
\bqn
\lb{eq3.76-1}
M_{+} &=& \mathcal{D}.
\eqn  
However, at $x = - \infty$ we have $b(-\infty) = 0$, and  the physical quantities, $\rho, \; p_r$ and $p_{\theta}$, all become unbounded,  so a spacetime curvature singularity
appears at $x = - \infty$.

  \subsection{$\mathcal{D}=0$}
  
    From Eq.(\ref{eq2.5}) we find that 
 \bqn
\lb{eq3.76-2}
{a}(x) = \frac{X^2 Y^2}{Z^2},\quad {b}(x) =  \frac{Z}{Y},
\eqn
where $X, \; Y$, and $Z$ are given by Eq.(\ref{eq2.6}).  From the above expressions, it can be shown that   there are two 
 asymptotically flat regions, corresponding to $x \rightarrow \pm \infty$,
 respectively. They are still connected by a throat located at $x_{m}$ given by Eq.(\ref{eq3.7}) [cf. Fig. \ref{fig1a-c}(a)]. But since $a(x) \not =0$ for any given $x \in (-\infty, \infty)$, as it can be seen
 from the above expression, horizons, either WHs or BHs, do not exist.

At the throat, the effective energy density $\rho$ and pressures $p_r$ and $p_{\theta}$ are given by
\bqn
 \lb{eq3.76-3}
 \rho = -\frac{5}{2^{8/3} \mathcal{\hat C} ^2},\;\;
  p_r  =-p_{\theta} =  -\frac{1}{2^{8/3} \mathcal{\hat C} ^2}, 
  \eqn
 so none of the   WEC, SEC, and DEC is  satisfied.

At the spatial  infinities $x \rightarrow \pm \infty$, we find that the effective energy density $\rho$ and pressures $p_r$ and $p_{\theta}$ take the forms,  
\bqn
\lb{eq3.76-4}
\rho(x) &=& 
\begin{cases}
-\frac{5 \mathcal{C} ^6}{64 x^8}+ {\cal{O}}\left(\epsilon^9\right), & x \rightarrow \infty,\cr
-\frac{5 x_{0}^{16}}{64 x^8 \mathcal{C} ^{10}}+ {\cal{O}}\left(\epsilon^9\right), &  x \rightarrow - \infty,\cr
\end{cases}\nb\\
p_r(x) &=&
\begin{cases}
-\frac{x_{0}^2}{4 x^4}+ {\cal{O}}\left(\epsilon^6\right), &  x \rightarrow \infty,\cr
-\frac{x_{0}^6}{4 x^4 \mathcal{C} ^4}+ {\cal{O}}\left(\epsilon^6\right), &  x \rightarrow - \infty,\cr
\end{cases}\nb\\
p_{\theta}(x) &=&
\begin{cases}
\frac{x_{0}^2}{4 x^4}+ {\cal{O}}\left(\epsilon^6\right), &  x \rightarrow \infty,\cr
\frac{x_{0}^6}{4 x^4 \mathcal{C} ^4}+ {\cal{O}}\left(\epsilon^6\right), &  x \rightarrow - \infty,\cr
\end{cases}
\eqn
 from which we can see that {\it none of the three energy conditions is satisfied   at both $x = -\infty$ and $x = \infty$}.

In addition, we also have
\bqn
\lb{eq3.76-5}
a(x) &=& \begin{cases}
 \frac{1}{4}\left(1+\frac{2x_0^2 }{ b^2}\right) +  {\cal{O}}\left(\epsilon^3\right), & x  \rightarrow  \infty,\cr
\frac{x_{0}^4}{4 \mathcal{C} ^4}\left(1+\frac{2 \mathcal{C}^4 }{x_0^2 b^2}\right)+{\cal{O}}\left(\epsilon^2\right), & x  \rightarrow  -\infty,\cr
\end{cases}\nb\\
b(x) &\simeq&  \begin{cases}
2x+{\cal{O}}\left(\epsilon\right), & x  \rightarrow  \infty,\cr
-\frac{2 x \mathcal{C} ^2}{x_{0}^2}+{\cal{O}}\left(\epsilon\right). & x  \rightarrow  -\infty,\cr
\end{cases}
\eqn
from which we can see that the space-time is asymptotically flat as $x  \rightarrow \pm \infty$.

Similar to the last subcase,  the quantum gravitational effects are concentrated in the region near the throat, and are rapidly vanishing as away from the throat in each side of it for the proper choice of the free parameters, as in the corresponding case $\Delta > 0, \; {\cal{D}} = 0$ and $x_0{\cal{C}} \not= 0$.

In addition,  the above analysis is valid only for $x_0 \mathcal{C} \not= 0$. Otherwise, we have the following limiting case. 

  {\bf $\bullet \; x_0 \not= 0,\; \mathcal{C} = 0$:}  Then, we have
 \bqn
\lb{eq3.76-6}
{a}(x) = \frac{X^2 }{Y^2},\quad {b}(x) =  Y.
\eqn
Since $a(x) \not =0$ for any given real value of $x$, as it can be seen
 from the above expression, horizons, either WHs or BHs, do not exist, but $b(x)$ is still a monotonically increasing function with $b(x = -\infty) =0$, as shown in Fig. \ref{fig1a-c}(c).

At the spatial  infinities $x \rightarrow \pm \infty$, we find that the effective energy density $\rho$ and pressures $p_r$ and $p_{\theta}$ are given by 
\bqn
\lb{eq3.76-7}
\rho(x) &=& 
\begin{cases}
0, & x \rightarrow \infty,\cr
0, &  x \rightarrow - \infty,\cr
\end{cases}\nb\\
p_r(x) &=&
\begin{cases}
-\frac{x_{0}^2}{4 x^4}+ {\cal{O}}\left(\epsilon^6\right), &  x \rightarrow \infty,\cr
-\frac{16 x^2}{x_{0}^4}-\frac{4}{x_{0}^2}+\frac{x_{0}^2}{4 x^4}+{\cal{O}}\left(\epsilon^6\right), &  x \rightarrow - \infty,\cr
\end{cases}\nb\\
p_{\theta}(x) &=&
\begin{cases}
\frac{x_{0}^2}{4 x^4}+ {\cal{O}}\left(\epsilon^6\right), &  x \rightarrow \infty,\cr
\frac{16 x^2}{x_{0}^4}+\frac{4}{x_{0}^2}-\frac{x_{0}^2}{4 x^4}+{\cal{O}}\left(\epsilon^6\right), &  x \rightarrow - \infty,\cr
\end{cases}\nb\\
\eqn
 from which we can see that {\it none of the three energy conditions is satisfied to the leading order of $(1/x)$ at both $x = -\infty$ and $x = \infty$}.

In addition, we also have
\bqn
\lb{eq3.76-8}
a(x) &=& \begin{cases}
 \frac{1}{4}\left(1+\frac{2x_0^2 }{ b^2}\right) +  {\cal{O}}\left(\epsilon^3\right), & x  \rightarrow  \infty,\cr
\frac{4 x^4}{x_{0}^4}+\frac{6 x^2}{x_{0}^2}+\frac{7}{4}+{\cal{O}}\left(\epsilon^2\right), & x  \rightarrow  -\infty,\cr
\end{cases}\nb\\
b(x) &\simeq&  \begin{cases}
2x+{\cal{O}}\left(\epsilon\right), & x  \rightarrow  \infty,\cr
-\frac{x_{0}^2}{2 x}+{\cal{O}}\left(\epsilon^3\right), & x  \rightarrow  -\infty,\cr
\end{cases}
\eqn
from which we can see that the space-time is asymptotically flat as $x  \rightarrow + \infty$, but a spacetime curvature singularity appears at  $x  = - \infty$,
where $b(x  = - \infty) = 0$, as it can be seen from the above expressions.

  \subsection{$\mathcal{D}<0$}

  From Eq.(\ref{eq2.5}) we find that 
 \bqn
\lb{eq3.47}
{a}(x) = \frac{\left(X + \left|{\cal{D}}\right|\right)XY^2}{Z^2},\quad {b}(x) =  \frac{Z}{Y},
\eqn
where $X, \; Y$, and $Z$ are given by Eq.(\ref{eq2.6}). From the above expressions, it can be shown that   there are two 
 asymptotically flat regions, corresponding to $x \rightarrow \pm \infty$,
 respectively. They are still connected by a throat located at $x_{m}$ given by Eq.(\ref{eq3.7}) [cf. Fig. \ref{fig1a-c}(a)]. But since $a(x) \not =0$ for any given $x$, 
 horizons, either WHs or BHs, do not exist.
  
   At the throat, the effective energy density $\rho$ and pressures $p_r$ and $p_{\theta}$ are given by Eq.(\ref{eq3.8}). Then,
 it can be easily shown that none of the three energy conditions, WEC, SEC, and DEC,  can be satisfied in the current case.
 
 Similarly,  the quantum gravitational effects are concentrated in the region near the throat for only when   the free parameters are properly chosen, and are rapidly vanishing as away from the throat in each side of it.

  At the spatial  infinities $x \rightarrow \pm \infty$, we find that the expression of $\rho, p_r, p_{\theta}$ are still given by Eq.(\ref{eq3.20}), from which we can see that {\it none of the three energy conditions is satisfied to the leading order of $(1/x)$}. 

The asymptotic expressions of $a(x)$ and $b(x)$ are  given by Eq.(\ref{eq3.21}), and the total mass at $x \rightarrow \pm \infty$ is still given by Eq.(\ref{eq3.22}), but since we now have $\mathcal{D} < 0$, the total mass becomes negative.

Similar to the last case, the above analysis holds only for  $x_0 \mathcal{C} \not= 0$. When $x_0 \mathcal{C} = 0$, we find that only the possibility,  $x_0 \not= 0,\; \mathcal{C} = 0$, is allowed. 

  {\bf $\bullet \; x_0 \not= 0,\; \mathcal{C} = 0$:}   From Eq.(\ref{eq2.5}) we find that 
 \bqn
\lb{eq3.49}
{a}(x) = \frac{\left(X + \left|{\cal{D}}\right|\right)X}{Y^2},\quad {b}(x) =  Y,
\eqn
where $X, \; Y$, and $Z$ are given by Eq.(\ref{eq2.6}). Clearly, $a(x) = 0$ has no real roots,  thus  no horizons exist. On the other hand, $b(x)$ is still a monotonically increasing function with $b(x = -\infty) =0$, as shown in Fig. \ref{fig1a-c}(c).

At the spatial  infinities $x \rightarrow \pm \infty$, we find that  the effective energy density and pressures are still given by Eq.(\ref{eq3.39}), from which we find that none of  the three energy conditions, WEC, SEC, and DEC, is satisfied at the spatial infinities. In addition, the asymptotic behaviors of $a(x)$ and $b(x)$ are still given by Eq.(\ref{eq3.40}). Therefore, the total mass at $x = \infty$   is still given by 
Eq.(\ref{eq3.34d}), which is always negative.
 
However, at $x = - \infty$ we have $b(-\infty) = 0$, and  the physical quantities, $\rho, \; p_r$, and $p_{\theta}$, all become unbounded,  so a spacetime curvature singularity appears at $x = - \infty$.

This completes our analysis for the solutions with $\Delta < 0$, and the main properties of these solutions are summarized in Table \ref{Table3}.

\begin{widetext}

 \begin{table}
\caption{\label{tab:table1}%
The main properties of the solutions given by Eqs.(\ref{eq2.4})-(\ref{eq2.3b}) with $\Delta < 0$, for which no horizons exist in all these solutions. Here
bhH $\equiv$ black hole horizon, whH $\equiv$ white hole horizon,  ECs $\equiv$ energy conditions, 
SAF $\equiv$ spacetime is asymptotical flat,   and SCS $\equiv$ spacetime curvature singularity.  
In addition, ``$\checkmark$" means yes,  ``$\times$" means no, while  ``N/A"  means not applicable.  }
\begin{tabular}{|l|c|c|c|c|c|c|c|c|c|c|c|c|}  \hline
\multicolumn{1}{|c|}{\bf  Properties}
& \multicolumn{6}{c|}{\bf  $\Delta<0$}\\ \cline{2-7}
 & \multicolumn{2}{c|}{$\mathcal{D}>0$} & \multicolumn{2}{c|}{$\mathcal{D}=0$}& \multicolumn{2}{c|}{$\mathcal{D}<0$}   \\ \cline{2-7}
 & $\mathcal{C} x_0 \not= 0$  & $\mathcal{C} = 0, \; x_0 \not=0$  & $  x_0 \mathcal{C} \not= 0$  & $\mathcal{C} = 0, \; x_0 \not=0$  & $\mathcal{C} x_0 \not= 0,$  & $\mathcal{C} = 0, \; x_0 \not=0$     \\ \hline 
bhH/whH exists? & $\times$     &  $\times$  & $\times$ & $\times$ &  $\times$   & $\times$    \\ \hline  
Throat exists? &  \checkmark  & $\times$   & \checkmark  & $\times$ & \checkmark  & $\times$   \\ \hline
ECs at throat   & Eq.(\ref{eq3.72})    & N/A  &  $\times$ &  N/A  & $\times$ &  N/A   \\ \hline
ECs at $x = \infty$   &  $\times$    &  $\times$  &  $\times$   & $\times$ & $\times$  &  $\times$  \\ \hline
Mass at $x = \infty$  & $\mathcal{D}$   & $\mathcal{D}$  &  $-x_0^2$     & $-x_0^2$ (SAF)  &   $\mathcal{D}$    & $\mathcal{D}$   \\ \hline 
ECs at $x = -\infty$   & $\times$     &  $\times$  &  $\times$    &  $\times$ & $\times$ &  $\times$  \\ \hline
Mass at $x = -\infty$  &  $\frac{\mathcal{D} \mathcal{C} ^2}{x_0^2 }$   &  SCS & $-\frac{{\cal{C}}^4}{x_0^2}$   &  SCS  & $\frac{\mathcal{D} \mathcal{C} ^2}{x_0^2 }$ & SCS  \\ \hline     
\end{tabular}
\lb{Table3}
\end{table}

\end{widetext}

\section{Conclusions}
\label{Section.VI}
 \renewcommand{\theequation}{6.\arabic{equation}}\setcounter{equation}{0}

 In this paper, we have studied in detail  the   main properties of spherically symmetric black/white hole solutions, found recently by Bodendorfer, Mele, and M\"unch  \cite{BMM19}, inspired by the effective loop quantum gravity,
and paid particular attention to their local and global properties, as well as to the energy conditions of the effective energy-momentum tensor of the spacetimes.  Although this effective energy-momentum tensor 
is purely due to the quantum geometric effects, and is  not related to any real matter fields,  it does provide important information on how the spacetime singularity is avoided, and the deviations of the spacetimes 
from the classical one (the Schwarzschild solution). In particular,  spacetime singularities   inevitably occur in general relativity, as longer as  matter fields satisfy some  energy conditions, as follows directly from 
the Hawking-Penrose singularity theorems \cite{HE73}. In addition, due to the Birkhoff theorem, the spacetime is uniquely described by the Schwarzschild black hole solution in general relativity. Therefore, the presence
of this  effective energy-momentum tensor also characterizes the deviations of the quantum solutions from the classical one. 

 The most general  metric for static spherically symmetric spacetimes, without loss of the generality,   can be always cast in the form,  
 \bq
 \lb{6.0b}
 ds^2 = - a(x) dt^2 + \frac{dx^2}{a(x)} + b^2(x)\left(d^2\theta + \sin^2\theta d^2\phi\right),\nb
 \eq
 subjected to the following  {additional gauge transformations (gauge residuals), }
  \bq
 \lb{6.0c}
 t = \alpha \tilde{t} + t_0, \quad x = \xi(\tilde{x}), 
 \eq
 where $\alpha$ and $t_0$ are constant, and $\xi(\tilde{x})$ is an arbitrary function of $\tilde{x}$. Therefore, in general the phase space are four-dimensional, spanned by ($a, b, p_a, p_b$), but with one constraint,  
  the Hamiltonian constraint, $H_{\text{eff}} = 0$.
  {So,    the phase space is actually   three-dimensional, and the trajectories of the system are uniquely determined once the  three ``initial" conditions are given. However, due to the
  polymerization (\ref{eq2.12}), two new parameters are introduced, so the phase space is enlarged to five-dimensional, due to the polymerization quantizations.  Nevertheless, the trajectories of the system are also 
  gauge-invariant under the  transformations } (\ref{6.0c}), which reduce the dimensions of the phase space from five to three again. Therefore, {\it the phase space in  this model is 
  generically three-dimensional}. 
  
  The above general arguments  can be seen clearly from
  the particular solutions   considered in this paper, and the  three physically independent free parameters now can be chosen as (${\cal{C}}, \; {\cal{D}}, \; x_0$), defined explicitly by Eq.(\ref{eq2.3}), 
  \bqn
\lb{6.0a}
 {\cal{D}} &\equiv& \frac{3CD}{2\sqrt{n}}, \quad  {\cal{C}} \equiv \left(16C^2\lambda_1^2\right)^{1/6}, \quad x_0 \equiv \frac{\lambda_2}{\sqrt{n}},
\eqn
out of the five parameters,
 $\lambda_1, \; \lambda_2, \; n,\; C$, and $D$,  introduced  in  \cite{BMM19}.
  Thus, in comparison with the relativistic case,  the polymerizations  introduce two more free 
 parameters, and only when they vanish, i.e.,   $\lambda_1 = \lambda_2 = 0$ (or ${\cal{C}} = x_0 = 0$), can the solutions reduce to the Schwarzschild one with its mass $M_{BH} =  {\cal{D}}$, 
 and a spacetime curvature singularity located at the center ($b = 0$) appears. 
 If any of them vanishes, the corresponding  moment conjugate, $P_1$ or $P_2$, can become unbounded at some points (or in some regions) of the spacetime.
 As a result, spacetime curvature singularities  {will appear. From Tables II - IV it can be seen that in the current model 
 the condition for such singularities to appear is  indeed $\lambda_1 = 0$ (or ${\cal{C}} = 0$), the cases corresponding to Fig. 1(c). }

  The asymptotical properties of the spacetimes also depend on the choices of the two parameters ${\cal{C}}$ and $ x_0$.
In particular, when ${\cal{C}} x_0 \not = 0$,  we have $x \in(-\infty, \infty)$, and a minimal point (throat) of $b(x)$ always exists, with $b(\pm\infty) = \infty$ [cf. Fig. \ref{fig1a-c}(a)]. 
When ${\cal{C}} \not=0$ but $x_0 = 0$, the range of $x$ is
restricted to  $x \in(0, \infty)$ with  $b(0) = \infty$ and $b(\infty) = \infty$. In this case, a minimum of $b(x)$ also exists [cf. Fig. \ref{fig1a-c}(b)]. 
When ${\cal{C}} =0$ and $x_0 \not = 0$,
the range of $x$ is also $x \in(-\infty, \infty)$, but now $b(x)$ is a monotonically increasing function of $x$ with $b(-\infty) = 0$ and $b(\infty) = \infty$ [cf.  Fig. \ref{fig1a-c}(c)].

 In \cite{BMM19,BMM20,BL20}, the authors considered the case 
 \bq
 \lb{6.1}
 \Delta \equiv {\cal{D}}^2 - x_0^2 >0, \; {\cal{D}} > 0, \;{\cal{C}}x_0 \not= 0, 
 \eq
 for which the black and white hole horizons always exist, located at
 $$
 x_H^{\pm} = \pm \sqrt{\Delta}, 
 $$
 respectively, as shown in Sec. \ref{sec3-1} [See also Table \ref{Table1}].  The corresponding spacetime has two asymptotically flat regions 
 $x \rightarrow \pm \infty$, which are connected by a throat located at 
 $$
 x_m = \frac{1}{2{\cal{C}}}\left({\cal{C}}^2 -x_0^2\right),
 $$
 as can be seen from  Eq.(\ref{eq3.7}) and Fig. \ref{fig1a-c}(a)]. It is remarkable to note that in this case  the surface gravity at the black hole horizon $x = x_H^{+}$ is always
 positive, while at the white hole horizon $ x= x_H^{-}$, it is   always negative, as the latter represents an antitrapped surface.
  In the asymptotical region  $x \rightarrow + \infty$, the ADM mass reads 
 \bq
 \lb{6.2}
 M_{BH} =  {\cal{D}},  
 \eq
  while in the asymptotical region  $x \rightarrow - \infty$, it reads 
  \bq
  \lb{6.3}
  M_{WH} =   \frac{{\cal{D}}{\cal{C}}^2}{x_0^2},  
  \eq
  as given explicitly in Eq.(\ref{eq3.22}), which are all positive, too.    {All the above properties are mainly due to the fact that
   the Komar energy density \cite{Komar59} $\left(\rho + \sum_{i}{p_i}\right)$ remains positive over a large region of the spacetime, despite  that
 all  the three energy conditions are violated in most part of the spacetime, including the regions near the throat and horizons, as well as in the two asymptotically flat regions.}

    In addition, the quantum gravitational effects are mainly  concentrated in the neighborhood of the throat.
  However, in the current model,  such effects can be still large at the two horizons even for solar mass black/white hole spacetimes, depending on the choice of the free parameters.
  They   become negligible  near the black/white hole horizons only for some particular choices of these free parameters [cf. Eq.(\ref{eq3.25e})]. 
  
 Moreover,   the ratio $M_{BH}/M_{WH}$ can take in principle any value,  $M_{BH}/M_{WH} \in (0, \infty)$, as the three parameters ${\cal{C}}, \; {\cal{D}}, \; x_0$ now are all arbitrary (subjected only to the constraint $\mathcal{C} \ge 0$ as can be seen from Eq.(\ref{6.0a})) \cite{BMM20}.
 
 It should be also noted that the region defined by Eq.(\ref{6.1}) is quite small  in the whole three-dimensional phase space, spanned by 
 $ ({\cal{C}}, \; {\cal{D}}, \; x_0)$,   {where}
 \bq
 \lb{6.4}
 {\cal{D}}, \; x_0 \in (-\infty, \infty), \quad  {\cal{C}} \in [0, \infty),
 \eq
although the cases with ${\cal{D}} = 0$, or ${\cal{C}} = 0$, or $x_{0}= 0$ can be obtained only by  taking certain proper limits of the five free parameters,  $\lambda_1, \; \lambda_2, \; n,\; C$, and $D$, 
 as explained explicitly in the content. 
   
 With all the above in mind, we have explored the whole three-dimensional phase space of the three free parameters (${\cal{C}}, \; {\cal{D}}, \; x_0$), and found that the solutions have very rich physics. In particular, 
 the existence of  the black/white horizons crucially depends on the values of $\Delta$. When  $\Delta > 0$, they always exist and are located at $x_H^{\pm} = \pm \sqrt{\Delta}$, respectively.  {The spacetime
 in the region $ x_H^- < x < x_H^+$ becomes trapped.} When  $\Delta = 0$, 
 they also exist, but now become degenerate,  $x_H^{\pm} = 0$, that is, $a(x) = 0$ now has a double root,  { the trapped region ($a(x) < 0$) disappears, }
 and the surface gravity at the horizon is always zero now, quite similar to the extreme case of the charged RN solution
 with $|e| = m$. When  $\Delta < 0$, the equation $a(x) = 0$  has no real roots, and, as a result,  { in this case no horizons exist at all, neither a trapped region.}

Thus, depending on the choices of the three free parameters,  ${\cal{C}}, \; {\cal{D}}, \; x_0$, there are various cases that all have different (local and global) properties. In Secs. \ref{Section.III} - \ref{Section.V}, we have studied the cases
$\Delta > 0$, $\Delta = 0$, and $\Delta < 0$, separately, and in each of which all the three possible choices of ${\cal{C}}$ and $x_0$, as illustrated in Fig. \ref{fig1a-c},   {raise and} have been studied in detail. Their main properties are summarized in the three tables, Tables \ref{Table1} - \ref{Table3}. From these tables, the following interested cases are worthwhile of particularly mentioning: 

\begin{itemize}

\item {\bf $\Delta > 0,\; {\cal{D}} > 0, \; {\cal{C}}x_0 \not = 0$:} As mentioned above, in this case the solutions were first studied in \cite{BMM19,BMM20,BL20}, and in the present paper we have studied them in detail, and found the remarkable features stated above. In particular, we have shown explicitly that the quantum geometric efforts are mainly concentrated in the region near the throat (transition surface). However,  {in the current model } such effects can be still large at the black/white hole
horizons even for   solar mass black/white holes. They become negligible only in a restricted region of the 3D phase space, defined by Eq.(\ref{eq3.25e}).

\item {\bf $\Delta = 0,\; {\cal{D}} > 0, \; {\cal{C}}x_0 \not = 0$:} In this case, the black/white horizons coincide and all are located at $x_H^{\pm} = 0$, so the surface gravity at the horizon is zero, quite similar to the extreme case $|e| = m$ of the 
RN solution in general relativity. But, it is fundamentally different from the RN solution, as now there are no spacetime curvature singularities, and the spacetime becomes asymptotically flat in  both of the regions $x \rightarrow \pm \infty$. 

  In addition, all the three energy
conditions are  satisfied at the horizon, but at the throat $x = x_m$, they are satisfied only when ${\cal{D}} = {\cal{C}}$, for which the throat coincides with the horizon, i.e., $x_m = x_H^{\pm} = 0$. 

Similar to the last case (in fact, in all the cases, including $\Delta > 0$ and $\Delta <0$), none of the three energy conditions is satisfied at the spatial infinities $b(\pm \infty) = \infty$, although the quantum gravitational effects are also mainly concentrated at the throat, as shown in Fig. \ref{fig12-1}. In this case, the black/white hole masses are also given by Eqs.(\ref{6.2}) and (\ref{6.3}) but now with $\left|x_0\right| = {\cal{D}}$.

\item {\bf $\Delta < 0,\; {\cal{D}} > 0, \; {\cal{C}}x_0 \not = 0$:} In this case, the function $a(x)$ is always positive, and no horizons exist, although a throat does exist, as shown in Fig. \ref{fig1a-c}(a), at which all the three energy
conditions are satisfied, as long as the conditions (\ref{eq3.72}) hold. By properly choosing the free parameters, the quantum geometric effects can be made to be mainly concentrated at the throat, and the spacetime is asymptotically flat at both of the two limits,
$x \rightarrow \pm \infty$, with the ADM masses, given, respectively, by  Eqs.(\ref{6.2}) and (\ref{6.3}), which can be all positive. However, since no horizons exist, the  { spacetimes now represent  wormholes without any spacetime curvature singularities. } Again, this is not in conflict to the Hawking-Penrose singularity theorems \cite{HE73}, as none of the three energy conditions holds at the asymptotically flat regions, $x = \pm \infty$.

\end{itemize}

The main properties of other  {interesting} cases can be found in Tables \ref{Table1} - \ref{Table3}. 

It should be noted that, although in this paper we have studied only the solutions found recently in  \cite{BMM19}, we expect that  quantum 
black hole solutions share similar properties. In particular, due to the quantum geometric effects, an effective energy-momentum tensor inevitably appears, which generically violates the weak/strong energy conditions at the throat, so the 
spacetime is opened up by such repulsive forces. As a result,  the throat will connect two asymptotically flat regions. For spherical spacetimes \cite{Ashtekar20}, such effects are uniquely characterized by the two  quantum parameters  $\lambda_1$ and $\lambda_2$. The classical limit is obtained by setting $\lambda_1 = \lambda_2 = 0$. Therefore,  {the singularities inside the classical black holes are }
resolved by the polymerization \cite{Thiemann08}, given by  Eq.(\ref{eq2.12}), provided that 
\bq
 \lb{6.5}
\lambda_1  \lambda_2 \not= 0. 
\eq
If any of these two parameters vanishes, a spacetime curvature singularity can appear, as it is shown explicitly by the current model. 

Therefore, spherically  quantum   black holes should generically also contain three free parameters, which uniquely determine the location of the throat and the two masses, 
measured by observers located in the two asymptotically flat regions. Here  we use ``black holes" to emphasize the fact that in such resultant spacetimes white/black hole horizons are not necessarily always present, and 
spacetimes with  wormhole structures (without horizons) can be equally possible, unless the two free parameters $\lambda_1$ and $  \lambda_2$
are fixed by some physical considerations \cite{AOS18a,AOS18b,Ashtekar20}. It is also equally true that the two (Komar) masses are independent and can be assigned arbitrary values, unless additional physics
is taken into account \cite{AOS18a,AOS18b,BMM19,BMM20}. To understand these  { issues further,} one way is to consider the formation of such spacetimes from gravitational collapse of realistic matter fields 
\cite{BGMS05,LM08,ZZMB15,LMMB15,ABU20a,ABU20b,KSWe20b,ABMU20}.

Finally, we would like to mention that to get a universal curvature upper bound in these  polymer black holes, we need to impose specific relations between black and white hole masses \cite{BMM19},
 which amounts to impose further constraint in the parameter space. In this paper, we did not impose this condition in order to study properties in the whole parameter space.
  To overcome this problem,  recently BMM proposed another set of canonical variables in which one of the canonical momentum is precisely  the square root of the Kretschmann scalar \cite{BMM20}. In this new model, 
  a universal curvature upper bound can be obtained without any further constraint on the relation between black and white hole masses.

 
\section*{Acknowledgements}

 We would like to thank  Profs. Pisin Chen and Parampreet Singh for  their valuable comments and suggestions, which lead to various modifications.   
 W-C.G. is supported by Baylor University through the Baylor Physics graduate program. 
  This work is also partially supported by the National Natural Science Foundation of China with the Grants No. 11665016, No. 11675145, No. 11975116, and No. 11975203, 
  and Jiangxi Science Foundation for Distinguished Young Scientists under the Grant No. 20192BCB23007.

\section*{Appendix A: The general expressions of the energy density and pressures}
\label{Appendix:A}
 \renewcommand{\theequation}{A.\arabic{equation}}\setcounter{equation}{0}

Inserting the solutions given by Eq.(\ref{eq2.5}) into Eq.(\ref{eq3.5}), we find that
\begin{widetext} 
\bqn
\lb{A.1}
\rho(x)&=&\frac{Y^3}{X^2 Z^8}\Bigg[\Big(10 \mathcal{D}  x_{0}^{10} x+160 \mathcal{D}  x_{0}^8 x^3-20 x_{0}^6 \mathcal{C} ^6  
+672 \mathcal{D}  x_{0}^6 x^5+1024 \mathcal{D}  x_{0}^4 x^7-260 x_{0}^4 \mathcal{C} ^6 x^2\nb\\
&& +110 \mathcal{D}  x_{0}^4 \mathcal{C} ^6 x+512 \mathcal{D}  x_{0}^2 x^9-560 x_{0}^2 \mathcal{C} ^6 x^4 
+440 \mathcal{D}  x_{0}^2 \mathcal{C} ^6 x^3-320 \mathcal{C} ^6 x^6+352 \mathcal{D}  \mathcal{C} ^6 x^5 \Big) X \nb\\ 
&&
+\mathcal{D}  \mathcal{C} ^{12}+\mathcal{D}  x_{0}^{12}+50 \mathcal{D}  x_{0}^{10} x^2+400 \mathcal{D}  x_{0}^8 x^4 
+22 \mathcal{D}  x_{0}^6 \mathcal{C} ^6+1120 \mathcal{D}  x_{0}^6 x^6-100 x_{0}^6 \mathcal{C} ^6 x\nb\\
&& +1280 \mathcal{D}  x_{0}^4 x^8  -500 x_{0}^4 \mathcal{C} ^6 x^3+286 \mathcal{D}  x_{0}^4 \mathcal{C} ^6 x^2 
+512 \mathcal{D}  x_{0}^2 x^{10}-720 x_{0}^2 \mathcal{C} ^6 x^5+616 \mathcal{D}  x_{0}^2 \mathcal{C} ^6 x^4\nb\\
&& -320 \mathcal{C} ^6 x^7+352 \mathcal{D}  \mathcal{C} ^6 x^6\Bigg],
\eqn
\bqn
\lb{A.2}
p_r(x)&=&-\frac{Y^3}{X^2 Z^8} \Bigg[\Big( 2 x_{0}^{12}+100 x_{0}^{10} x^2-10 \mathcal{D}  x_{0}^{10} x 
+800 x_{0}^8 x^4-160 \mathcal{D}  x_{0}^8 x^3+2240 x_{0}^6 x^6\nb\\
&& -672 \mathcal{D}  x_{0}^6 x^5 +2560 x_{0}^4 x^8-1024 \mathcal{D}  x_{0}^4 x^7
+10 \mathcal{D}  x_{0}^4 \mathcal{C} ^6 x +1024 x_{0}^2 x^{10}-512 \mathcal{D}  x_{0}^2 x^9\nb\\
&& +40 \mathcal{D}  x_{0}^2 \mathcal{C} ^6 x^3+2 \mathcal{C} ^{12}+32 \mathcal{D}  \mathcal{C} ^6 x^5  \Big)X 
 -\mathcal{D}  \mathcal{C} ^{12}-\mathcal{D}  x_{0}^{12}+20 x_{0}^{12} x+340 x_{0}^{10} x^3\nb\\
&& -50 \mathcal{D}  x_{0}^{10} x^2+1664 x_{0}^8 x^5-400 \mathcal{D}  x_{0}^8 x^4 
+2 \mathcal{D}  x_{0}^6 \mathcal{C} ^6+3392 x_{0}^6 x^7-1120 \mathcal{D}  x_{0}^6 x^6\nb\\
&& +3072 x_{0}^4 x^9-1280 \mathcal{D}  x_{0}^4 x^8+26 \mathcal{D}  x_{0}^4 \mathcal{C} ^6 x^2
+1024 x_{0}^2 x^{11}-512 \mathcal{D}  x_{0}^2 x^{10}+56 \mathcal{D}  x_{0}^2 \mathcal{C} ^6 x^4\nb\\
&& +32 \mathcal{D}  \mathcal{C} ^6 x^6\Bigg],
\eqn
and
\bqn
\lb{A.3}
p_{\theta}(x)&=& \frac{Y^2}{2X^3 Z^8} \Bigg[\Big( 4 x_{0}^{14}+244 x_{0}^{12} x^2-34 \mathcal{D}  x_{0}^{12} x 
+2480 x_{0}^{10} x^4-720 \mathcal{D}  x_{0}^{10} x^3+9408 x_{0}^8 x^6\nb\\
&& -4256 \mathcal{D}  x_{0}^8 x^5+16384 x_{0}^6 x^8-10240 \mathcal{D}  x_{0}^6 x^7 
+12 \mathcal{D}  x_{0}^6 \mathcal{C} ^6 x+13312 x_{0}^4 x^{10}-10752 \mathcal{D}  x_{0}^4 x^9\nb\\
&& +88 \mathcal{D}  x_{0}^4 \mathcal{C} ^6 x^3+4 x_{0}^2 \mathcal{C} ^{12}+4096 x_{0}^2 x^{12} 
- 4096 \mathcal{D}  x_{0}^2 x^{11}+192 \mathcal{D}  x_{0}^2 \mathcal{C} ^6 x^5+128 \mathcal{D}  \mathcal{C} ^6 x^7\nb\\
&& +4 \mathcal{C} ^{12} x^2-2 \mathcal{D}  \mathcal{C} ^{12} x  \Big) X -3 \mathcal{D}  x_{0}^{14}+44 x_{0}^{14} x 
+924 x_{0}^{12} x^3-194 \mathcal{D}  x_{0}^{12} x^2+5808 x_{0}^{10} x^5\nb\\
&& -2080 \mathcal{D}  x_{0}^{10} x^4+2 \mathcal{D}  x_{0}^8 \mathcal{C} ^6+16192 x_{0}^8 x^7 
-8288 \mathcal{D}  x_{0}^8 x^6+22528 x_{0}^6 x^9-15104 \mathcal{D}  x_{0}^6 x^8\nb\\
&&  +40 \mathcal{D}  x_{0}^6 \mathcal{C} ^6 x^2+15360 x_{0}^4 x^{11}-12800 \mathcal{D}  x_{0}^4 x^{10} 
+168 \mathcal{D}  x_{0}^4 \mathcal{C} ^6 x^4-3 \mathcal{D}  x_{0}^2 \mathcal{C} ^{12}+4096 x_{0}^2 x^{13}\nb\\
&& -4096 \mathcal{D}  x_{0}^2 x^{12}+256 \mathcal{D}  x_{0}^2 \mathcal{C} ^6 x^6+4 x_{0}^2 \mathcal{C} ^{12} x 
+128 \mathcal{D}  \mathcal{C} ^6 x^8+4 \mathcal{C} ^{12} x^3-2 \mathcal{D}  \mathcal{C} ^{12} x^2\Bigg].
\eqn

\end{widetext}


\begin{thebibliography}{199}


\bibitem{Singh:2009mz}
P.~Singh,
Are loop quantum cosmos never singular?
Class. Quant. Grav. \textbf{26} (2009), 125005.

\bibitem{Ashtekar:2011ni}
A.~Ashtekar and P.~Singh,
Loop Quantum Cosmology: A Status Report,
Class. Quant. Grav. \textbf{28} (2011), 213001.



\bibitem{Ashtekar:2005qt}
A.~Ashtekar and M.~Bojowald,
Quantum geometry and the Schwarzschild singularity,
Class. Quant. Grav. \textbf{23} (2006) 391.

\bibitem{Modesto:2005zm}
L.~Modesto,
Loop quantum black hole,
Class. Quant. Grav. \textbf{23} (2006), 5587.


\bibitem{Campiglia:2007pb}
M.~Campiglia, R.~Gambini and J.~Pullin,
Loop quantization of spherically symmetric midi-superspaces : The Interior problem,
AIP Conf. Proc. \textbf{977} (2008)  52.

\bibitem{Bohmer:2007wi}
C.~G.~Boehmer and K.~Vandersloot,
Loop Quantum Dynamics of the Schwarzschild Interior,
Phys. Rev. D \textbf{76} (2007), 104030.

\bibitem{Chiou:2008eg}
D.~W.~Chiou,
Phenomenological dynamics of loop quantum cosmology in Kantowski-Sachs spacetime,
Phys. Rev. D \textbf{78} (2008), 044019.


\bibitem{GP08} R. Gambini, J. Pullin, Black Holes in Loop Quantum Gravity: The Complete Space-Time, Phys. Rev. Lett. {\bf  101}, 161301 (2008).

\bibitem{Modesto:2008im}
L.~Modesto,
Semiclassical loop quantum black hole,
Int. J. Theor. Phys. \textbf{49} (2010), 1649.

\bibitem{Chiou:2008nm}
D.~W.~Chiou,
Phenomenological loop quantum geometry of the Schwarzschild black hole,
Phys. Rev. D \textbf{78} (2008), 064040.


\bibitem{Brannlund:2008iw}
J.~Brannlund, S.~Kloster and A.~DeBenedictis,
The Evolution of Lambda Black Holes in the Mini-Superspace Approximation of Loop Quantum Gravity,
Phys. Rev. D \textbf{79} (2009), 084023.



\bibitem{GP13} R. Gambini, J. Pullin,  An introduction to spherically symmetric loop quantum gravity black holes, AIP Conf. Proc. \textbf{1647}, 19 (2015). arXiv:1312.5512.

\bibitem{Joe:2014tca}
A.~Joe and P.~Singh,
Kantowski-Sachs spacetime in loop quantum cosmology: bounds on expansion and shear scalars and the viability of quantization prescriptions,
Class. Quant. Grav. \textbf{32} (2015)  015009.

\bibitem{Corichi:2015xia}
A.~Corichi and P.~Singh,
Loop quantization of the Schwarzschild interior revisited,
Class. Quant. Grav. \textbf{33} (2016)  055006.




\bibitem{Dadhich:2015ora}
N.~Dadhich, A.~Joe and P.~Singh,
Emergence of the product of constant curvature spaces in loop quantum cosmology,
Class. Quant. Grav. \textbf{32} (2015)  185006.

\bibitem{Cortez:2017alh}
J.~Cortez, W.~Cuervo, H.~A.~Morales-Técotl and J.~C.~Ruelas,
Effective loop quantum geometry of Schwarzschild interior,
Phys. Rev. D \textbf{95} (2017) 064041.

\bibitem{Olmedo:2017lvt}
J.~Olmedo, S.~Saini and P.~Singh,
From black holes to white holes: a quantum gravitational, symmetric bounce,
Class. Quant. Grav. \textbf{34} (2017)  225011.








\bibitem{AP17} A. Perez, Black holes in loop quantum gravity, Rep. Prog. Phys. {\bf 80} (2017) 126901. 

\bibitem{BMM18} A. Barrau, K. Martineau and F. Moulin, A Status Report on the Phenomenology of Black Holes in Loop Quantum Gravity: Evaporation, Tunneling to White Holes, Dark Matter and Gravitational Waves, 
Universe {\bf 4} (2018) 102.


\bibitem{BCDHR18} E. Bianchi, M. Christodoulou, F. D'Ambrosio, H. M Haggard, and C. Rovelli, White holes as remnants: a surprising scenario for the end of a black hole, Class. Quantum Grav. {\bf 35} (2018) 225003.

\bibitem{AOS18a} A. Ashtekar, J. Olmedo, P. Singh, Quantum Transfiguration of Kruskal Black Holes, Phys. Rev. Lett. {\bf 121}, 241301 (2018).

\bibitem{AOS18b} A. Ashtekar, J. Olmedo, P. Singh, Quantum extension of the Kruskal spacetime, Phys. Rev. D{\bf 98}, 126003 (2018).


  \bibitem{CR18} C. Rovelli, Black Hole Evolution Traced Out with Loop Quantum Gravity, Physics {\bf 11} (2018) 127. 
 
 
 

\bibitem{Alesci:2019pbs}
E.~Alesci, S.~Bahrami and D.~Pranzetti,
Quantum gravity predictions for black hole interior geometry,
Phys. Lett. B \textbf{797} (2019), 134908.

\bibitem{Assanioussi:2019twp}
M.~Assanioussi, A.~Dapor and K.~Liegener,
Perspectives on the dynamics in a loop quantum gravity effective description of black hole interiors,
Phys. Rev. D \textbf{101} (2020)  026002.

\bibitem{Bodendorfer:2019nvy}
N.~Bodendorfer, F.~M.~Mele and J.~Munch,
(b,v)-Type Variables for Black to White Hole Transitions in Effective Loop Quantum Gravity,
arXiv:1911.12646.




 
 \bibitem{CDLV19} R. Carballo-Rubio, F. Di Filippo, S. Liberati, M. Visser, Geodesically complete black holes, Opening the Pandora's box at the core of black holes, Classical Quantum Gravity \textbf{37}, 145005 (2020). arXiv:1908.03261.


\bibitem{MBM19}  F. Moulin, A. Barrau, and K. Martineau, 
An overview of quasinormal modes in modified and extended gravity,
Universe \textbf{5} (2019)  202.

\bibitem{AAN20}  D. Arruga, J. Ben Achour, and K. Noui, 
Deformed General Relativity and Quantum Black Holes Interior,
Universe \textbf{6} (2020)  39.



\bibitem{Liu:2020ola}
C.~Liu, T.~Zhu, Q.~Wu, K.~Jusufi, M.~Jamil, M.~Azreg-Anou and A.~Wang,
Shadow and Quasinormal Modes of a Rotating Loop Quantum Black Hole,
Phys. Rev. D \textbf{101} (2020)  084001.


 \bibitem{Agullo20}  I.Agullo, V. Cardoso, A. del Rio, M. Maggiore, and  J. Pullin, Gravitational-wave signatures of quantum gravity, arXiv:2007.13761.
 


 \bibitem{AO20} A. Ashtekar, J. Olmedo, Properties of a recent quantum extension of the Kruskal geometry, Int. J. Mod. Phys. D \textbf{29}, 2050076 (2020). arXiv:2005.02309.
 
 
 \bibitem{ZMSZ20} C. Zhang, Y.-G. Ma, S.-P. Song X.-D. Zhang, Loop quantum Schwarzschild interior and black hole remnant, Phys. Rev. D \textbf{102}, 041502 (2020). arXiv:2006.08313.


\bibitem{GOP20} R. Gambini, J. Olmedo, J. Pullin, Spherically symmetric loop quantum gravity: Analysis of improved dynamics, Classical Quantum Gravity \textbf{37}, 205012 (2020). arXiv:2006.01513.


 \bibitem{KSWe20}  J. G. Kelly, R. Santacruz, E. Wilson-Ewing, Effective loop quantum gravity framework for vacuum spherically symmetric space-times, arXiv:2006.09302.

\bibitem{Bojowald:2018xxu}
M.~Bojowald, S.~Brahma and D.~h.~Yeom,
Effective line elements and black-hole models in canonical loop quantum gravity,
Phys. Rev. D \textbf{98}, no.4, 046015 (2018).
 
 \bibitem{CR17} C. Rovelli and F. Vidotto, Small Black/White Hole Stability and Dark Matter, Universe {\bf 4} (2018) 127.

\bibitem{RMD18} C. Rovelli and P. Martin-Dussaud, Interior metric and ray-tracing map in the firework black-to-white hole transition, Class. Quantum Grav. {\bf 35} (2018) 147002.


 \bibitem{MDR19}  P. Martin-Dussaud and C. Rovelli, Evaporating black-to-white hole, Class. Quantum Grav. {\bf 36} (2019) 245002.
 
 
 
 
 
 
 
 \bibitem{KM04} K.A. Meissner, Black-hole entropy in loop quantum gravity, Classical Quantum Gravity {\bf 21}, 5245 (2004).


\bibitem{Thiemann08} T. Thiemann, Modern canonical quantum general relativity (Cambridge University Press, 2008).

\bibitem{Ashtekar20} A. Ashtekar, Black Hole evaporation: A Perspective from Loop Quantum Gravity, Universe {\bf 6} (2020) 21.

\bibitem{Bouhmadi-Lopez:2019hpp}
M.~Bouhmadi-Lopez, S.~Brahma, C.~Y.~Chen, P.~Chen and D.~h.~Yeom,
Asymptotic non-flatness of an effective black hole model based on loop quantum gravity, 
Phys. Dark Universe {\bf 30}, 100701 (2020).
arXiv:1902.07874.

\bibitem{Bojowald:2019dry}
M.~Bojowald,
Comment (2) on ``Quantum Transfiguration of Kruskal Black Holes",
arXiv:1906.04650.

 \bibitem{Bojowald20} M. Bojowald, A no-go result for covariance in models of loop quantum gravity,
 Phys. Rev. D {\bf 102}, 046006 (2020). arXiv:2007.16066.
 
 
  \bibitem{GPS02} H. Goldstein, C. Poole, and J. Safko, Classical Mechanics, Third Edition (Addison Wesley, New York, 2002), pp. 368-424.
  

\bibitem{BMM19} N. Bodendorfer, F.M. Mele, and J. M\"unch, Effective quantum extended spacetime of polymer Schwarzschild black hole, Class. Quantum Grav. {\bf 36} (2019) 195015.

\bibitem{BMM20} N. Bodendorfer, F.M. Mele, and J. M\"unch, Mass and Horizon Dirac Observables in effective Models of
Quantum Black-to-White Hole Transition, arXiv:1912.00774.

\bibitem{APSV07} A. Ashtekar, T. Pawlowski, P. Singh, K. Vandersloot, Loop quantum cosmology of $k = 1$ FRW models, Phys.  Rev. D{\bf 75}, 024035 (2007).
 
 \bibitem{HE73} S.W. Hawking, G.F.R. Ellis, {\em The Large Scale Structure of Spacetime} (Cambridge University Press, Cambridge, 1973).
 
   \bibitem{Komar59} A. Komar, Covariant Conservation Laws in General Relativity, Phys. Rev. {\bf 113} (1959) 934.

\bibitem{Birkhoff23} G. D. Birkhoff, Relativity and Modern Physics (Cambridge, Massachusetts, Harvard University Press, 1923).


 
 
  \bibitem{BL20}  M. Bouhmadi-Lopez, S. Brahma, C.-Y. Chen, P. Chen, D.-h. Yeom, A consistent model of non-singular Schwarzschild black hole in loop quantum gravity and its quasinormal modes,
  J. Cosmol. Astropart. Phys. 07 (2020) 066.
 arXiv:2004.13061.
 
 
 
 
 
 
 \bibitem{BGMS05}  M. Bojowald, R. Goswami, R. Maartens, and P. Singh, Black hole mass threshold from nonsingular quantum gravitational collapse,
 Phys. Rev. Lett. {\bf 95}, 091302 (2005).
 
\bibitem{LM08}  L. Modesto, Gravitational Collapse in Loop Quantum Gravity,  Int. J. Theor. Phys. {\bf 47} (2008) 357.
 

 
 
  \bibitem{ZZMB15} Y. Zhang, Y. Zhu, L. Modesto, C. Bambi, Can static regular black holes form from gravitational collapse? Eur. Phys. J. C {\bf 75} (2015) 96.
 
 
  \bibitem{LMMB15} Y. Liu, D. Malafarina, L. Modesto, and C. Bambi, Singularity avoidance in quantum-inspired inhomogeneous dust collapse, 
   Phys. Rev. D{\bf 90}, 044040 (2014).
 
 \bibitem{ABU20a}  J. Ben Achour, S. Brahma,  J-P. Uzan, Bouncing compact objects. Part I. Quantum extension of the Oppenheimer-Snyder collapse,
 JCAP {\bf 03} (2020) 041. 
 
 
  \bibitem{ABU20b}  J. Ben Achour, S. Brahma,  J-P. Uzan, Bouncing compact objects II: Effective theory of a pulsating Planck star, 	arXiv:2001.06153.
  
    
\bibitem{ABMU20}  J. Ben Achour, S. Brahma,  S. Mukohyama,  J-P. Uzan, Towards consistent black-to-white hole bounces from matter collapse, J. Cosmol. Astropart. Phys. 09 (2020) 020. arXiv:2004.12977.


  
   \bibitem{KSWe20b}  J. G. Kelly, R. Santacruz, E. Wilson-Ewing, Black hole collapse and bounce in effective loop quantum gravity, arXiv:2006.09325.






 





\end{thebibliography}
\end{document}